\pdfoutput=1
\documentclass[twocolumn,prd,floatfix,nofootinbib,longbibliography,notitlepage,superscriptaddress]{revtex4-2}

\usepackage{amsmath}
\usepackage{amssymb}
\usepackage{bm}
\usepackage{subfigure}
\usepackage[usenames,svgnames,dvipsnames]{xcolor}

\newcommand{\ay}[1]{{\bf \color{magenta}{{#1}}}}

\DeclareMathOperator{\sech}{sech}
\newcommand{\Ren}{\text{R}}
\newcommand{\Sec}{\text{sec}}
\newcommand{\pd}{\partial}

\usepackage{graphicx}

\graphicspath{{./Figures/}}

\usepackage[T1]{fontenc}
\usepackage[utf8]{inputenc}
\usepackage{lmodern}

\usepackage[unicode, colorlinks, allcolors=blue!70!black, linktocpage, pdfusetitle]{hyperref}
\hypersetup{
    colorlinks=true,
    urlcolor=SteelBlue,
    linkcolor=red,
    citecolor=blue,
}
\usepackage[all]{hypcap}
\usepackage{orcidlink}
\usepackage{microtype}

\usepackage{tikz}
\usetikzlibrary{calc, decorations.markings, arrows.meta, intersections}
\ifdefined\myext
\usetikzlibrary{external}
\fi

\newcommand{\paramsolsmap}{%
\begin{figure*}[t!]
\begin{center}
\ifdefined\myext
  \tikzsetnextfilename{map-tikz}
  \input{Figures/map-tikz}
\else
  \ifx\relsstandalone\undefined
    \input{Figures/map-tikz}
  \else
    \includegraphics[width=.95\textwidth]{map-tikz}
  \fi
\fi
\end{center}
\vspace{-1em}
\caption{%
  Flows in parameter space $\Lambda$ are mapped to the flows in
solution space $\mathcal{S}$ via the solution map $\varphi^{(0)}$.  The flow
is an integral curve of the vector field $\vec{\beta}$ (in blue),
whereas the vector field $\vec{\alpha}$ (in orange) generates an 
infinitesimal transformation (diffeomorphism) redefining the initial 
coordinate.
  Every point $\varphi^{(0)}(\lambda) \in \mathcal{S}$ corresponds to an entire
solution, expanded at right.  The tangent space
$T_{\varphi^{(0)}(\lambda)}\mathcal{S}$ consists of homogeneous solutions
$\varphi^{(1)}$ to the differential equation linearized about
$\varphi^{(0)}(\lambda)$, which correspond to shifts in parameter space.\label{fig:map}}
\end{figure*}
}

\newcommand{\diffmap}{%
\begin{figure}[t!]
\begin{center}
\ifdefined\myext
  \tikzsetnextfilename{diff-map-tikz}
  \colorlet{betacolor}{NavyBlue}
\colorlet{alphacolor}{Orange}

\def \globalscale {0.6000}
\begin{tikzpicture}[y=0.80pt, x=0.80pt,
  yscale=\globalscale, xscale=\globalscale,
  inner sep=0pt, outer sep=0pt,
  basicstyle/.style={draw=black, thick},
  tangentstyle/.style={draw=red, thick},
  midarrow/.style={decoration={
      markings,
      mark=at position 0.45 with {\arrow{Latex[length=3.5mm]}}},
    postaction={decorate}},
  trajstyle/.style={draw=MidnightBlue, very thick, midarrow},
  mapstyle/.style={draw=black, thick, midarrow},
  targetstyle/.style={draw=ForestGreen, thick},
  lambda1style/.style={>=stealth,->, thick, color=purple},
  lambda2style/.style={>=stealth,->, thick, color=orange},
  varphistyle/.style={>=stealth,->, thick, color=blue},
  circstyle/.style={basicstyle, circle, minimum size=2mm, fill=white}
  ]
  
  \coordinate (center lambda) at (70.0,139.);

  \node at (70., 130.) [basicstyle,
    shape=rectangle,
    minimum width=100.,
    minimum height=117.,
    label=south:{\raisebox{-4mm}{$\Lambda$}}] {};

 \draw[style=dashed, color=red, thick] (70.,139.) circle (1.5cm);
 \draw(35.,98.,) node {$T_{\lambda_0}\Lambda$};
 \draw[lambda1style] (center lambda) -- ++(90.0:55.0);
 \draw[lambda2style] (center lambda) -- ++(0.0:55.0);
 \draw(48.0,211.5) node {$\frac{\vec{\partial}}{\partial\lambda^{1}}$};
 \draw(143.0,143.5) node {$\frac{\vec{\partial}}{\partial\lambda^{2}}$};

  \path[targetstyle, path picture={
    \draw[step=20,draw=ForestGreen!10]
    (300,-100) grid[rotate=10]
    (500,200);}]
    (255.6537,103.5555) .. controls (255.4804,83.3586) and (256.4160,61.0075) ..
    (267.5195,44.1357) .. controls (279.4817,25.9594) and (299.5403,8.4473) ..
    (321.2563,7.0733)
    node[below] {\raisebox{-4mm}{$\mathcal{S}$}}
    .. controls (345.1447,5.5620) and (371.5161,20.3156) ..
    (385.0357,40.0681) .. controls (400.1433,62.1409) and (384.3863,94.7447) ..
    (393.6087,119.8523) .. controls (405.4553,141.2143) and (429.9938,153.9116) ..
    (431.9884,176.3157) .. controls (435.6639,196.9300) and (433.3768,224.4705) ..
    (416.8061,240.2716) .. controls (398.5005,251.4130) and (369.1419,244.4354) ..
    (347.4156,236.4959) .. controls (316.3430,225.1409) and (288.0711,201.5991) ..
    (270.8426,173.3569) .. controls (258.4421,153.0291) and (255.8580,127.3663) ..
    (255.6537,103.5555) -- cycle;

  \path[trajstyle]
    (47.3,50.5) .. controls (47.3,50.5) and (51.3,88.5) ..
    (57.,104.) .. controls (73.,152.) and (92.5,185.5) ..
    (123,221.);

  \path[trajstyle]
    (321.2600,72.4113) .. controls (313.7709,106.4117) and (340.7678,160.3171) ..
    (370.8319,191.5466) .. controls (371.2932,190.8900) and (369.6757,190.3137) ..
    (369.0600,189.7313);

  \coordinate (top map start) at (center lambda);
  \coordinate (top map end)   at (336.5,141.7853);
  \coordinate (bot map start) at (97.5,139.);
  \coordinate (bot map end)   at (351.5,143.7853);
  \coordinate (end first vec) at (366.0442,146.995);

 \draw[style=dashed, color=red, thick] (336.,144.) circle (1.72cm);
 \draw(378.,229.,) node {$T_{\varphi(\vec{\lambda})}\mathcal{S}$};
 \draw[varphistyle] (336.5,141.7853) -- (358.5,193.7853);
 \draw(334.5,170.7853) node {\scriptsize $\varphi^{(1)}_{\parallel}$};
 \draw[lambda2style] (top map end) -- ++(10.0:30.0);
 \draw[lambda1style] (end first vec) -- ++(100.0:45.7853);
 \draw(360.5,125.7853) node {\scriptsize $V^{2}\frac{\delta\varphi^{(0)}}{\delta\lambda^{2}}$};
 \draw(399.5,173.7853) node {\scriptsize $V^{1}\frac{\delta\varphi^{(0)}}{\delta\lambda^{1}}$};

   \path[mapstyle]
    (bot map start) .. controls (129.6593,84.3606) and (188.6083,82.5344) ..
    (232.8016,86.3871)
    node[below=4mm, pos=0.7, fill=white] {$d\varphi^{(0)}: T_{\vec{\lambda}}\Lambda \to T_{\varphi(\vec{\lambda})}\mathcal{S} $}
    .. controls (262.5108,90.4800) and (289.5982,92.8928) ..
    (bot map end);

  \node[circstyle, label=left:{%
    \begin{minipage}{1cm}%
      \setlength{\jot}{-1pt}
      \begin{align*}%
        \scriptstyle ~( t_{0}&\scriptstyle{},
        \scriptstyle\vec{\lambda}+\varepsilon \vec{\alpha})%
      \end{align*}%
    \end{minipage}%
  }] at (top map start) {};
  \node[circstyle, label=left:{%
    \scriptsize$\scriptstyle\varphi(t_{0})$
  }]
  (top map end circ) at (top map end)   {};

\end{tikzpicture}
\else

  \ifx\relsstandalone\undefined
    \colorlet{betacolor}{NavyBlue}
\colorlet{alphacolor}{Orange}

\def \globalscale {0.6000}
\begin{tikzpicture}[y=0.80pt, x=0.80pt,
  yscale=\globalscale, xscale=\globalscale,
  inner sep=0pt, outer sep=0pt,
  basicstyle/.style={draw=black, thick},
  tangentstyle/.style={draw=red, thick},
  midarrow/.style={decoration={
      markings,
      mark=at position 0.45 with {\arrow{Latex[length=3.5mm]}}},
    postaction={decorate}},
  trajstyle/.style={draw=MidnightBlue, very thick, midarrow},
  mapstyle/.style={draw=black, thick, midarrow},
  targetstyle/.style={draw=ForestGreen, thick},
  lambda1style/.style={>=stealth,->, thick, color=purple},
  lambda2style/.style={>=stealth,->, thick, color=orange},
  varphistyle/.style={>=stealth,->, thick, color=blue},
  circstyle/.style={basicstyle, circle, minimum size=2mm, fill=white}
  ]
  
  \coordinate (center lambda) at (70.0,139.);

  \node at (70., 130.) [basicstyle,
    shape=rectangle,
    minimum width=100.,
    minimum height=117.,
    label=south:{\raisebox{-4mm}{$\Lambda$}}] {};

 \draw[style=dashed, color=red, thick] (70.,139.) circle (1.5cm);
 \draw(35.,98.,) node {$T_{\lambda_0}\Lambda$};
 \draw[lambda1style] (center lambda) -- ++(90.0:55.0);
 \draw[lambda2style] (center lambda) -- ++(0.0:55.0);
 \draw(48.0,211.5) node {$\frac{\vec{\partial}}{\partial\lambda^{1}}$};
 \draw(143.0,143.5) node {$\frac{\vec{\partial}}{\partial\lambda^{2}}$};

  \path[targetstyle, path picture={
    \draw[step=20,draw=ForestGreen!10]
    (300,-100) grid[rotate=10]
    (500,200);}]
    (255.6537,103.5555) .. controls (255.4804,83.3586) and (256.4160,61.0075) ..
    (267.5195,44.1357) .. controls (279.4817,25.9594) and (299.5403,8.4473) ..
    (321.2563,7.0733)
    node[below] {\raisebox{-4mm}{$\mathcal{S}$}}
    .. controls (345.1447,5.5620) and (371.5161,20.3156) ..
    (385.0357,40.0681) .. controls (400.1433,62.1409) and (384.3863,94.7447) ..
    (393.6087,119.8523) .. controls (405.4553,141.2143) and (429.9938,153.9116) ..
    (431.9884,176.3157) .. controls (435.6639,196.9300) and (433.3768,224.4705) ..
    (416.8061,240.2716) .. controls (398.5005,251.4130) and (369.1419,244.4354) ..
    (347.4156,236.4959) .. controls (316.3430,225.1409) and (288.0711,201.5991) ..
    (270.8426,173.3569) .. controls (258.4421,153.0291) and (255.8580,127.3663) ..
    (255.6537,103.5555) -- cycle;

  \path[trajstyle]
    (47.3,50.5) .. controls (47.3,50.5) and (51.3,88.5) ..
    (57.,104.) .. controls (73.,152.) and (92.5,185.5) ..
    (123,221.);

  \path[trajstyle]
    (321.2600,72.4113) .. controls (313.7709,106.4117) and (340.7678,160.3171) ..
    (370.8319,191.5466) .. controls (371.2932,190.8900) and (369.6757,190.3137) ..
    (369.0600,189.7313);

  \coordinate (top map start) at (center lambda);
  \coordinate (top map end)   at (336.5,141.7853);
  \coordinate (bot map start) at (97.5,139.);
  \coordinate (bot map end)   at (351.5,143.7853);
  \coordinate (end first vec) at (366.0442,146.995);

 \draw[style=dashed, color=red, thick] (336.,144.) circle (1.72cm);
 \draw(378.,229.,) node {$T_{\varphi(\vec{\lambda})}\mathcal{S}$};
 \draw[varphistyle] (336.5,141.7853) -- (358.5,193.7853);
 \draw(334.5,170.7853) node {\scriptsize $\varphi^{(1)}_{\parallel}$};
 \draw[lambda2style] (top map end) -- ++(10.0:30.0);
 \draw[lambda1style] (end first vec) -- ++(100.0:45.7853);
 \draw(360.5,125.7853) node {\scriptsize $V^{2}\frac{\delta\varphi^{(0)}}{\delta\lambda^{2}}$};
 \draw(399.5,173.7853) node {\scriptsize $V^{1}\frac{\delta\varphi^{(0)}}{\delta\lambda^{1}}$};

   \path[mapstyle]
    (bot map start) .. controls (129.6593,84.3606) and (188.6083,82.5344) ..
    (232.8016,86.3871)
    node[below=4mm, pos=0.7, fill=white] {$d\varphi^{(0)}: T_{\vec{\lambda}}\Lambda \to T_{\varphi(\vec{\lambda})}\mathcal{S} $}
    .. controls (262.5108,90.4800) and (289.5982,92.8928) ..
    (bot map end);

  \node[circstyle, label=left:{%
    \begin{minipage}{1cm}%
      \setlength{\jot}{-1pt}
      \begin{align*}%
        \scriptstyle ~( t_{0}&\scriptstyle{},
        \scriptstyle\vec{\lambda}+\varepsilon \vec{\alpha})%
      \end{align*}%
    \end{minipage}%
  }] at (top map start) {};
  \node[circstyle, label=left:{%
    \scriptsize$\scriptstyle\varphi(t_{0})$
  }]
  (top map end circ) at (top map end)   {};

\end{tikzpicture}
  \else
    \includegraphics[width=.45\textwidth]{diff-map-tikz}
  \fi
\fi
\end{center}
\vspace{-1em}
\caption{%
  Illustration of the differential map $d\varphi^{(0)}:T\Lambda\to T\mathcal{S}$.
The parameter space coordinate basis vectors $\pd/\pd\lambda^{i}$ can
be pushed forward to span the tangent space at
$T_{\varphi(\lambda)}S$, giving the basis functions
$\delta\varphi^{(0)}/\delta\lambda^{i}$.
The first order solution has a projection into the tangent space,
$\varphi^{(1)}_{\parallel} = \varphi^{(1)}-\varphi^{(1)}_{\perp}$;
this projection is decomposed with the basis functions, yielding the
components of $\vec{\alpha}$ and $\vec{\beta}$.
\vspace{-1em}
\label{fig:diff-map}}
\end{figure}
}

\newcommand{\figbackfull}{%
\begin{figure*}[t!]
\centering
\subfigure{
\includegraphics[width=.36\textwidth]{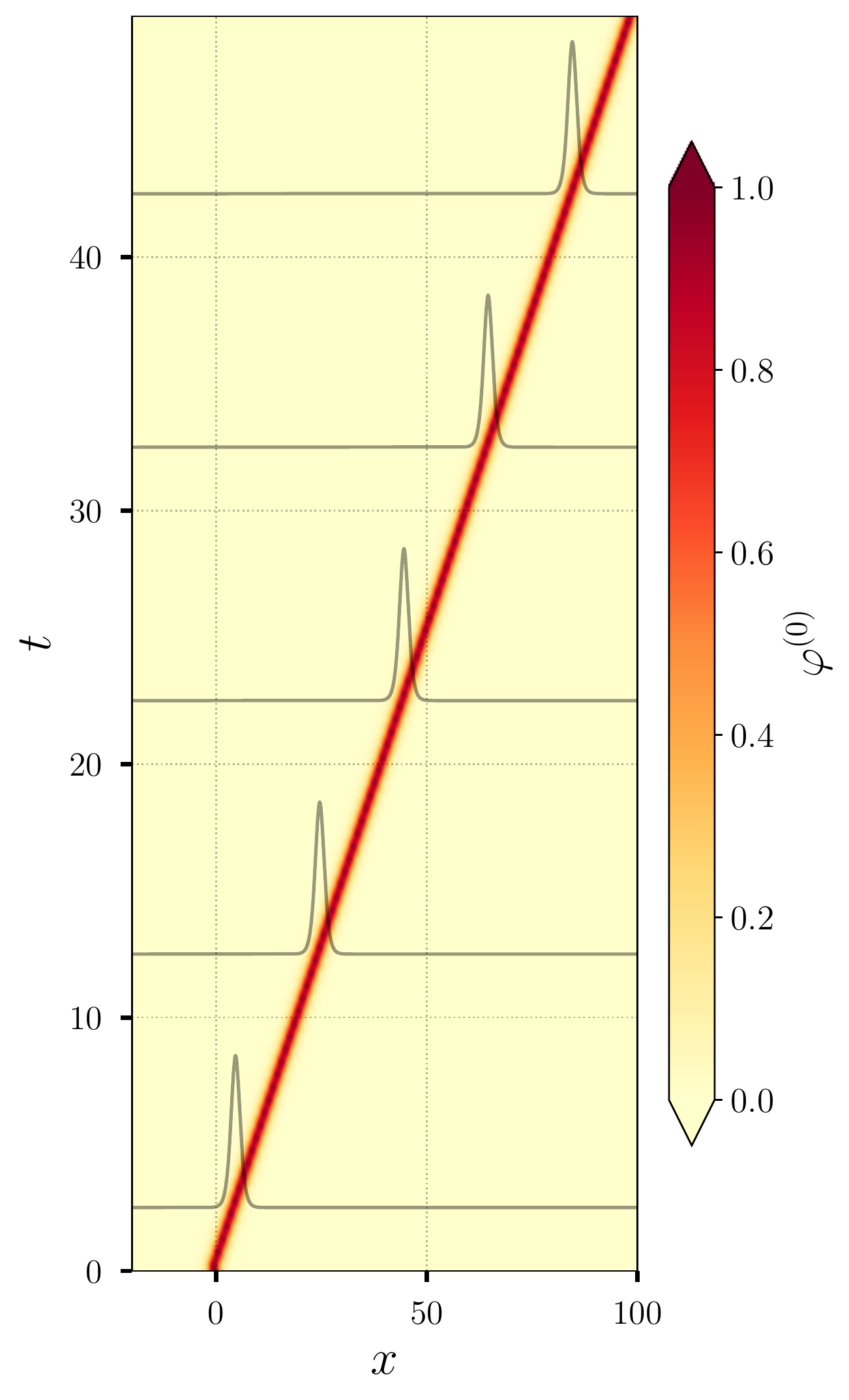}} \,
\subfigure{
\includegraphics[width=.355\textwidth]{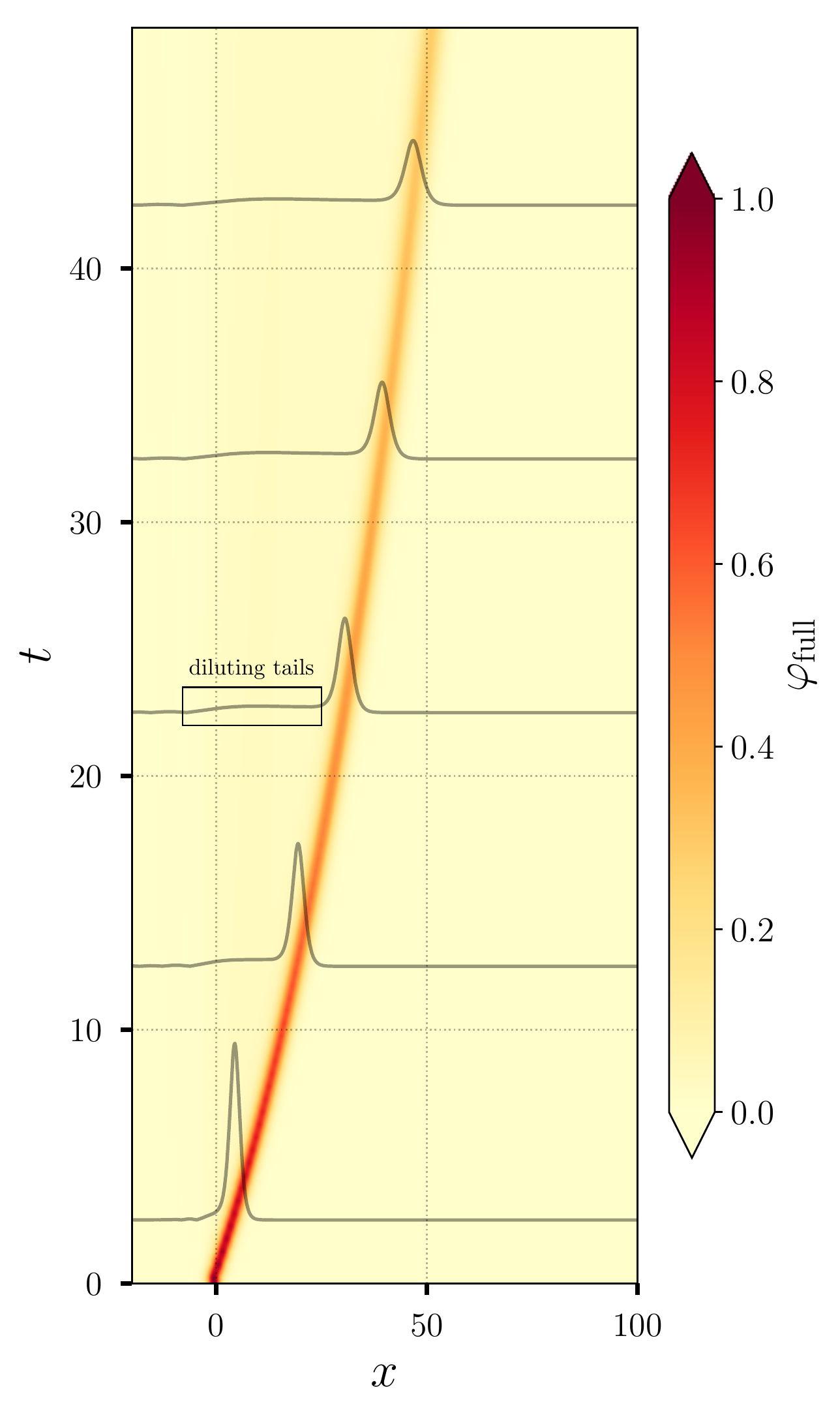}}
\caption{\label{fig:kdv_sols} Left panel: Density plot and
  constant-time profiles (in black) of the analytic soliton
  solution in Eq.~\eqref{eq:kdv_soliton} for $v=2$. Right panel: Numerical
  solution of the KdVB equation for $\varepsilon=0.1$ using as an
  initial condition the same $v=2$ KdV soliton plotted in the left
  panel.
  The full solution can be approximated by
  a decelerating soliton with decreasing amplitude and increasing
  width, which can be captured by a renormalized solution
  $\varphi^{(0)}(\vec{\lambda}_{\Ren})$.
  There is also a small step in the full
  solution (dubbed as ``diluting tails'' in the black rectangle), which 
  makes the solution asymmetric.  This tail is present in the
  residual $\varphi^{(1)}_{\perp}$ [see Eq.~\eqref{eq:difference_op}
  and Fig.~\ref{fig:fitting_error_pert}], and could be used to improve
  the renormalized solution.%
  \vspace{-1em}
  }
\end{figure*}
}

\newcommand{\tabAlphaBetaExtracted}{%
\begin{table*}[t!]
\centering
\scalebox{1.05}{
  \setlength{\tabcolsep}{4pt}
  \begin{tabular}{l|ccc|ccc|r}
    \hline\hline \noalign{\smallskip}
  {$v$} & {$\beta^v$} & {$\sigma_{\beta^v}$ (conv.)} &{$\sigma_{\beta^v}$ (RE)}& {$\alpha^v$} & {$\sigma_{\alpha^v}$ (conv.)} &{$\sigma_{\alpha^v}$ (RE)} & {$T_{\max}$}\\
  \noalign{\smallskip} \hline \noalign{\smallskip}
    0.0625 & $-1.010 \times 10^{-3}$          & $4.352 \times 10^{-8}$ & $3.050 \times 10^{-5}$ & $4.482 \times 10^{-2}$ & $4.287 \times 10^{-5}$ & $6.484 \times 10^{-3}$ & 2400\\
    0.125  & $-4.184 \times 10^{-3}$          & $2.976 \times 10^{-7}$ & $2.170 \times 10^{-6}$ & $8.828 \times 10^{-2}$ & $1.809 \times 10^{-4}$ & $2.780 \times 10^{-4}$ & 1500\\
    0.25   & $-1.667 \times 10^{-2}$          & $9.255 \times 10^{-7}$ & $2.064 \times 10^{-5}$ & $1.334 \times 10^{-1}$ & $7.459 \times 10^{-4}$ & $2.688 \times 10^{-3}$ & 2000\\
    0.5    & $-6.672 \times 10^{-2}$          & $1.236 \times 10^{-6}$ & $6.887 \times 10^{-6}$ & $1.887 \times 10^{-1}$ & $2.980 \times 10^{-4}$ & $4.469 \times 10^{-3}$ & 600\\
    0.75   & $-1.500 \times 10^{-1}$          & $9.266 \times 10^{-7}$ & $3.667 \times 10^{-4}$ & $2.311 \times 10^{-1}$ & $1.117 \times 10^{-4}$ & $5.953 \times 10^{-3}$ & 300\\
    1.0    & $-2.667 \times 10^{-1}$          & $1.654 \times 10^{-6}$ & $2.250 \times 10^{-5}$ & $2.668 \times 10^{-1}$ & $1.991 \times 10^{-4}$ & $4.499 \times 10^{-3}$ & 300\\
    1.25   & $-4.167 \times 10^{-1}$          & $2.587 \times 10^{-6}$ & $1.729 \times 10^{-5}$ & $2.982 \times 10^{-1}$ & $3.116 \times 10^{-4}$ & $3.629 \times 10^{-3}$ & 300\\
    1.5    & $-6.000 \times 10^{-1}$          & $3.724 \times 10^{-6}$ & $1.406 \times 10^{-5}$ & $3.237 \times 10^{-1}$ & $4.490 \times 10^{-4}$ & $3.098 \times 10^{-3}$ & 300\\
    1.75   & $-8.167 \times 10^{-1}$          & $5.064 \times 10^{-6}$ & $1.160 \times 10^{-5}$ & $3.530 \times 10^{-1}$ & $6.115 \times 10^{-4}$ & $2.666 \times 10^{-3}$ & 300\\
    2.0    & $-1.067 \times 10^{0\phantom{-}}$ & $6.599 \times 10^{-6}$ & $9.526 \times 10^{-6}$ & $3.774 \times 10^{-1}$ & $7.989 \times 10^{-4}$ & $2.338 \times 10^{-3}$ & 300\\
  \noalign{\smallskip} \hline\hline
\end{tabular}}
\caption{\label{tab:beta_v} Values of $\beta^v$ and $\alpha^v$
  extracted at different
  values of $v$, as depicted in Fig.~\ref{fig:beta_v}, and their
  corresponding uncertainties.
  Convergence errors are denoted $\sigma(\text{conv.})$ (see
  Appendix~\ref{app:numerical}), and Richardson extrapolation errors
  are denoted $\sigma(\text{RE})$, as depicted in
  Fig.~\ref{fig:2d_alpha_beta}.  The values of $T_{\max}$ satisfy
  the condition for the soliton to translate much more than one width,
  $T_{\max}\gg v^{-3/2}$.}
\end{table*}
}
\newcommand{\tabResolutions}{%
\begin{table}[t!]
\renewcommand{\arraystretch}{1.1}
\setlength{\tabcolsep}{0.3em} 
\centering
\begin{tabular}{ c|c|c|c|c }
 \hline\hline \noalign{\smallskip}
  & \multicolumn{2}{c|}{$v\geq 0.5$} & \multicolumn{2}{c}{$v< 0.5$}\\
  \hline
  Resolutions & $\Delta t$ & {\# of nodes} & $\Delta t$ & {\# of nodes} \\
  \hline
  high (hi)           & {$0.00005$} & {$2^{14}$} & {$0.0001$} & {$2^{14}$}\\
  mid (mi)            & {$0.0001$}  & {$2^{14}$} & {$0.001$}  & {$2^{13}$}\\
  low (low)           & {$0.001$}   & {$2^{13}$} & {$0.002$}  & {$2^{13}$}\\
  ultra-low (ult.low) & {$0.002$}   & {$2^{13}$} & {$0.01$}   & {$2^{12}$}\\
  \noalign{\smallskip}\hline\hline
\end{tabular}
\caption{\label{tab:resolutions} Resolutions and their corresponding
  values for the time step and the number of collocation points in
  Fourier grid. These are the same resolution levels used to produce
  the convergence errors represented in Fig.~\ref{fig:beta_v}.}
\end{table} 
}

\newcommand{\figextinvTmax}{%
\begin{figure}[t!]
\centering
\includegraphics[width=.47\textwidth]{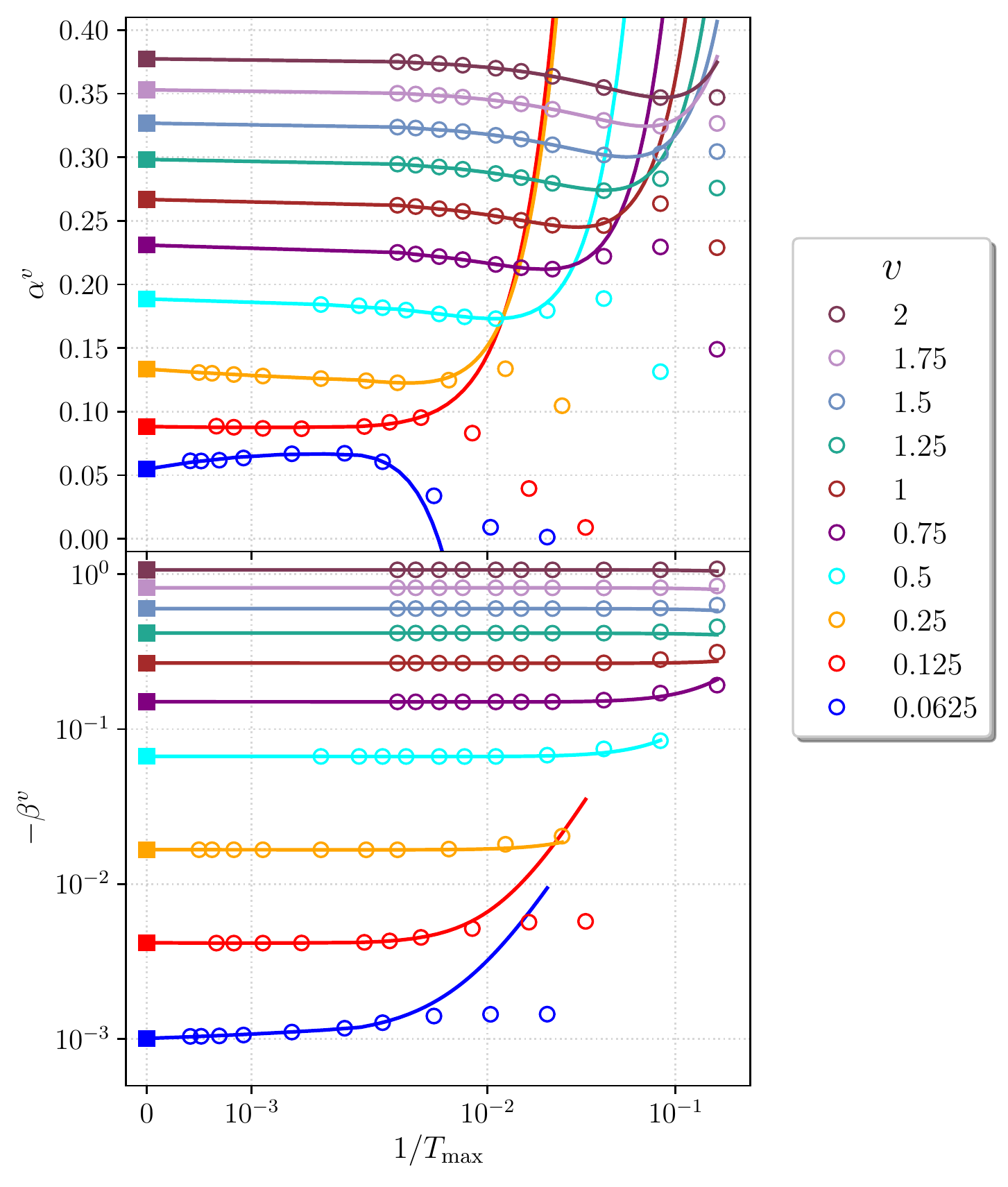}
\caption{\label{fig:2d_alpha_beta} Quadratic RE in the limit $1/T_{\max}\to 0$.
The extrapolated values (filled squares) correspond to the values of
$\alpha^v$ and $\beta^v$ in Table~\ref{tab:beta_v}.
The extrapolating polynomial (solid line) is a quadratic in powers of
$T_{\max}^{-1}$, hence the unphysical blowup to the right.}
\end{figure}
}

\newcommand{\tabAlphaBetaFit}{%
  \begin{table}
  \scalebox{1.05}{
  \renewcommand{\arraystretch}{1.1}
\setlength{\tabcolsep}{0.4em} 
   \begin{tabular}{c|cc}
      \hline\hline \noalign{\smallskip}
      & $m$ & $b$ \\
      \noalign{\smallskip} \hline \noalign{\smallskip}
      $\beta^v$ & $1.99976 \pm 9.3\times 10^{-5}$  & $-1.32158 \pm 6.0\times 10^{-5}$ \\
      $\alpha^v$ & $0.52537 \pm 2.2\times 10^{-3}$ & $-1.33342 \pm 3.6\times 10^{-3}$ \\
      \noalign{\smallskip} \hline\hline
    \end{tabular}
  }
  \caption{\label{tab:alpha_beta_fit}%
    Linear regression coefficients and errors for $\ln |\beta^v|$ and
    $\ln \alpha^v$ as functions of $\ln v$.}
  \end{table}
}

\newcommand{\fignaivepert}{%
\begin{figure*}
\centering
\subfigure{
\includegraphics[height=4.in]{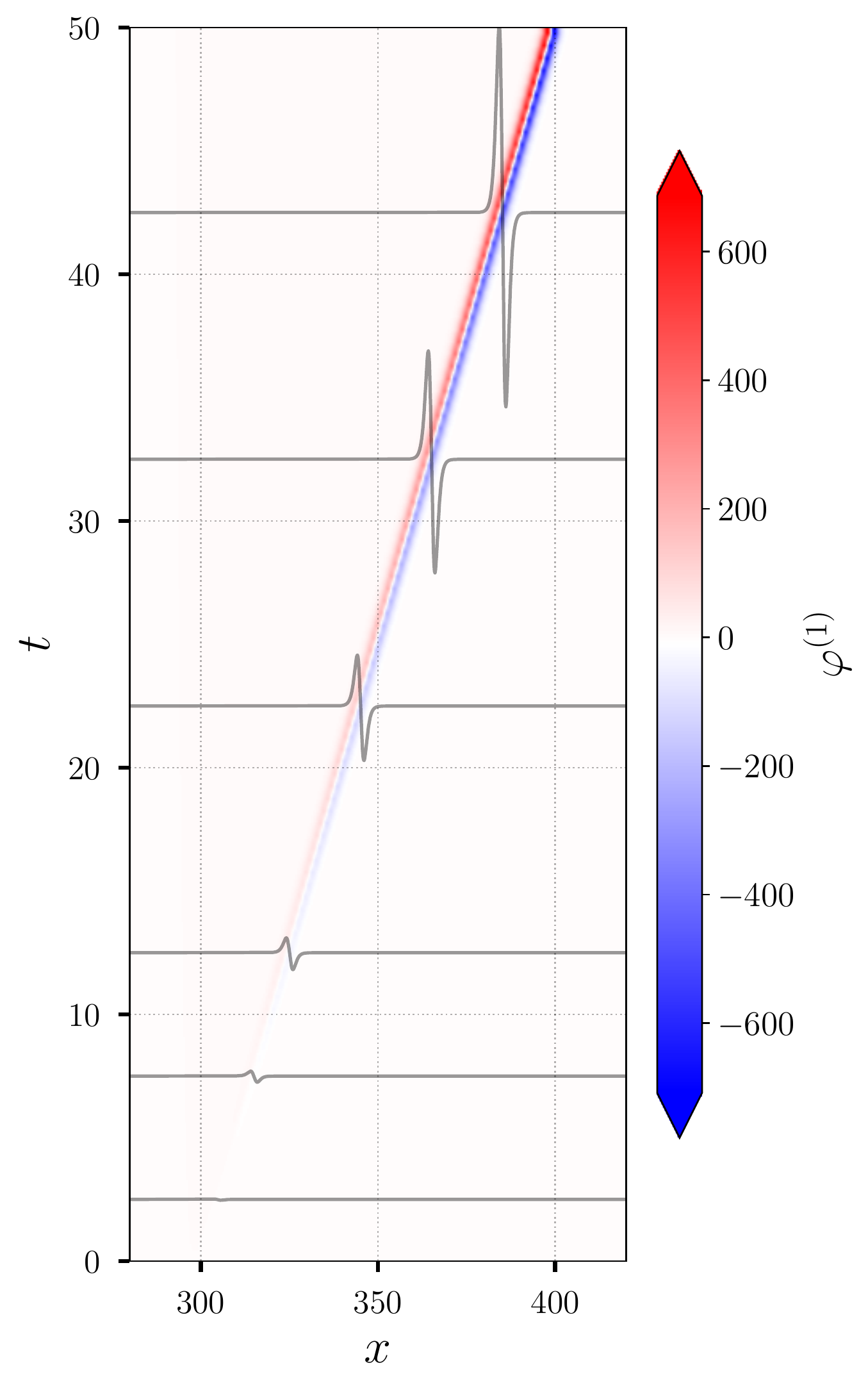}} \,
\subfigure{
\includegraphics[height=4.in]{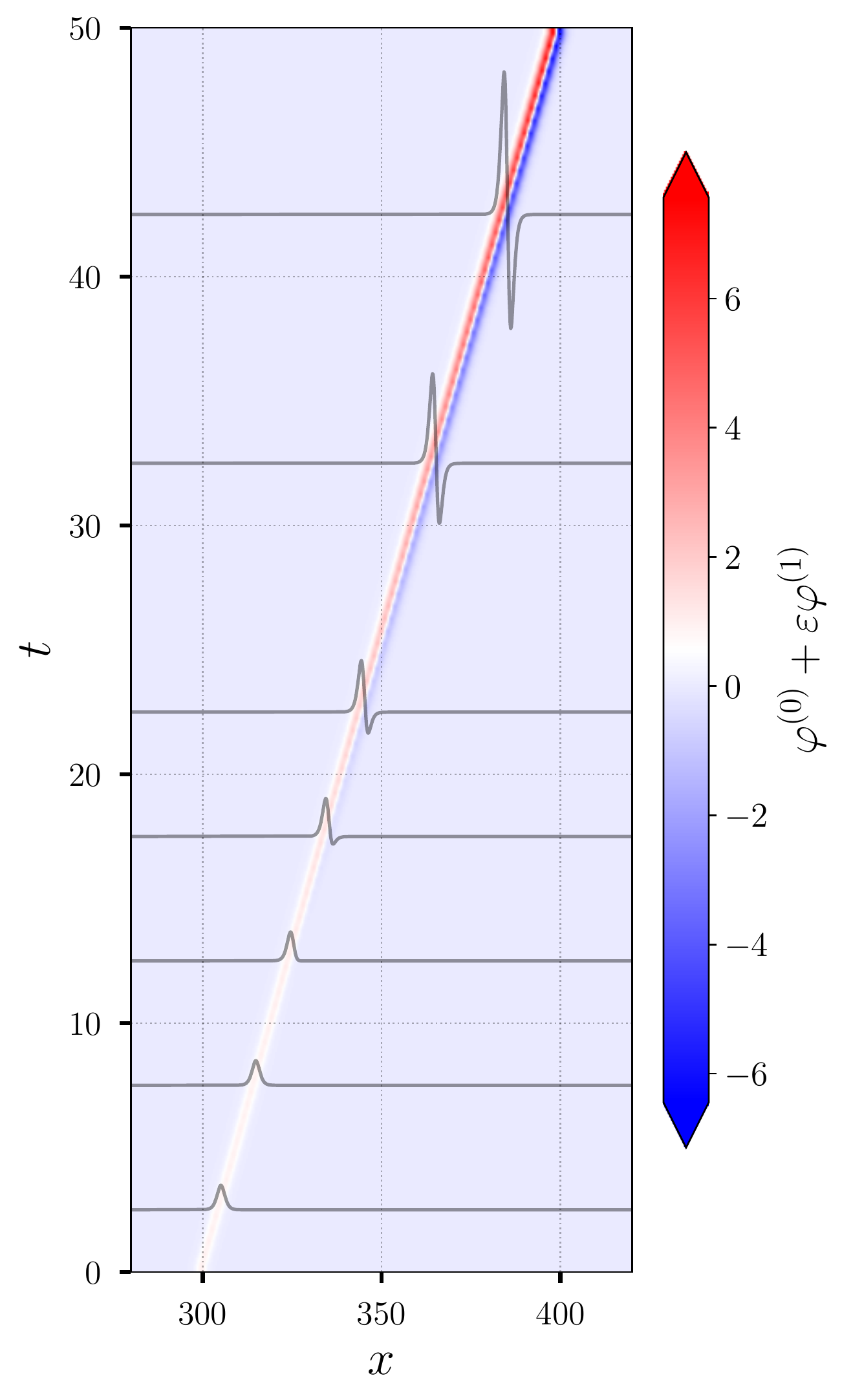}}
\caption{\label{fig:kdv_pert} Left panel: Density plot and
  constant-time profiles (in black) with the solution of the
  perturbation equation in Eq.~\eqref{eq:kdv_pert} for $v=2$. Right panel:
  Calculation of field at the first-order expansion in
  Eq.~\eqref{eq:sol_pert_m} for $\varepsilon=0.01$ following the standard
  prescription for perturbation theory. Secular divergences are visible
  in the amplitude at times $t\sim 1/\varepsilon$. In this figure, the
  range of the $x$-axis
  is different from Fig.~\ref{fig:kdv_sols} (this is simply a
  shift, allowed by translation invariance of the KdVB equation).
  }
\end{figure*}
}

\newcommand{\figalphabetav}{%
\begin{figure*}[t!]
\centering
\subfigure{
\includegraphics[width=.47\textwidth]{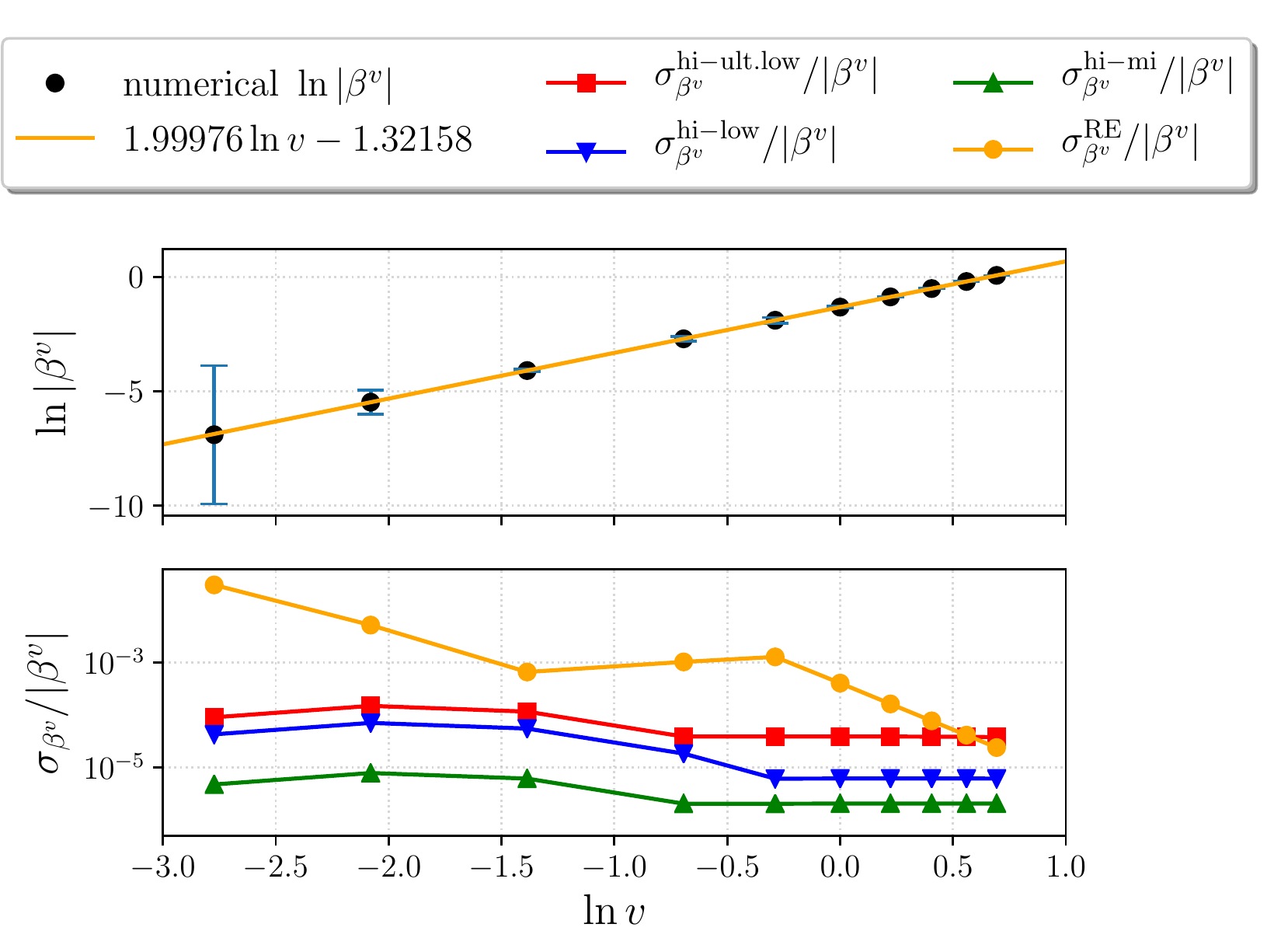}}\,
\subfigure{
\includegraphics[width=.45\textwidth]{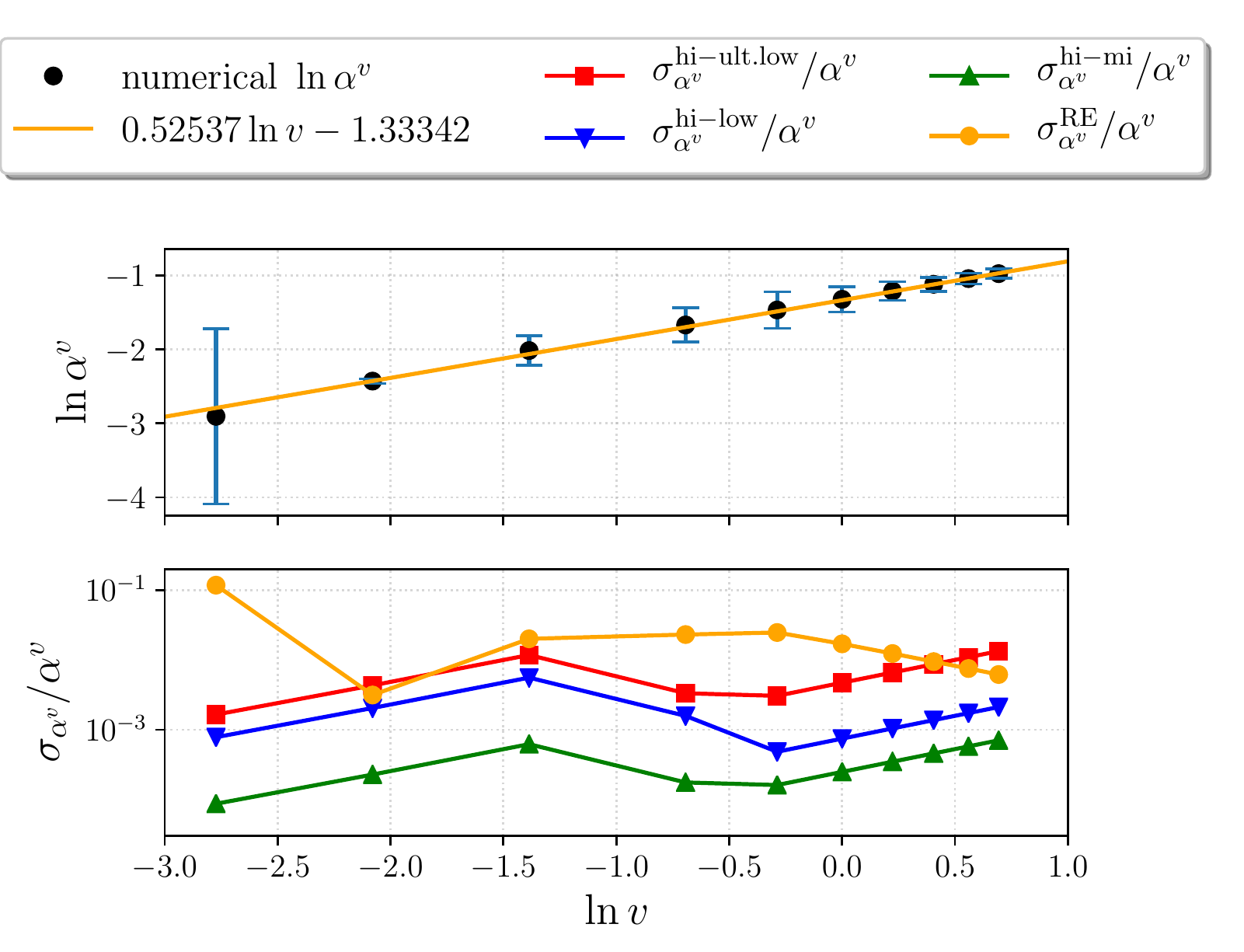}}
\caption{\label{fig:beta_v}
  Power law fits to the Richardson-extrapolated values of $\beta^v$ and
  $\alpha^v$, as a function of $v$.
  Upper panels: Best fit for the alpha and beta functions (in orange)
  with enlarged error bars (in blue), estimated as due to RE.
  Error bars were enlarged by a factor of $10^2$ for $\ln|\beta^v|$,
  and by a factor of $10$ for $\ln \alpha^v$.
  Lower panels: Magnitude of the fractional errors from different
  sources. $\sigma^{\text{RE}}$ is the difference between the values
  of $\alpha^v$ and $\beta^v$ extracted by RE, and the values at
  finite $T_{\max}$ (the last column of
  Table~\ref{tab:beta_v}).  Convergence errors are obtained by
  comparing the extracted quantities at four different resolutions,
  named high (hi), mid (mi), low (low), and ultra-low (ult.low). The
  error bars in the upper panels are $\sigma^{\text{RE}}$,
  which are larger than convergence errors.}
\end{figure*}
}

\newcommand{\figlnpart}{%
\begin{figure}[t!]
\centering
\includegraphics[width=.38\textwidth]{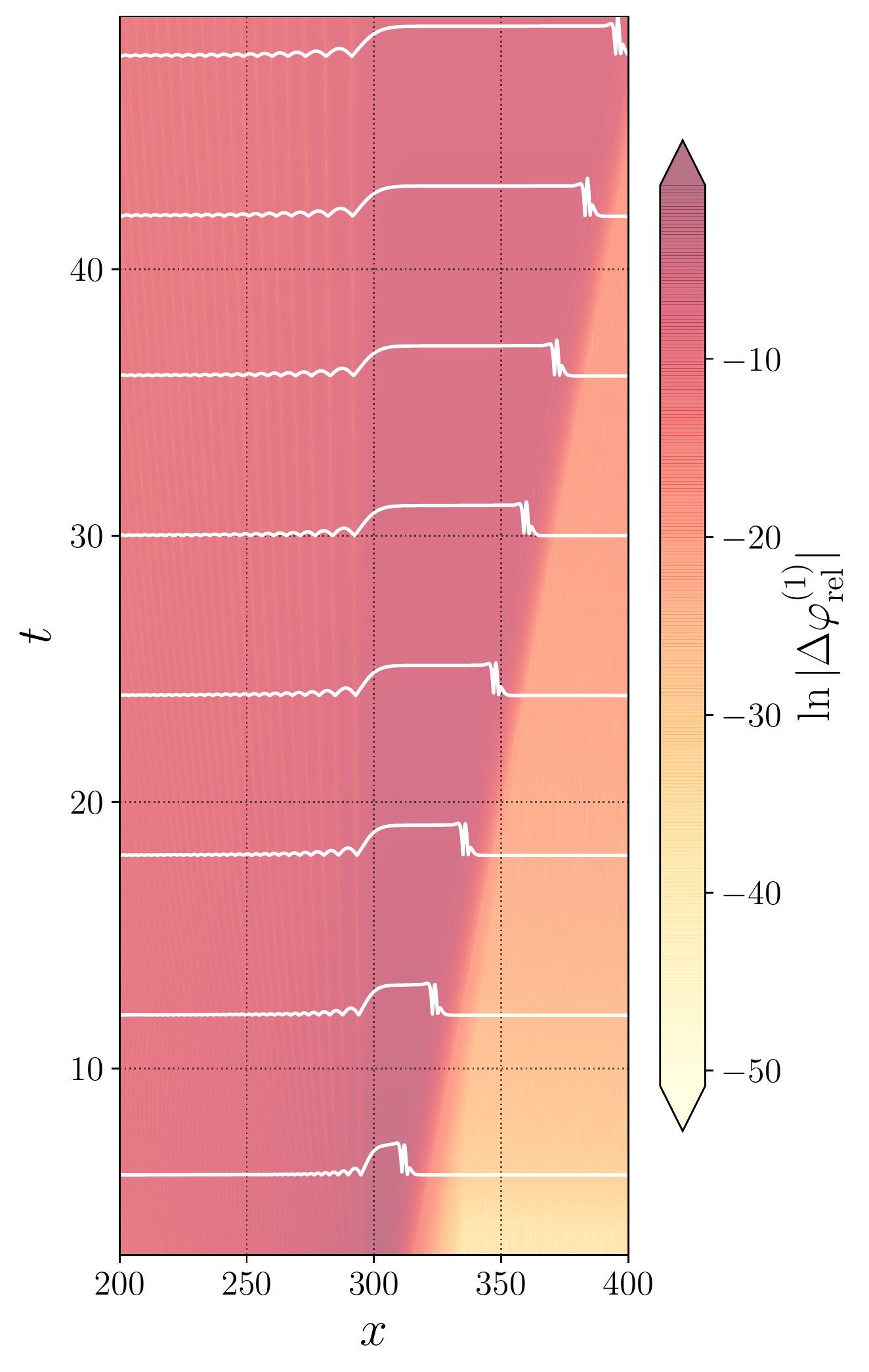}
\caption{\label{fig:fitting_error_pert} Color: The relative difference
  $\Delta\varphi^{(1)}_{\mathrm{rel}}$ for $v=2$ remains 
  small for the duration of the simulation. White curves:
  The residual $|\varphi^{(1)}_{\perp}|$ (not scaled by the max)
  at different times, growing in spatial extent.
  This feature coincides with the bump due to the ``diluting tails''
  in the right panel of Fig.~\ref{fig:kdv_sols}.}
\end{figure}
}

\newcommand{\figdeltaphirecon}{%
\begin{figure}[t!]
\centering
\includegraphics[width=.42\textwidth]{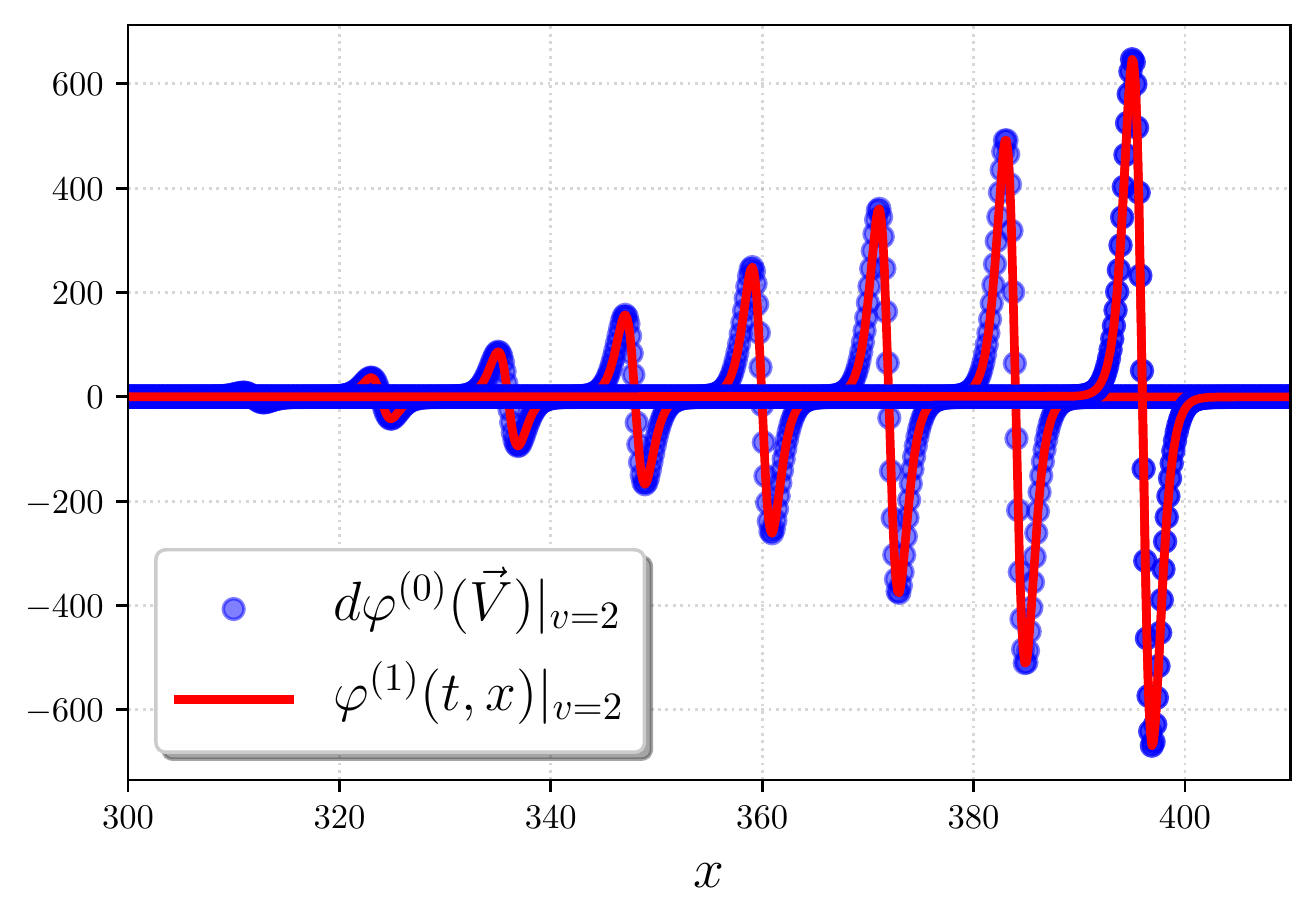}
\caption{\label{fig:fitting_pert}
  The perturbative solution $\varphi^{(1)}$ for $v=2$, and extracted
  $d\varphi^{(0)}(\vec{V})$,
  as functions of $(x,t)$, shown at different
  instants of time. The extracted values of $\alpha^v$ and $\beta^v$
  make $d\varphi^{(0)}(\vec{V})$ a good fit to the
  perturbative solution.}
\end{figure}
}

\newcommand{\figphirenres}{%
\begin{figure*}[t!]
\centering
\subfigure{
\includegraphics[width=.38\textwidth]{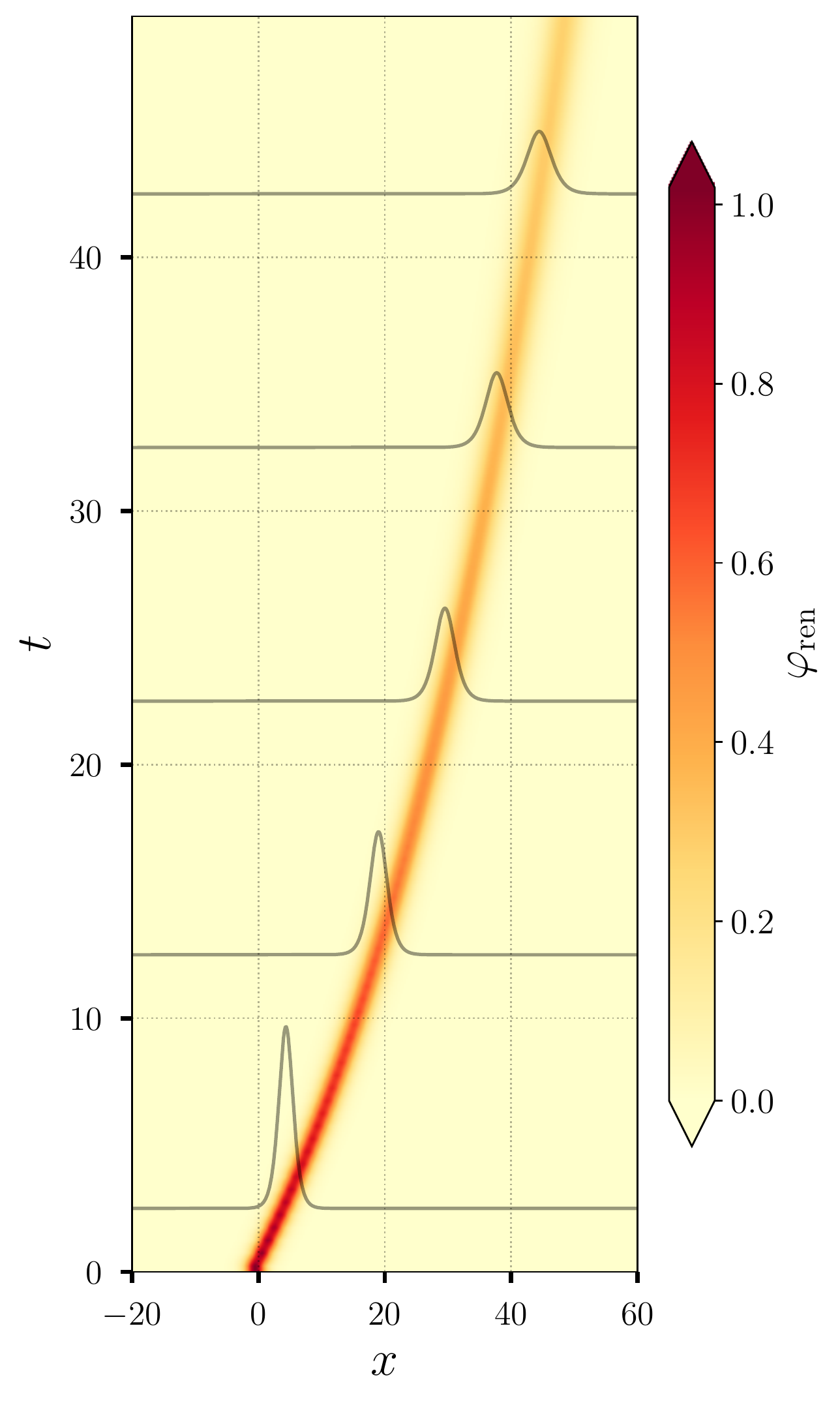}} \,
\subfigure{
\includegraphics[width=.38\textwidth]{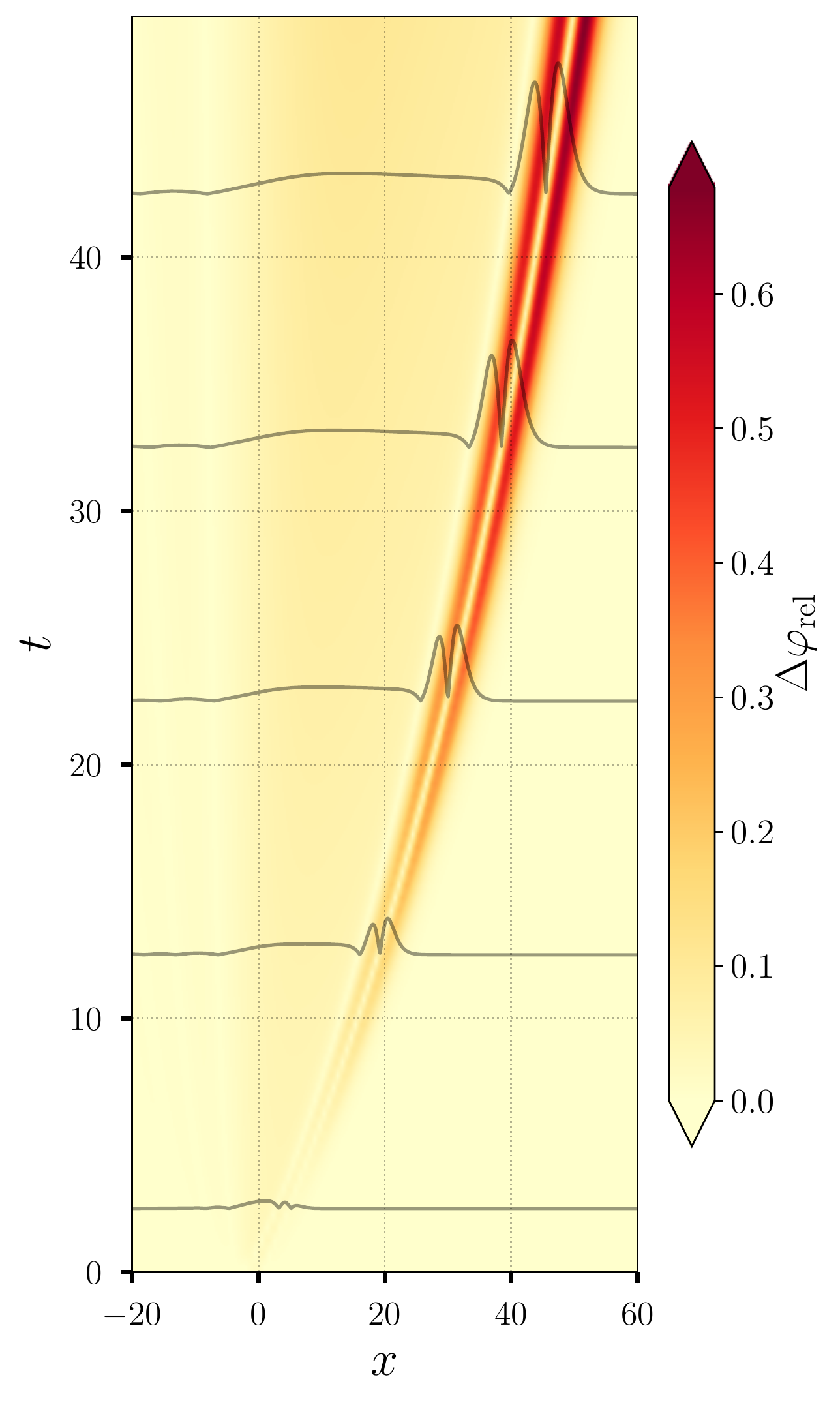}}
\caption{\label{fig:phi_ren_vs_full} Left panel: Renormalized solution
  in Eqs.~\eqref{eq:ren} for $v=2$ and $\varepsilon=0.1$
  built from the solution of the renormalized flow of
  Eqs.~(\ref{eq:flow_eq_v}--\ref{eq:new_peak}).
  Here the damping effects can be clearly noticed as the peak
  decelerates, its amplitude attenuates, and it
  becomes broader. Right panel: Fractional difference between the
  renormalized and full
  solutions.  The small difference between the
  position of the peaks is visible in the separation of the two red
  density profiles.  The peaks of
  the renormalized and full solutions have very similar
  amplitudes.  At $t=50$ (much longer than $\varepsilon^{-1/2}$),
  the distance between the peaks is roughly half a peak width.}
\end{figure*}
}

\newcommand{\figalphasbetasTs}{%
\begin{figure*}[t!]
\centering
\includegraphics[width=.85\textwidth]{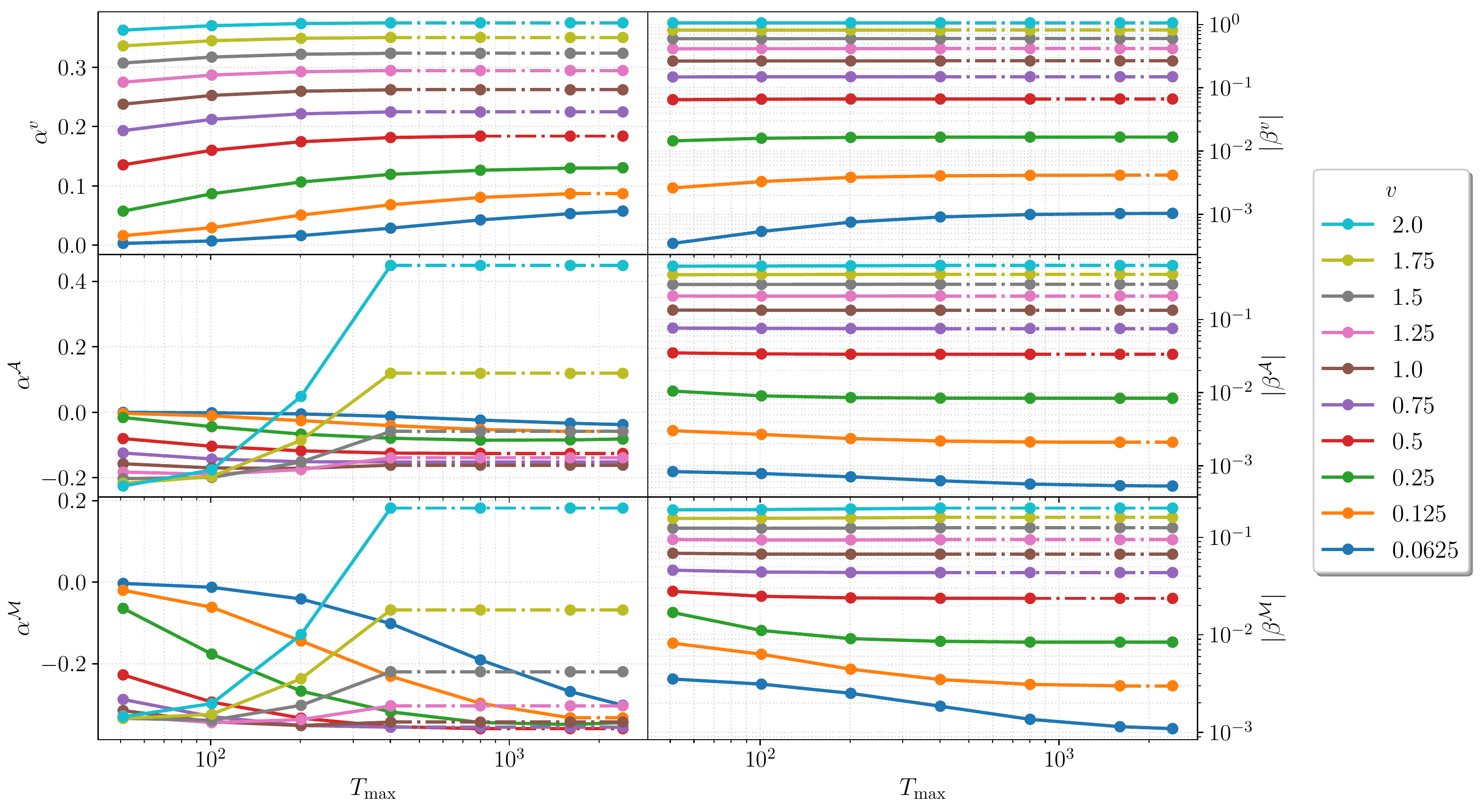}
\caption{\label{fig:all_alpha_beta}
  All the values of $\alpha$ and
  $\beta$ for different values of $T_{\max}$.  Dashed lines represent
  the intervals of $T_{\max}$ where the values of $\alpha$ or $\beta$
  have not been computed.  All the beta functions have converged to a fixed
  value before the dashed lines, but $\alpha^{\mathcal{A}}$ and
  $\alpha^{\mathcal{M}}$ have not converged to stable values at large
  $T_{\max}$.  This indicates that we have not found their correct
  time dependence at the infinitesimal level.}
\end{figure*}
}

\newcommand{\figsolparamevol}{%
\begin{figure*}[t!]
\centering
 \includegraphics[width=.80\textwidth]{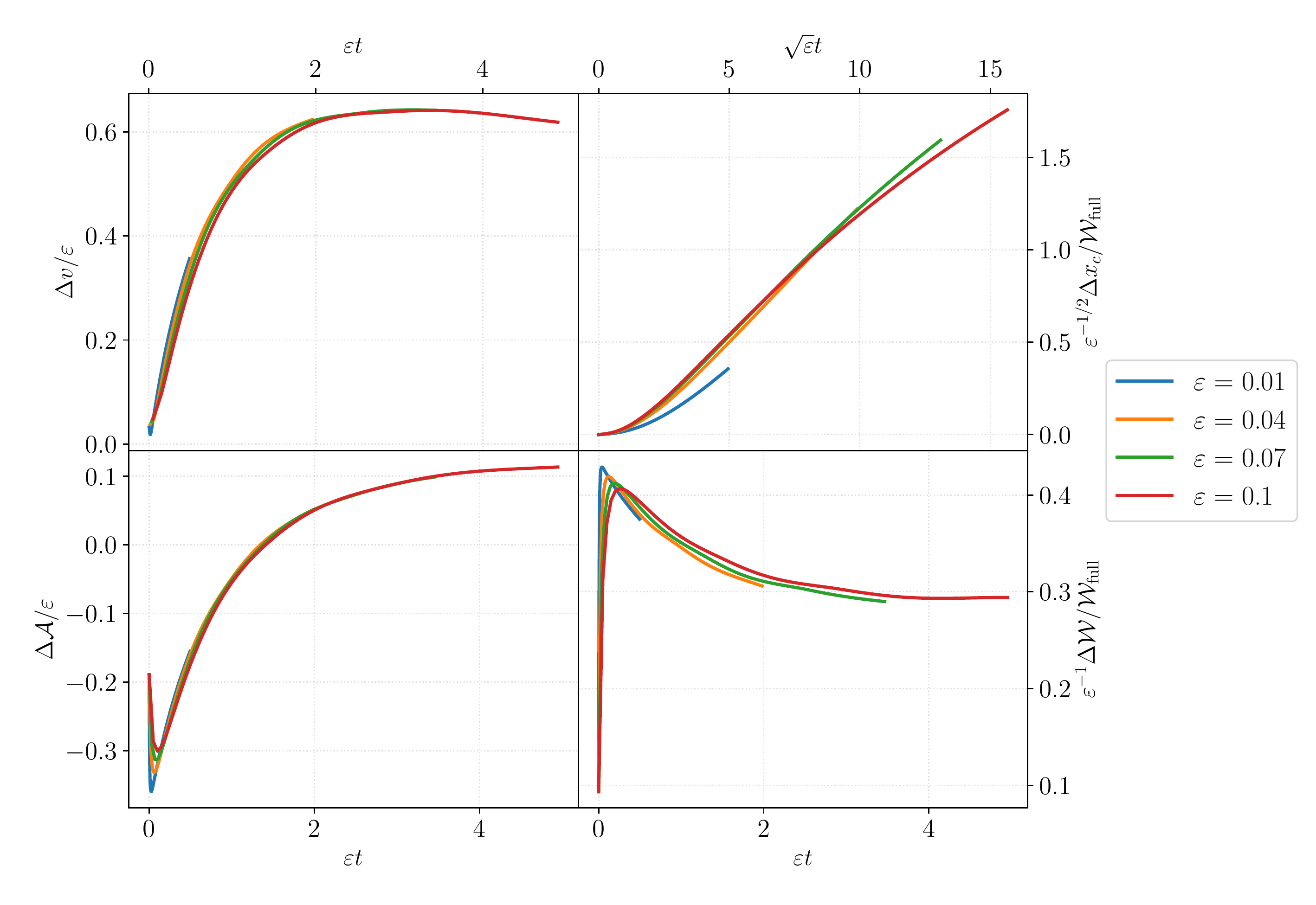}
\caption{\label{fig:error_param_unif}
  Comparing the shape parameters of the KdVB solution (for four
  $\varepsilon$ values) versus the parameters of the
  renormalized solution $\varphi_{\text{ren}}$, as functions of time.  The
  errors in peak velocity $\Delta v$ and amplitude
  $\Delta \mathcal{A}$ are bounded in time, and proportional to $\varepsilon$, even for times
  significantly larger than $1/\sqrt{\varepsilon}$.  The error in peak
  position $\Delta x_{c}$ (in units of width) is linear in time and linear in
  $\varepsilon$ at late times, due to the bounded velocity error.
  $\Delta \mathcal{W}/\mathcal{W}_{\text{full}}$ is linear in $\varepsilon$,
  bounded, and approaches a constant for $t\gg 1/\sqrt{\varepsilon}$.}
\end{figure*}
}

\newcommand{\figbetaspread}{%
\begin{figure}[t!]
\centering
\includegraphics[width=.45\textwidth]{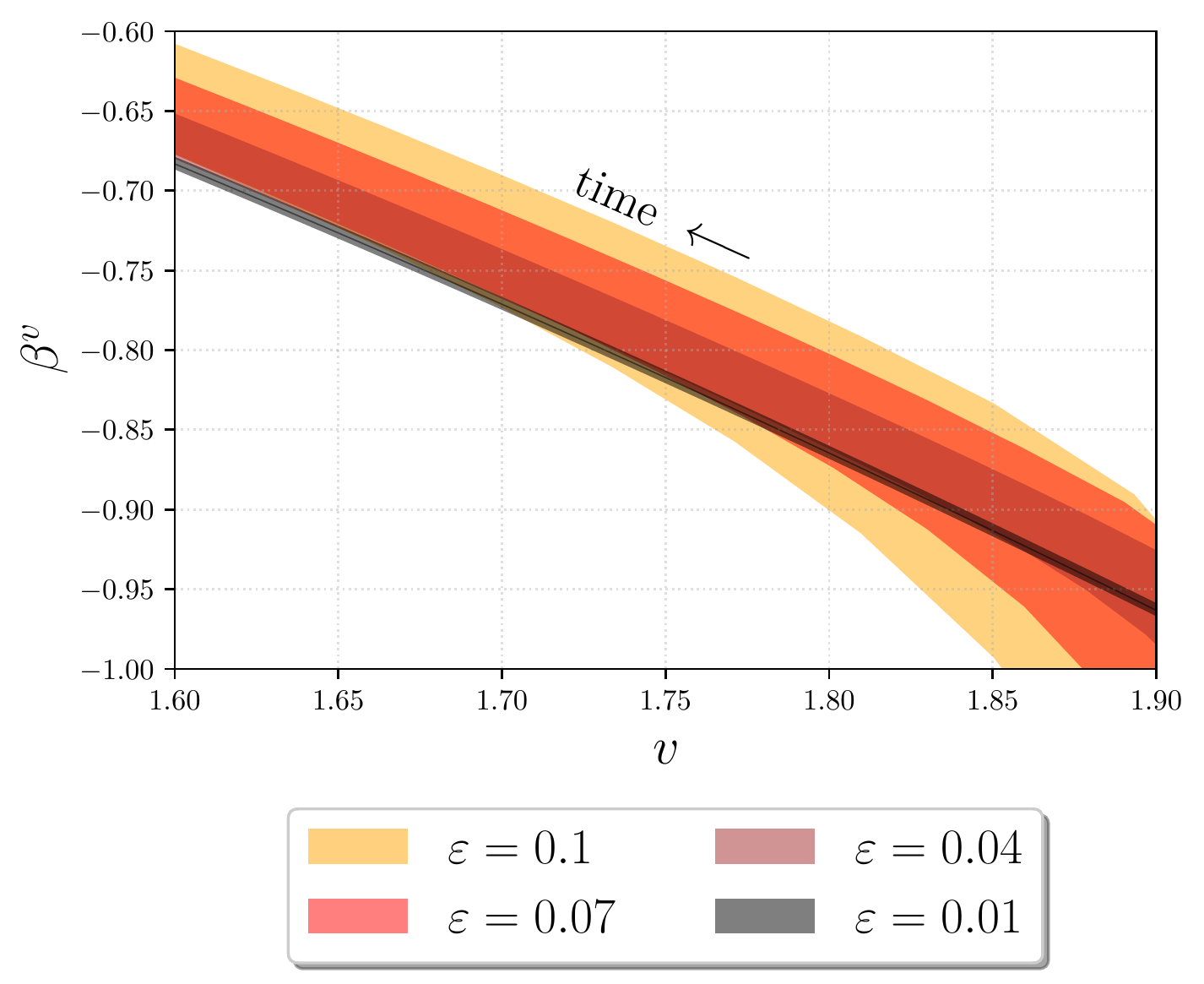}
\caption{\label{fig:beta_spread} Convergence of the beta function curves in 
  Eqs.~(\ref{eq:beta_v_cent}-\ref{eq:beta_v_a}) built from the different 
  shape parameters. The colored regions, containing all of the alternative 
  forms of the beta function, expand as $\varepsilon$ becomes larger.\vspace{-1em}}
\end{figure} 
}

\newcommand{\figbetaconv}{%
\begin{figure}[t!]
\centering
\includegraphics[width=.45\textwidth]{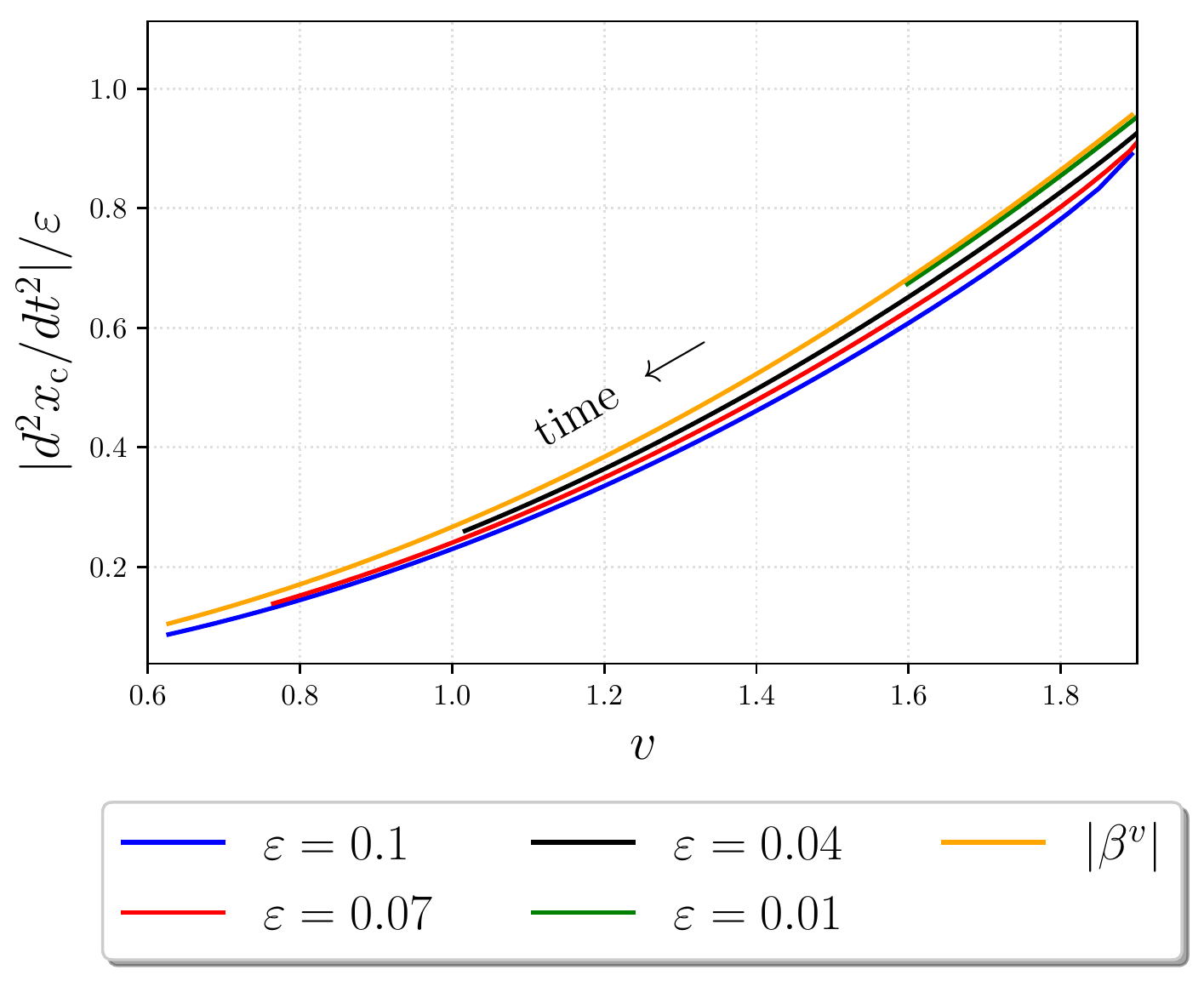} \,
\caption{\label{fig:beta_const}
  Using the kinematic definition of the
  peak velocity, we reconstruct the beta function from the peak
  position of the full solution $\varphi_{\text{full}}$, at different
  values of $\varepsilon$ (using Eq.~\eqref{eq:beta_v_cent}).  These
  converge to the DRG-computed $|\beta^v|$ (extracted in
  Sec.~\ref{subsec:single}) as $\varepsilon$ becomes smaller.
  We note that if the horizontal axis is reparameterized to be $v -
  \varepsilon \alpha^{v}_{\text{full}}(v)$, all of the curves coincide.
}
\end{figure} 
}

\newcommand{\figalphadiv}{%
\begin{figure}[t!]
\includegraphics[width=.45\textwidth]{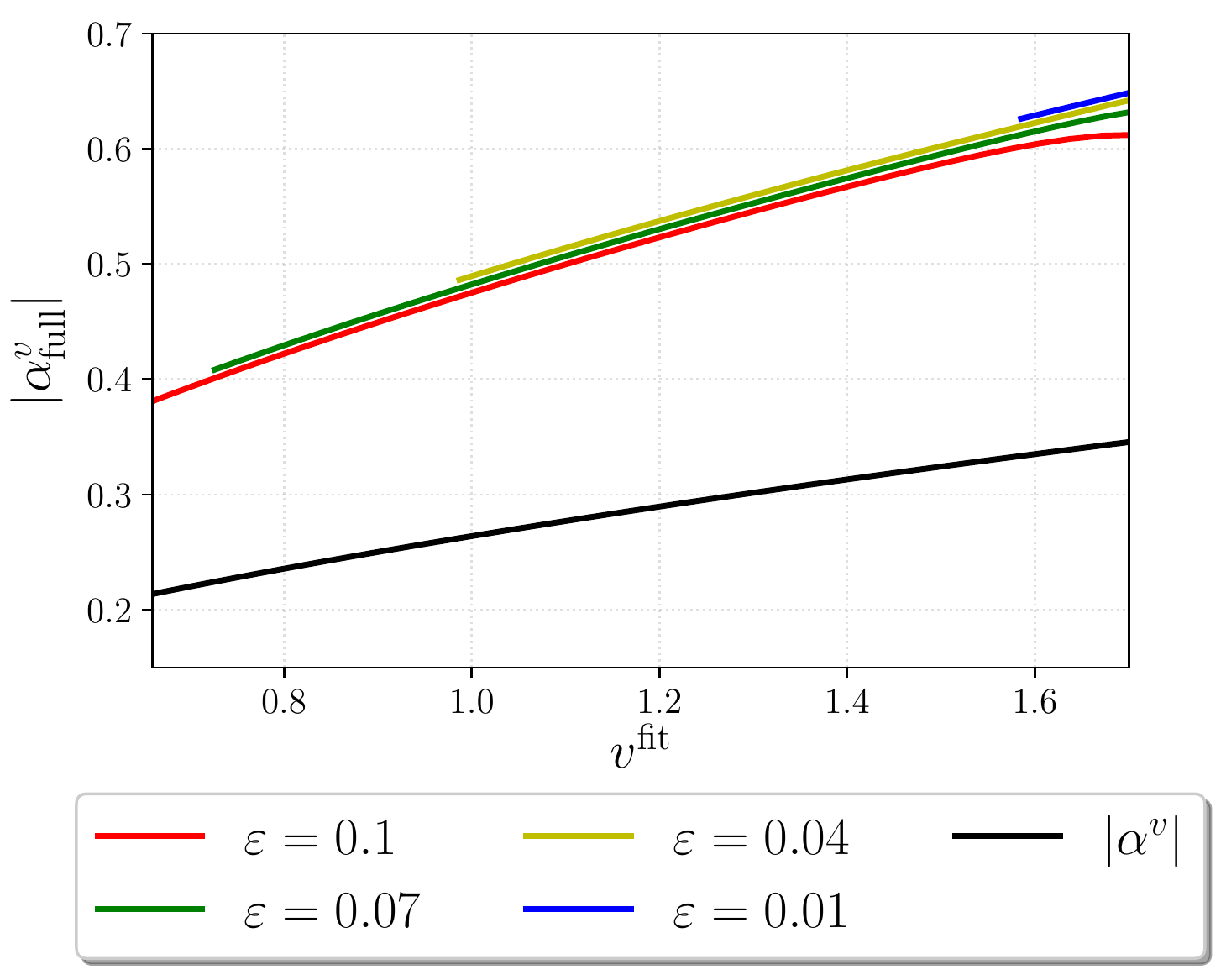}
\caption{\label{fig:alpha_err}%
  Mismatch between $\alpha^{v}$, extracted using the DRG, and
  $\alpha^{v}_{\text{full}}$, extracted by fitting the full solution
  (at four different values of $\varepsilon$).
  There is an unexplained ratio of approximately 2 between them.
  }
\end{figure}
}

\newcommand{\figevolscheme}{%
\begin{figure}[t!]
\centering
\includegraphics[width=.35\textwidth]{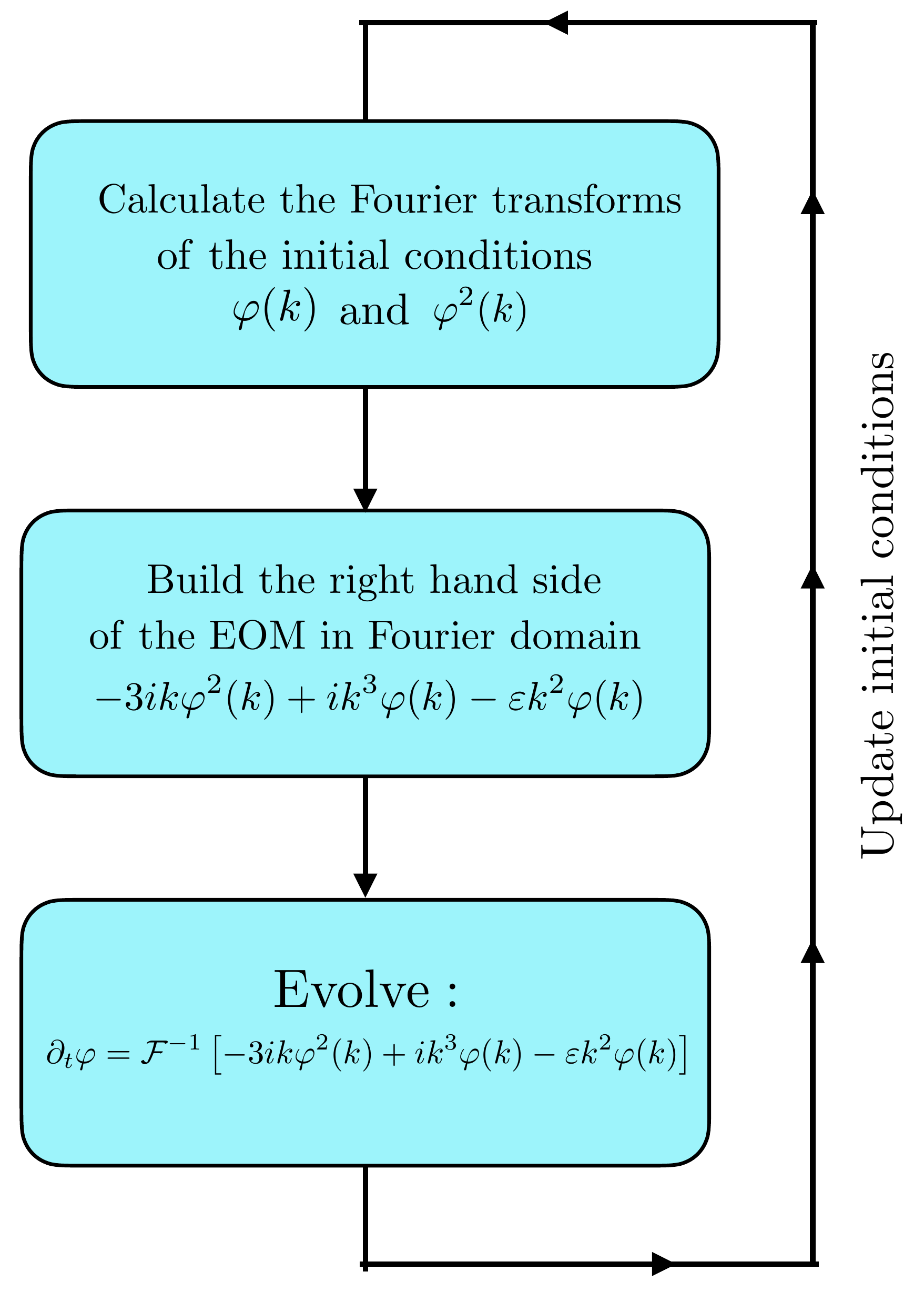}
\caption{\label{fig:scheme} Simplified scheme showing the solution
  algorithm for the KdVB equation in Eq.~\eqref{eq:kdvb}, the solution for
  the perturbation $\varphi^{(1)}$ follows a similar process. The
  symbol $\mathcal{F}^{-1}$ denotes the inverse Fourier transform.}
\end{figure}
}

\newcommand{\figlownoise}{%
\begin{figure*}
\centering
\subfigure{
\includegraphics[width=.46\textwidth]{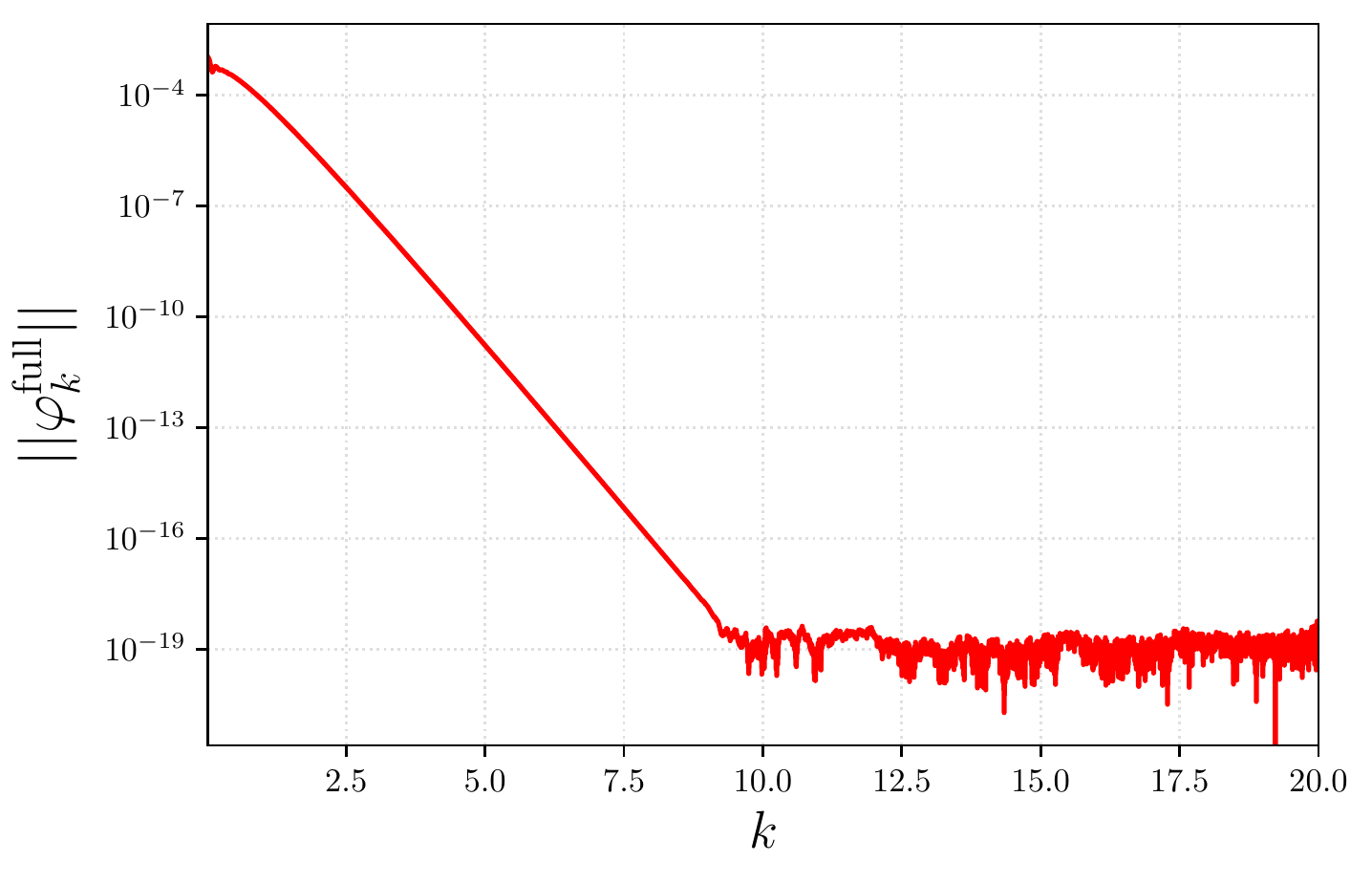}} \,
\subfigure{
\includegraphics[width=.468\textwidth]{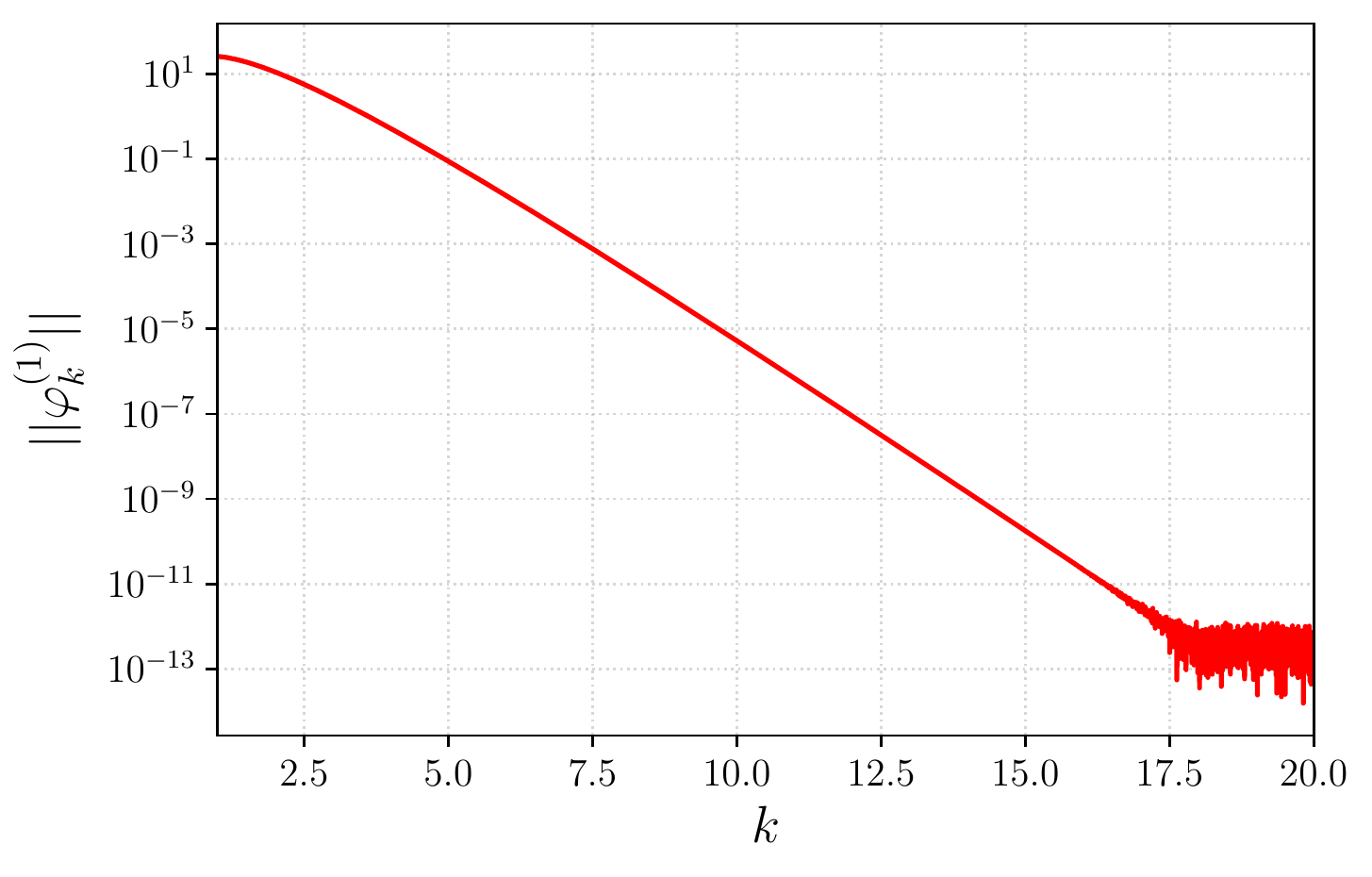}}
\caption{\label{fig:spec_conv} Left panel: Power spectrum of
  $\varphi_{\mathrm{full}}$ at $t=50$, corresponding to
  the final snapshot of the numerical evolution from
  Eq.~\eqref{eq:kdvb}. We considered a KdV soliton as an initial condition
  with $v=2$. Right panel: Power spectrum of the perturbation solution
  of Eq.~\eqref{eq:kdv_pert} at $T_{\max}=300$, where the background
  is a $v=2$ soliton.  In both panels, we observe
  that the high frequency contribution remains in the levels of
  round-off errors in double precision.}
\end{figure*} 
}

\newcommand{\figconvalphabetav}{%
\begin{figure}[t!]
\centering
\includegraphics[width=.38\textwidth]{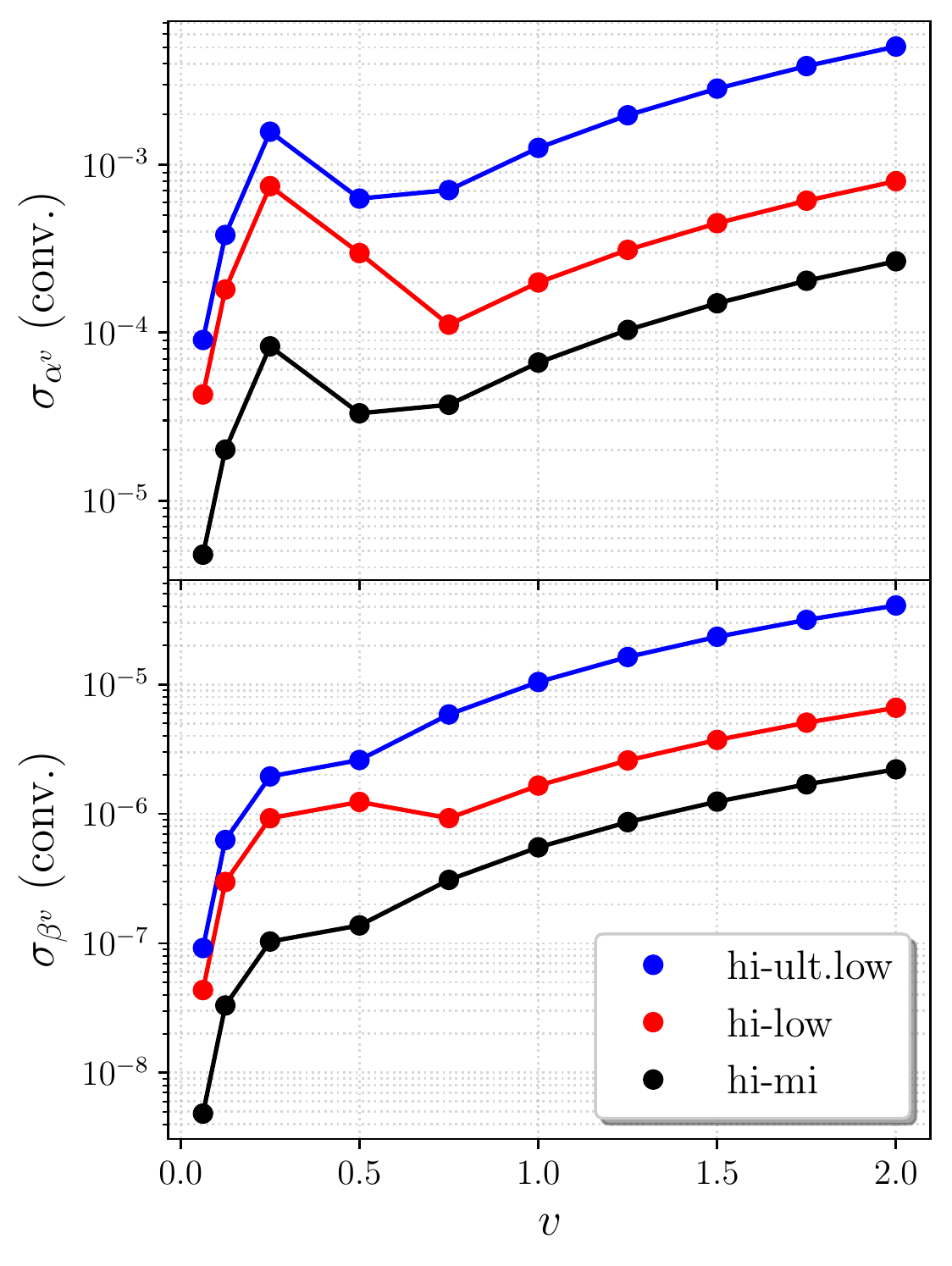}
\caption{\label{fig:conv_errors} Convergence plots for $\alpha^v$ and
  $\beta^v$ as functions of the solution parameter $v$. The differences
  between the values of $\beta^v$ and $\alpha^v$ reduce as the quantities are 
  extracted from better resolved datasets. The differences $\sigma_{\alpha^v}$ 
  and $\sigma_{\beta^v}$, reported in Table~\ref{tab:beta_v}, correspond to the 
  error curves in red.}
\end{figure}
}

\newcommand{\figpeakevolren}{%
\begin{figure}[t!]
\centering
\includegraphics[width=.45\textwidth]{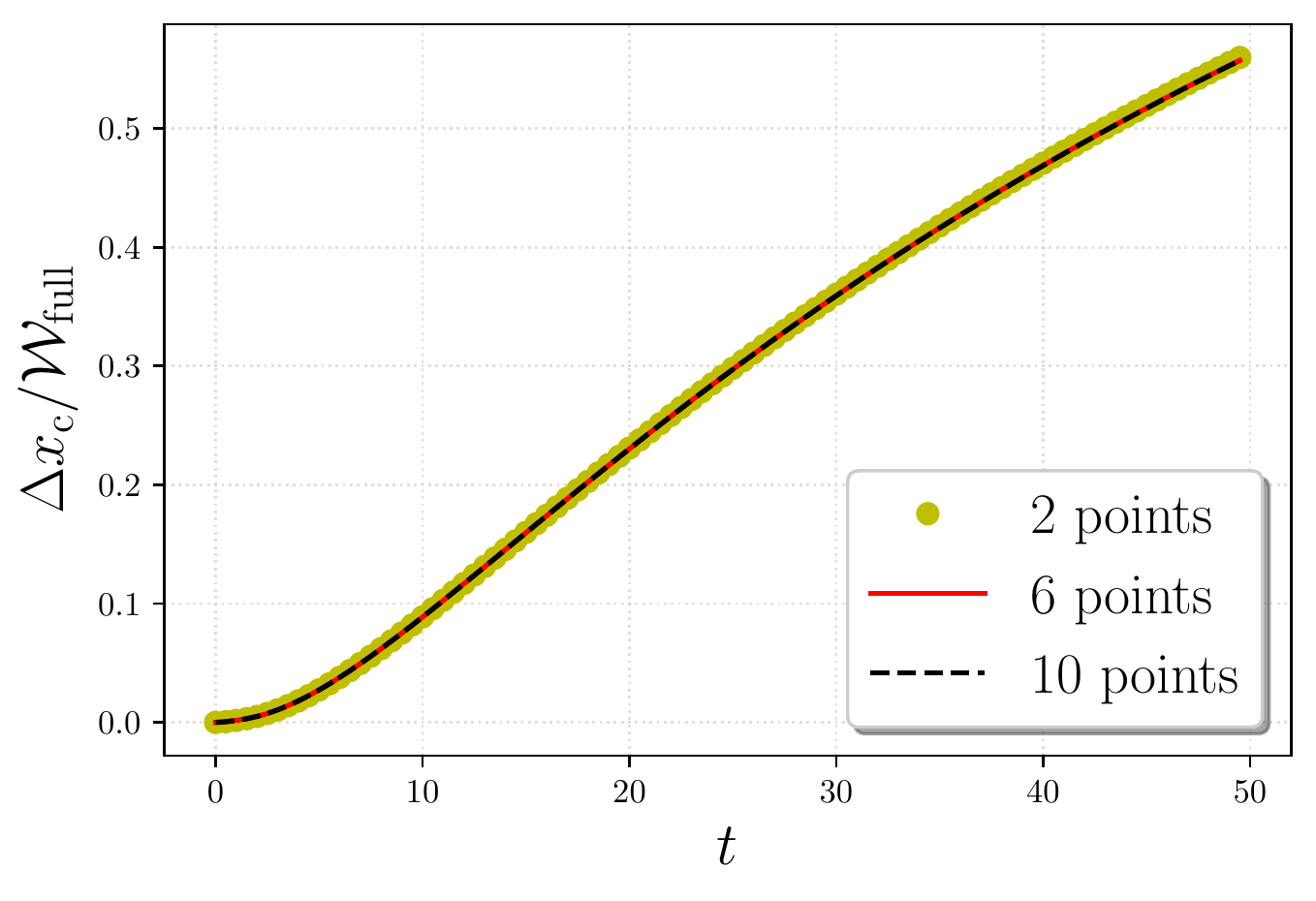}
\caption{\label{fig:error_xcs}
  Comparing the peak position of the
  renormalized solution with the peak position of
  $\varphi_{\mathrm{full}}$ in a ratio with the width. The
  renormalized solution is generated by picking 2, 6, and 10 different
  values of $\beta^v$ from Table~\ref{tab:beta_v} and
  $\alpha^v|_{v=2}$ at $\varepsilon=0.1$ to reparameterize the initial 
  velocity. Changing the number of points does not introduce any 
  significant difference in the error variable $\Delta x_{\text{c}}/\mathcal{W}_{\text{full}}$.
  \vspace{-1em}}
\end{figure}
}

\newcommand{\figtesttang}{%
\begin{figure}[t!]
\centering
\includegraphics[width=.385\textwidth]{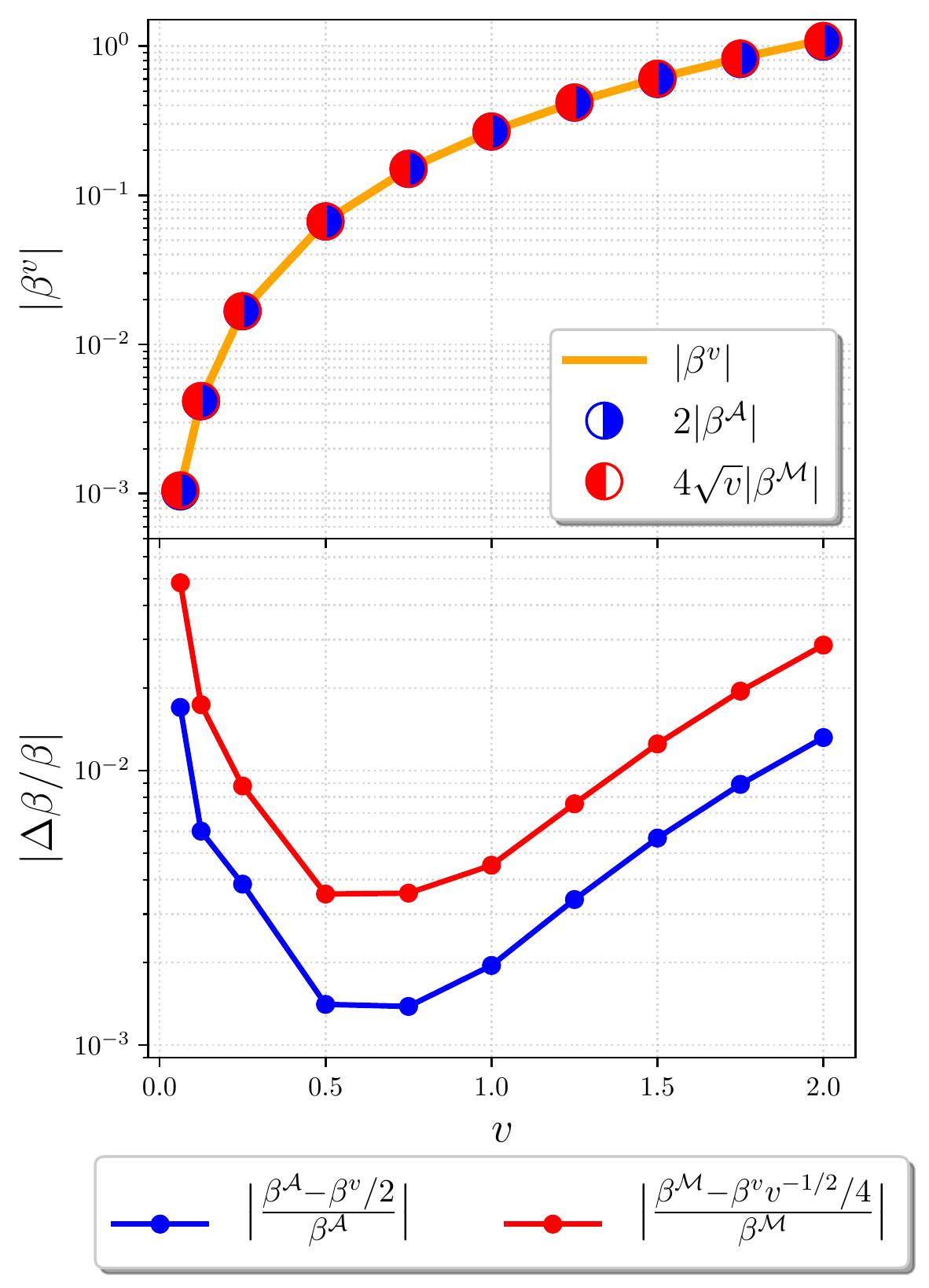}\vspace{-.7em}
\caption{\label{fig:a_w_tang} Testing tangent flows in parameter space
  as a function of $v$. In the upper panel, we observe that $\beta^v$,
  $2\beta^{\mathcal{A}}$, and $4\sqrt{v}\beta^{\mathcal{M}}$ are
  consistent with the relations in Eq.~\eqref{eq:check_1}.
  In the lower panel, we plot the relative deviations in the
  reconstructed $\beta^\mathcal{A}$ and $\beta^\mathcal{M}$ against
  the appropriate function of $\beta^v$ if the flow is tangent; the
  deviations are of the order of 1\%.
  \vspace{-1.3em}
}
\end{figure}
}

\newcommand{\figlnalpha}{%
\begin{figure}[t!]
\centering
\includegraphics[width=.45\textwidth]{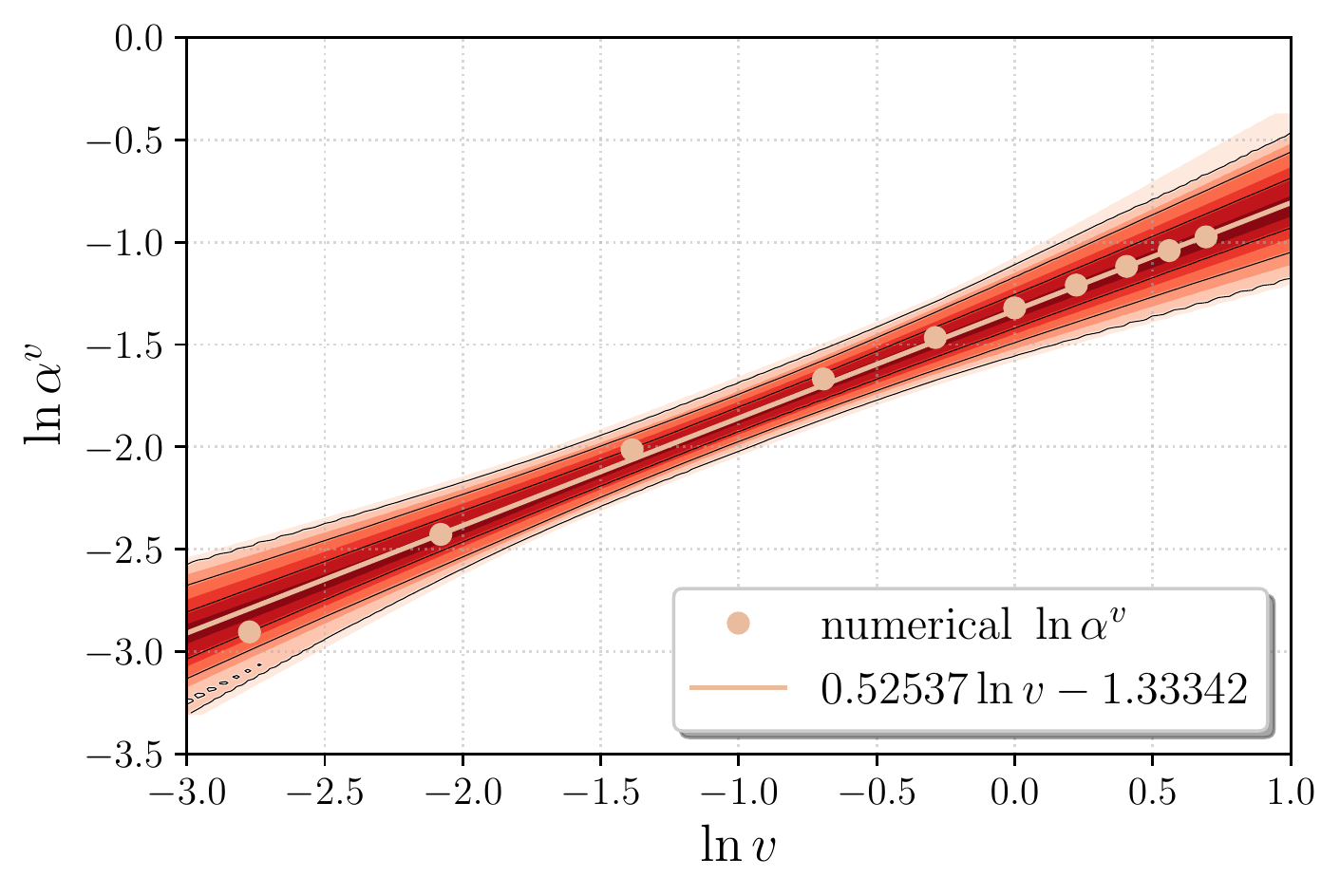} \,
\caption{\label{fig:beta_props} Using an enlarged version
  of the covariance matrix in Eq.~\eqref{eq:num_cov_beta} 
  (multiplied by $500$), we represent the region where the 
  extracted alpha function can be found.
  \vspace{-1em}}
\end{figure} 
}
\newcommand{\UMiss}{\affiliation{Department of Physics and Astronomy, The University of Mississippi, University, MS 38677, USA}}

\begin{document}

\title{Numerical renormalization group-based approach to secular perturbation theory}

\author{Jos\'e T.\ \surname{G\'alvez Ghersi}\,\orcidlink{0000-0001-7289-3846}}
\email{jgalvezg@cita.utoronto.ca}
\affiliation{Canadian Institute for Theoretical Astrophysics,
 University of Toronto, 60~St.~George Street, Toronto, ON M5S 3H8, Canada}
\UMiss
\author{Leo C.\ Stein\,\orcidlink{0000-0001-7559-9597}}
\email{lcstein@olemiss.edu}
\UMiss

\date{\today}

\hypersetup{pdfauthor={G\'alvez Ghersi and Stein}}

\begin{abstract}
  Perturbation theory is a crucial tool for many physical systems,
  when exact solutions are not available, or nonperturbative numerical
  solutions are intractable.
  Naive perturbation theory often fails on long timescales, leading
  to secularly growing solutions.
  These divergences have been treated with a variety of techniques,
  including the powerful dynamical renormalization group (DRG).
  Most of the existing DRG approaches rely on having analytic solutions
  up to some order in perturbation theory.
  However, sometimes the equations can only be solved numerically.
  We reformulate the DRG in the language of
  differential geometry, which allows us to apply it to numerical solutions of the 
  background and perturbation equations. This formulation also enables us to
  use the DRG in systems with background parameter flows, and therefore, extend 
  our results to any order in perturbation theory.
  As an example, we apply this method
  to calculate the soliton-like solutions of the Korteweg-de~Vries equation
  deformed by adding a small damping term.
  We numerically construct DRG solutions which are valid on
  secular time scales, long after naive perturbation theory has broken
  down.
\end{abstract}
\maketitle

\section{Introduction}
The career of a physicist consists of treating the harmonic
oscillator in ever-increasing levels of abstraction, according to
Sidney Coleman~\cite{Chen:2018cts}.  Although a joke, the truth is that
perturbation theory is an indispensable tool in physics.  Perturbation
theory allows us to gain insights into problems that are too difficult
to solve exactly, too expensive to solve numerically, or we demand
more control than is afforded by numerical simulations.  Entire
textbooks focus just on various methods in perturbation
theory~\cite{bender1999advanced, kevorkian1981perturbation, goldstein,
  jose_saletan_1998}.
In the literature, one can find a plethora of applications of the
perturbative approach which include critical phenomena in condensed 
matter systems~\cite{ABRIKOSOV1957199, Goldenfeld:1992qy, PhysRevE.51.5577, 
Chen:1995ena, bellac1991quantum, doi:10.1119/1.11019}, particle physics~\cite{Dyson:1952tj,
Kadanoff:1976wy, Weinberg:1995mt}, and gravitation and cosmology~\cite{Misner:1974qy,
Mukhanov:1990me, Ma:1995ey, Burgess:2009bs}.

However, caution is always warranted when applying naive perturbation
theory.  There are many ways in which traditional perturbation theories can fail.
In this paper, we are interested in breakdown on secularly long
timescales (typically proportional to an inverse power of a control
parameter), even when the dynamical system is known to be
bounded~\cite{jose_saletan_1998, bender1999advanced,
  kevorkian1981perturbation}.  There are many approaches to secular
perturbation theory, tailored to specific situations, for example the
Poincar\'e-Lindstedt method for problems with periodic
solutions~\cite{jose_saletan_1998, bender1999advanced}.  Many of these
disparate approaches have been subsumed by the method of the dynamical
renormalization group (DRG)~\cite{Chen:1995ena, Kunihiro:1995zt,
  10.1143/PTPS.99.244, Ei:1999pk}.  In the DRG, constant parameters of
the background solutions are promoted to time-dependent functions,
which satisfy so-called \emph{beta function} flow equations.  By
making the ``constants'' vary with time, the secular growth can be
exactly canceled.

Although DRG includes RG in the name, this is renormalization in a
Gell-Mann-Low sense~\cite{Higashijima:1980fm}, which is still
perturbative, unlike the non-perturbative Wilsonian or
Callan-Symanzik~\cite{Callan:1970yg, Symanzik:1971vw, Symanzik:1970rt}
point of view.  DRG relies on the existence of an attractor manifold
of an unperturbed problem to control the calculation of a deformed
problem.
Despite being perturbative, DRG can still re-sum solutions
that include non-perturbative effects.

For systems that have self-similar solutions, there has been work on
the RG approach~\cite{Goldenfeld:1992qy, Bricmont}, including some
numerical work~\cite{PhysRevE.51.5577, Betelu_2000,
  braga2020numerical}.  However, the majority of the existing DRG
literature (that we are aware of) has been applied to analytical
problems, and there has not been a general numerical approach to the
DRG outside of self-similarity.
This creates a limitation: one does not always have the luxury of a
self-similar solution, or an analytical solution at background or at
linear order.  In this case, neither the analytical DRG nor the previous
numerical approaches can be applied.

In this paper, we propose a general numerical approach to the DRG.  To do so,
we have reformulated the DRG in the language of differential geometry.
This extends the envelope picture of~\cite{Kunihiro:1995zt,
  Ei:1999pk}.  The key insight is this: The details of the secular
growth of the naive perturbation solution encode the time dependence
(beta functions) and reparameterizations (alpha functions) of the
solution parameters.  Because of our geometric formulation, we can
equally well apply the DRG to problems which already experience a
background parameter flow.  Our formulation also makes it mechanical
to see how to continue to arbitrary perturbation order.

As a proof-of-concept, we apply this procedure to solve a deformation
to the Korteweg-de\ Vries (KdV) equation~\cite{CANOSA1977393,
  olver1993applications}.
We find solutions which are valid over
secular timescales, long after naive perturbation theory has broken
down.  To do so, we promote the velocity of the one-soliton KdV
solution into a time-dependent function.  We extract its
reparameterization and flow functions directly from diverging, naive
perturbative solutions.  We present several checks: of the DRG
approach itself, without reference to the true (nonperturbative)
numerical solution; and also against the true solution.  The
renormalized solution's velocity, amplitude, and width all agree with
the true solution.

Although a deformation to the KdV equation could be treated
analytically, we find this system ideal as our proof of concept for
numerical DRG.  Our numerical simulations do not take advantage of the
existence of analytical solutions, so we are demonstrating the
``full'' case of numerical DRG on top of numerical background
solutions.  Meanwhile, we are also able to assess how well the
numerical DRG performs by comparing with analytics.

We expect this method to be applicable to a variety of problems.
One of our motivations is in gravitational physics, namely in
modeling small deformations of Einstein's theory of general relativity
(GR).
Like the KdV equation, GR has stable nonlinear solutions (black holes),
and an attractor manifold (the space of binary black hole inspirals).
Also like the KdV equation, adding a deformation will lead to effects
on secularly long timescales.
This similarity motivated our use of the KdV equation as a
model problem, before applying the numerical DRG to the more
complicated problem of beyond-GR calculations.

The organization of this manuscript is as follows.  In
Section~\ref{sec:RG_PT}, we first give an analytical example to
describe the DRG method.  We then give our geometric formulation,
which can be applied to numerical problems.  This procedure extracts
the RG flow and reparameterization functions directly from the naive
perturbative solution.
In Section~\ref{sec:KdV}, we introduce the KdV equation and perturb it
to its damped form, also known as the Korteweg-de\ Vries-Burgers (KdVB) 
equation, showing all the elements needed to extract the parameter
flow generators.
In Section~\ref{sec:results}, we present the results of our extraction
scheme, and solve the flow
equations to find the renormalized parameters' evolution.  Once the bare parameters are 
replaced by the flowing parameters in the one-soliton KdV solution, we reconstruct a 
renormalized solution and compare it with the nonperturbative KdVB solution.
We also approach the problem using an alternative parameterization, to
test if the dimensionality of the parameter space was increased by the perturbation.
In Section~\ref{sec:FW}, we discuss a potential application of this
renormalization-based method to the calculation of gravitational
waves from theories beyond GR.
Finally, in Section~\ref{sec:conclusion}, we discuss and conclude.

\section{RG flow and first-order perturbation theory}\label{sec:RG_PT}

In this section, we present a procedure to build solutions free
from secular divergences. This procedure only requires knowledge
of the naive perturbative solution.
Throughout, we use the Einstein summation convention for repeated
indices.

\subsection{Analytical example}
\label{sec:analytical-example}

To demonstrate the concepts and features of this procedure,
we condense and simplify the results of Galley and
Rothstein~\cite{Galley:2016zee} as an example.
Consider the equations of motion for a binary
system, where the leading order is Newtonian gravity, and the
perturbation at order $\varepsilon$ is due to post-Newtonian radiation
reaction.  The radial and angular equations of motion read
\begin{align}
  \label{eq:rad_ins}
  \ddot{r}-r^2\omega = & -\frac{M}{r^2} \\
  &+\varepsilon\left[\frac{64M^3\nu}{15r^4}\dot{r}+
    \frac{16M^2\nu}{5r^3}\dot{r}^3+\frac{16M^2\nu}{5r}\dot{r}\omega^2\right]\nonumber\\
  r\dot{\omega}+2\dot{r}\omega =& -\varepsilon\left[\frac{24M^3\nu}{5r^3}\omega+
    \frac{8M^2\nu}{5r^2}\dot{r}^2\omega+\frac{8M^2\nu}{5}\omega^3\right]
  \,.\nonumber
\end{align}
The background solutions are simply elliptic Keplerian orbits.  For
small eccentricity $e\ll 1$, these are given by
\begin{align}
r^{(0)}(t) &= R_0+A\sin \phi_{\circ}(t) \,,\\
\omega^{(0)}(t)&=\Omega_0-\frac{2\Omega_0 A}{R_0}\sin \phi_{\circ}(t) \,, \\
\phi_{\circ}(t) &= \Omega_0(t-t_0)+\Phi_0 \,, \\
\phi^{(0)}(t) &= \phi_{\circ}(t) + \frac{2A}{R_0}\cos \phi_{\circ}(t) \,,
\end{align}
where $\Omega_0^2\equiv M/R_0^3$ and $A=eR_0$.
Here we have introduced the auxiliary phase for a circular orbit,
$\phi_{\circ}(t)$, and an orbital phase $\phi(t)$.  We will 
collect the four solution parameters into a single ``vector'' 
$\vec{\lambda} \equiv (R_0,\Omega_0, A, \phi_{\circ})$; the 
reason for using $\phi_{\circ}$ rather than $\Phi_{0}$ as a flowing 
parameter will become apparent below. 

The effects of radiation reaction appear with leading coefficient
$\nu\Omega_0^5R_0^5 \ll 1$, which is counted by powers of
$\varepsilon$.  To solve perturbatively, we pose
\begin{align}
  r(t) &= r^{(0)}(t) + \varepsilon r^{(1)}(t) \,, \\
  \omega(t) &= \omega^{(0)}(t) + \varepsilon \omega^{(1)}(t) \,.
\end{align}
Plugging this in to the differential equation and collecting at order
$\varepsilon^{1}$, we get the linearized differential equations
\begin{align}
\ddot{r}^{(1)}-3\Omega_0^2 r^{(1)} &= 2R_0\Omega_0 \omega^{(1)} \,,\label{eq:delta_r}\\
R_0\dot{\omega}^{(1)}+2\Omega_0\dot{r}^{(1)} &=- \frac{32}{5}\nu R_0^6\Omega_0^7 \,.
\end{align}

The solutions to these equations have homogeneous and particular
pieces, and the total solution is~\cite{Galley:2016zee}
\begin{widetext}
\begin{align}
\label{eq:pert_r}
r(t) &= R_0+A\sin\left(\Omega_0(t-t_0)+\Phi_0\right)-
\varepsilon\left[\frac{64\nu}{5}\Omega_0^6R_0^6(t-t_0)-
\frac{64\nu}{5}\Omega_0^5R_0^6\sin\Omega_0(t-t_0)\right]\,,\\
\label{eq:pert_omega}
\omega(t) &= \Omega_0-
\frac{2\Omega_0 A}{R_0}\sin\left(\Omega_0(t-t_0)+\Phi_0\right)+\varepsilon\left[\frac{96\nu}{5}\Omega_0^7R_0^5(t-t_0)-
\frac{128\nu}{5}\Omega_0^6R_0^5\sin\Omega_0(t-t_0)\right]\,,\\
\label{eq:pert_phi}
\phi(t) &= \Phi_0+\Omega_0(t-t_0)+\frac{2A}{R_0}\cos\left(\Omega_0(t-t_0)+\Phi_0\right)+
\varepsilon\left[\frac{48\nu}{5}R_0^5\Omega_0^7(t-t_0)^2+\frac{128\nu}{5}\Omega_0^5R_0^5\cos\Omega_0(t-t_0)\right]
\,,
\end{align}
\end{widetext}
where the expression $\phi(t)$ comes by direct integration of $\omega(t)$. 
There are two important features to
observe in the $\mathcal{O}(\varepsilon)$ pieces of these solutions.
The first term in the square brackets is a linear-in-time divergence 
for $\omega(t)$ and $r(t)$, and a quadratic divergence for $\phi(t)$.
Those diverging terms suggest two new secular 
timescales: one from $\omega(t)$ (and $r(t)$), which scales as 
$\varepsilon^{-1}$, and another from $\phi(t)$ scaling as $\varepsilon^{-1/2}$. 
Nominally, $T_{\mathrm{sec}}\sim\varepsilon^{-1/2}$ is the shortest timescale 
where secular divergences need to be controlled; but it is essential to 
describe in which circumstances each of the two timescales appears.
A traditional approach to handling these new time scales would be the
method of multiple scales~\cite{kevorkian1981perturbation,
  bender1999advanced}.  However we will follow the DRG approach, which
does not require a~priori the knowledge of how ``slow'' and ``fast'' times
are related.

The second term in the square brackets can be absorbed by a redefinition 
of the \emph{initial} values which are collected in $\vec{\lambda}(t_{0}) \equiv 
(R_0,\Omega_0, A, \Phi_{0})$.
Absorbing the last term in Eqs.~(\ref{eq:pert_r}-\ref{eq:pert_phi}) 
is accomplished by making an infinitesimal diffeomorphism of the 
initial values according to
\begin{align}
  \vec{\lambda}(t_{0}) \to \vec{\lambda}(t_{0}) + \varepsilon \vec{\alpha}(\vec{\lambda})
  \,,
\end{align}
with the specific solution
\begin{align}
  \label{eq:alpha-G-R}
  \vec{\alpha} = \left(0;0;\frac{64\nu}{5}R^6_0\Omega^5_0\cos\Phi_0;-\frac{64\nu}{5A}R^6_0\Omega^5_0\sin\Phi_0\right)
  \,.
\end{align}

Now to control the secular divergence, we promote $\vec{\lambda}$ to a
function of time, renaming its components to be the ``renormalized''
solution parameters $\vec{\lambda}_{\Ren} = (R_\Ren(t),\Omega_\Ren(t), A_\Ren(t),
\Phi_\Ren(t))$.  We promote the solution,
\begin{align}
  \label{eq:reparam_r}
  r(t) &= R_{\mathrm{R}}(t)+ A_{\mathrm{R}}(t)\sin\Phi_{\mathrm{R}}(t)\,,\\
  \label{eq:reparam_omega}
  \omega(t) &= \Omega_{\mathrm{R}}(t)-\frac{2\Omega_{\mathrm{R}}(t)A_{\mathrm{R}}(t)}{R_{\mathrm{R}}(t)}\sin\Phi_{\mathrm{R}}(t)\,,\\
  \phi(t) &= \Phi_{\Ren}(t)+\frac{2A_{\Ren}(t)}{R_{\Ren}(t)}\cos\Phi_{\Ren}(t)
  \,.
\end{align}
The new $\vec{\lambda}$ satisfies a ``beta function'' flow equation,
\begin{align}
  \label{eq:beta-func-general}
  \frac{d\vec{\lambda}_{\Ren}}{dt} = \vec{\beta}(\vec{\lambda}_{\Ren})
  = \vec{\beta}^{(0)}(\vec{\lambda}_{\Ren}) +\varepsilon \vec{\beta}^{(1)}(\vec{\lambda}_{\Ren})
  \,.
\end{align}
In the background solution, $\phi_{\circ}(t)$ was already flowing, which is
why we included it in $\vec{\lambda}$ instead of the constant
$\Phi_{0}$.  The background beta function was simply
\begin{align}
  \label{eq:inspiral-bg-beta-func}
  \vec{\beta}^{(0)} = (0;0;0;\Omega_{\Ren})
  \,.
\end{align}
Ref.~\cite{Galley:2016zee} found that the first order beta function is
\begin{align}
  \label{eq:beta-G-R}
  \vec{\beta}^{(1)} = \left(-\frac{64\nu}{5}R^6_\Ren\Omega^6_\Ren;\frac{96\nu}{5}\Omega^7_\Ren R^5_\Ren;0;0\right)
  \,.
\end{align}
These can be integrated explicitly, finding simple algebraic solutions
for $(R_\Ren(t),\Omega_\Ren(t), A_\Ren(t), \Phi_\Ren(t))$ (see
Eqs.~(4.42)--(4.45) in~\cite{Galley:2016zee}).
Let us also point out here that the two nonzero components in
Eq.~\eqref{eq:beta-G-R} are not independent: their relationship can be
found by taking a differential of Kepler's law $\Omega_0^2\equiv
M/R_0^3$.  We will return to this feature in our numerical example in
Sec.~\ref{subsec:multi_KdV}.

There are two equivalent ways to find the first order beta functions.
Galley and Rothstein followed the typical Wilsonian approach of
introducing appropriate counterterms which absorb the secular
divergences.  A more pedestrian approach from the point of view of the
differential equation is as follows.  For sufficiently short times,
$\varepsilon(t-t_{0}) \ll 1$, the evolution of the parameters is
linear in time.  Including the $\vec{\alpha}$ reparameterization, this
is equivalent to replacing $\vec{\lambda}(t)$ with
\begin{align}
\label{eq:param_flow_ex}
\vec{\lambda}_{\mathrm{R}}(t) = \vec{\lambda}(t) + \varepsilon\vec{\alpha}
+ \varepsilon(t-t_0)\vec{\beta}^{(1)}
+ \mathcal{O}(\varepsilon^{2})
\,,
\end{align}
where $\vec{\lambda}(t)$ satisfies the background flow equation.
In our promoted solutions, Eqs.~\eqref{eq:reparam_r} and
\eqref{eq:reparam_omega}, insert these flowing quantities
(the treatment of $\phi(t)$ in Eq.~\eqref{eq:pert_phi} is more subtle,
because of the background flow of $\phi_{\circ}$, and will be explained 
in the next section). Next, re-expand in powers of $\varepsilon$.  
Finally, read off functions of $\vec{\alpha}$ and $\vec{\beta}^{(1)}$ 
that will match the homogeneous solutions and secularly-divergent
terms at $\mathcal{O}(\varepsilon)$ in Eqs.~\eqref{eq:pert_r} and
\eqref{eq:pert_omega}.  Performing this coefficient-matching
gives the same components as in Eqs.~\eqref{eq:alpha-G-R} and
\eqref{eq:beta-G-R}.

This example demonstrates the analytical approach to the dynamical
renormalization group, which we will promote to a numerical approach.
We will revisit the problem of secular divergence in a binary inspiral
in the discussion in Sec.~\ref{sec:FW}.

\subsection{General formalism}
\label{sec:general-formalism}

\paramsolsmap

We now present the general framework for the DRG, 
in a form that is amenable to a numerical implementation.  
The analytical approach has been treated extensively, see
e.g.~\cite{Kunihiro:1995zt, Ei:1999pk}. Suppose we want to 
solve the differential equation
\begin{align}
  \label{eq:full_eq_mot}
  \frac{d\varphi^{A}}{dt} = F^{A}[\varphi^{B},t]+\varepsilon P^{A}[\varphi^{B},t]
  \,,
\end{align}
which is an $\mathcal{O}(\varepsilon)$ deformation of an equation which
we already know how to solve (at $\varepsilon=0$).  Here capital Latin
indices label the degrees of freedom (or fields) in the differential
equation.  In the case of a partial differential equation (PDE), $F^{A}$ and
$P^{A}$ can also depend on spatial derivatives of the $\varphi^{A}$
fields.  In this work we will focus on the autonomous case, so there
is no explicit time dependence in $F$ or $P$.

In our approach, we rely on the existence of an ``attractor,''
``invariant,'' or ``slow'' manifold for the space of solutions.
We assume that the $\varepsilon P$ deformation is mild enough that it
does not affect the existence of a slow manifold (this can be rather
subtle for PDEs, for example if including $P$ changes the principal
part of the system).  The
solutions are labelled by some parameters (or collective coordinates)
$\lambda^{i}$ in a space $\Lambda$ of finite dimension $m$, which we may also
denote as $\vec{\lambda}$.  The solutions to the background
($\varepsilon=0$) equations are
\begin{align}
  \label{eq:back_param}
  \varphi^{A} = \varphi^{(0)A}(t, \lambda^{i})
  \,,
\end{align}
and possibly spatial dependence in the case of a PDE.
We can think of $\varphi^{(0)}:\Lambda \to \mathcal{S}$ as a map from
parameter space to the solution space $\mathcal{S}$, as seen in
Fig.~\ref{fig:map}.  As seen in the
previous section [Eq.~\eqref{eq:inspiral-bg-beta-func}], the
background parameters may have their own flow equations,
\begin{align}
\label{eq:bkg_flow}
  \frac{d\vec{\lambda}}{dt} = \vec{\beta}^{(0)}(\vec{\lambda})
  \,,
\end{align}
referred to as the ``beta functions'' of the system.
The $\vec{\beta}$ vector field is depicted as the blue field in the
left panel of Fig.~\ref{fig:map}.  These beta functions will be
corrected at order $\varepsilon$, leading to a secular divergence in
the integral curves of the background and foreground beta functions.

The naive perturbation theory treatment of Eq.~\eqref{eq:full_eq_mot}
would pose the ansatz
\begin{equation}
\label{eq:sol_pert_m}
\varphi^{A} = \varphi^{(0)A}+\varepsilon\varphi^{(1)A}
\,,
\end{equation}
which then leads to the system of differential equations
\begin{align}
  \label{eq:back}
  \frac{d\varphi^{(0)A}}{dt} - F^{A}[\varphi^{(0)}]={}& 0 \,,\\
  \label{eq:pert_eq_mov}
  \frac{d\varphi^{(1)A}}{dt} - F^{(1)A}[\varphi^{(1)}; \varphi^{(0)}] ={}&
  P^{A}[\varphi^{(0)}]
  \,.
\end{align}
Here, $F^{(1)}$ is a linear differential operator that is the
linearization of $F$, namely,
\begin{align}
  \label{eq:F1-def}
  F^{(1)A}[\varphi^{(1)}; \varphi^{(0)}]  =
  \frac{d}{d\varepsilon}
  F^{A}[\varphi^{(0)} + \varepsilon \varphi^{(1)}]
  \Bigg|_{\varepsilon=0}
  \,.
\end{align}

The linear differential equation Eq.~\eqref{eq:pert_eq_mov}
generically leads to secular divergences in $\varphi^{(1)}$ [as seen
for example in Eqs.~\eqref{eq:pert_r} and \eqref{eq:pert_omega}], and
it is these divergences that we seek to renormalize.

First, the solution $\varphi^{(1)}$ may contain pieces that live in
the space of homogeneous solutions to the perturbation
equation~\eqref{eq:pert_eq_mov}.  These homogeneous solutions can be
absorbed by perturbative shifts of the initial parameters
$\vec{\lambda}(t_{0})$ via
\begin{align}
  \vec{\lambda}(t_{0}) \to \vec{\lambda}(t_{0}) + \varepsilon \vec{\alpha}(\vec{\lambda})
  \,.
\end{align}
The $\vec{\alpha}$ vector field is depicted as the orange field in
Fig.~\ref{fig:map}.
A perturbative shift of the initial parameters yields another nearby
solution of the background system Eq.~\eqref{eq:back}, and therefore
the difference is a homogeneous solution of the first order
perturbation equation,
\begin{align}
  &\varphi^{(0)A}(t, \lambda^{i} + \varepsilon \alpha^{i}) =
  \varphi^{(0)A}(t, \lambda^{i}) + \varepsilon \alpha^{i} \frac{\delta\varphi^{(0)A}}{\delta\lambda^{i}} \,, \\
  &
  \frac{d}{dt}\varphi^{(0)A}(t, \lambda^{i} + \varepsilon \alpha^{i})
  - F^{A}[\varphi^{(0)A}(t, \lambda^{i} + \varepsilon \alpha^{i})]
  = 0
  \,,
  \\
  \label{eq:delta-phi-is-hom-sol-F1}
  &
  \varepsilon \alpha^{i}
  \left(
    \frac{d}{dt}\frac{\delta\varphi^{(0)A}}{\delta\lambda^{i}}
    -
    F^{(1)A}
    \left[
      \frac{\delta\varphi^{(0)A}}{\delta\lambda^{i}} ;
      \varphi^{(0)}
    \right]
  \right) = 0
  \,.
\end{align}
This first order shift is generated by the functions that we called
$\delta\varphi^{(0)A}/\delta\lambda^{i}$, which have a clear
interpretation in differential geometry (in terms of the
\emph{differential} of a map) that we discuss below.

Besides the homogeneous solutions, there is another source of secular
divergence in naive perturbation theory.  The true solution at finite
$\varepsilon$ need not stay on the background solution manifold
$\mathcal{S}$ seen in Fig.~\ref{fig:map}, but there is a curve within
$\mathcal{S}$ that is closest to the true solution.  When this closest
curve is pulled back to the parameter manifold $\Lambda$, its flow
need not coincide with the background flow generated by
$\vec{\beta}^{(0)}$.  Therefore we need to allow for the
possibility of the flow of $\vec{\lambda}$ changing at first order,
giving the renormalized $\vec{\lambda}_{\Ren}$ solution,\footnote{%
Notice that Eq.~\eqref{eq:flow_equations} does not specify the normal
form of the differential system, and in fact a singular perturbation
may require further generalization (e.g.~a negative power of
$\varepsilon$ on the right hand side).  For further details
see~\cite{kuehn2015multiple}.
}
\begin{align}
\label{eq:flow_equations}
  \frac{d\vec{\lambda}_{\Ren}}{dt} = \vec{\beta}^{(0)}(\vec{\lambda}_{\Ren})
  + \varepsilon \vec{\beta}^{(1)}(\vec{\lambda}_{\Ren})
  \,.
\end{align}
In the absence of a background flow, short timescales satisfy
$(t-t_{0})\ll T_{\Sec}$, where $T_{\Sec} \sim (\varepsilon \beta^{(1)})^{-1}$ 
is the timescale of secular divergence of naive perturbation theory.
Thus, for sufficiently short time intervals we write
\begin{align}
  \label{eq:Ren-flow-pedestrian}
  \vec{\lambda}_{\Ren} = \vec{\lambda} + \varepsilon \vec{\alpha}
  + \varepsilon (t-t_{0}) \vec{\beta}^{(1)} + \mathcal{O}(\varepsilon^{2})
  \,.
\end{align}

Now there are two ways to write the first order solution: one
following naive perturbation theory [from Eq.~\eqref{eq:sol_pert_m}],
and one renormalized, where the correct choice of $\beta^{(1)}$ will 
ensure that the first order solution is bounded in time.  For small
times, we equate these two,\footnote{%
Let us remark that here we take a vanishing background flow,
$\beta^{(0)}=0$.  The full case will be given below.
}
\begin{align}
  \label{eq:rec_phi1R}
  & \varphi^{(0)A}(\vec{\lambda}) + \varepsilon \varphi^{(1)A} =
  \varphi^{(0)A}(\vec{\lambda}_{\Ren}) + \varepsilon \varphi_{\perp}^{(1)A} \,, \\
  & \varphi^{(1)A} =
  \left[\alpha^{i}+ (t-t_{0}) \beta^{(1)i}\right] \frac{\delta \varphi^{(0)A}}{\delta \lambda^{i}}
  + \varphi_{\perp}^{(1)A} \,.
\end{align}
Let us emphasize that this matching is the key to our formulation of
DRG: the details of the secular growth in naive perturbation theory
encode the data for renormalization, $\vec{\alpha}$ and
$\vec{\beta}^{(1)}$.
This gives us the condition for finding $\vec{\alpha}$ and
$\vec{\beta}^{(1)}$: keep the residual $\varphi^{(1)A}_{\perp}$ bounded in time.  We
take this to mean minimizing its norm in an appropriate function
space, for example,
\begin{align}
  \label{eq:norm-to-minimize}
  \left\|
    \varphi_{\perp}^{(1)}
  \right\|^{2}  &= \int
  \left| \varphi_{\perp}^{(1)A}\right|^{2} dt \,,\\
  \left\|
    \varphi_{\perp}^{(1)}
  \right\|^{2}  &= \int
  \left|
    \varphi^{(1)A}-\left[\alpha^{i}+(t-t_{0})\beta^{(1)i}\right]\frac{\delta \varphi^{(0)A}}{\delta\lambda^{i}} \right|^{2} dt \,,\nonumber
\end{align}
where the norm $|\cdot|$ inside the integral can e.g.\ include a spatial
integral, when solving a PDE.

\subsubsection{Differential geometry formulation of DRG}
\label{sec:diff-geom-form}

Before providing details of such minimization, we give a geometrical
interpretation for this procedure.  Above we presented the procedure
only to first order in $\varepsilon$ and first order in a time
difference $\Delta t = t-t_{0} \ll T_{\Sec}$.  However, we can promote
this to all orders by recognizing that the $\vec{\alpha}$ and
$\vec{\beta}$ vector fields generate diffeomorphisms of parameter
space.  The geometric version of the reparametrization
$\vec{\lambda} \to \vec{\lambda} + \varepsilon \vec{\alpha}$ is a
diffeomorphism generated by flowing along the vector field
$\vec{\alpha}$ by parameter $\varepsilon$.
This $\alpha$ can be generalized to higher orders, for example
defining $\vec{A} = \varepsilon \vec{\alpha}^{(1)} + \varepsilon^{2}
\vec{\alpha}^{(2)} + \ldots$, and then flowing along the integral
curves of $\vec{A}$ by parameter 1.  Likewise, the
time-dependent flow under the beta function equation corresponds to a
flow along the $\vec{\beta}$ vector field by parameter $(t-t_{0})$.

Let us write $\Phi^{V}_{s}: \mathcal{M} \to \mathcal{M}$ to represent
the flow along integral curves of the tangent vector field $V \in
\mathcal{X}(\mathcal{M})$, by a parameter $s$~\cite{lee2013smooth}.
From Fig.~\ref{fig:map}, we see that the
desired flow in parameter space should be the composition
\begin{align}
  \lambda_{\Ren}(t) = \Phi^{{\beta}}_{t-t_{0}} \circ \Phi^{A}_{1}
  \left[
    \lambda(t_{0})
  \right]
  \,.
\end{align}
It will be convenient to represent $\lambda_{0} \equiv \lambda(t_{0})$ in terms of
undoing the background $\vec{\beta}^{(0)}$ flow, namely,
\begin{align}
  \lambda(t) &= \Phi^{\beta^{(0)}}_{t-t_{0}}
  \left[
    \lambda(t_{0})
  \right] \,, \\
  \lambda(t_{0}) &= \Phi^{-\beta^{(0)}}_{t-t_{0}}
  \left[
    \lambda(t)
  \right] \,.
\end{align}
Therefore, the renormalized flow, as a function of the background
flow, is stated as
\begin{align}
  \label{eq:lambda-R-composition-3}
  \lambda_{\Ren}(t) = \Phi^{{\beta}}_{t-t_{0}} \circ \Phi^{A}_{1}
  \circ \Phi^{-\beta^{(0)}}_{t-t_{0}}
  \left[
    \lambda(t)
  \right]
  = \Phi^{V}_{1}[\lambda(t)]
  \,,
\end{align}
and this will be the argument to the background solution map,
$\varphi^{(0)}$.  Here we used the fact that diffeomorphisms form a
group, so the composition can be rewritten as the flow under a single
vector field $\vec{V}$, which can be determined using the
Baker-Campbell-Hausdorff (BCH) theorem below.

\diffmap

There is also a clear geometric meaning for the functions
$\delta\varphi^{(0)A}/\delta \lambda^{i}$ which appear in the norm,
Eq.~\eqref{eq:norm-to-minimize}, which we will minimize.
The map $\varphi^{(0)}:\Lambda\to\mathcal{S}$
induces a map called the \emph{differential}, $d\varphi^{(0)}:
T\Lambda \to T\mathcal{S}$, from the tangent space at $\lambda$ to the
tangent space at the image $\varphi^{(0)}(\lambda)$.
This is illustrated in Fig.~\ref{fig:diff-map}.  The tangent
space at the image consists of solutions to the linearization of the
background differential equation (when linearized about the solution
$\varphi^{(0)}(\lambda)$), as demonstrated in traditional notation in
Eq.~\eqref{eq:delta-phi-is-hom-sol-F1}.
The matching performed in Eq.~\eqref{eq:rec_phi1R} can be written
geometrically as finding the decomposition
\begin{align}
\label{eq:decomp_pert}
  \varphi^{(1)} =
  d\varphi^{(0)}(\vec{V})
  + \varphi_{\perp}^{(1)}
  \,,
\end{align}
where $\varphi_{\perp}^{(1)}$ lies outside of the vector space
$T_{\varphi^{(0)}(\lambda)}\mathcal{S}$.
The differential
$d\varphi^{(0)}$ can be thought of as a matrix, where the $i$th
column, $\delta\varphi^{(0)A}/\delta \lambda^{i}$, is a vector in
$T_{\varphi^{(0)}(\lambda)}\mathcal{S}$, which corresponds to the change
in the solutions under an infinitesimal shift in the $\lambda^{i}$
direction in parameter space.  The solution to the linear perturbation
problem in Eq.~\eqref{eq:pert_eq_mov} is also a vector in
$T_{\varphi^{(0)}(\lambda)}\mathcal{S}$, and the minimization
procedure that we employ decomposes this vector as a linear
combination of these appropriate basis functions.
This procedure is essentially a fit of the data, $\varphi^{(1)}$, with
the functional form given by $d\varphi^{(0)}(\vec{V})$, and the fit
parameters being the values of $\vec{\alpha}$, $\vec{\beta}^{(1)}$,
and potentially higher order coefficients.
The orders of $t-t_{0}$ and $\varepsilon$ kept in calculating
$\vec{V}$ will affect the quality of this fit.

To determine the generator $\vec{V}$ of the composition, we apply the
BCH theorem~\cite{achilles2012early}.  If a
function $f$ is right-composed with a diffeomorphism $\Phi^{V}_{s}$,
this is equivalent to the left-action of the exponential of the Lie
derivative acting on it,
\begin{align}
  \exp(\mathcal{L}_{sV}) \cdot f = f \circ \Phi^{V}_{s}
  \,.
\end{align}
We want to find the vector field $V$ which generates
\begin{align}
  & \exp(\mathcal{L}_{V}) \cdot f = f \circ
  \Phi^{{\beta}}_{t-t_{0}} \circ \Phi^{A}_{1}
  \circ \Phi^{-\beta^{(0)}}_{t-t_{0}} \,, \\
  \label{eq:exp-composition}
  &\exp(\mathcal{L}_{V}) =
  \exp(\mathcal{L}_{-(t-t_{0})\beta^{(0)}}) \cdot
  \exp(\mathcal{L}_{A}) \cdot
  \exp(\mathcal{L}_{(t-t_{0})\beta})
  \,.
\end{align}
Here we will demonstrate with just the first few terms of the BCH
theorem, namely,
\begin{align}
  \exp(\mathcal{L}_{C}) ={}&
  \exp(\mathcal{L}_{A}) \cdot
  \exp(\mathcal{L}_{B}) \,, \\
  \label{eq:BCH3}
  C ={}& A + B + \frac{1}{2}[A,B] \nonumber\\
  &{}+ \frac{1}{12}[A,[A,B]] - \frac{1}{12}[B,[A,B]] + \ldots
  \,.
\end{align}
Applying the BCH formula to the two compositions in
Eq.~\eqref{eq:exp-composition} gives us
\begin{align}
  \label{eq:vec_v}
  \begin{split}
    \vec{V} ={}& \varepsilon \Big\{ \vec{\alpha}^{(1)} + (t-t_{0})
      \vec{\beta}^{(1)} + (t-t_{0}) \left[ \vec{\alpha}^{(1)},
        \vec{\beta}^{(0)} \right] \\
    & {}+ \frac{1}{2} (t-t_{0})^{2}
    \left[\vec{\beta}^{(0)} ,
      \left[\vec{\beta}^{(0)}, \vec{\alpha}^{(1)}\right]
      -\vec{\beta}^{(1)}
    \right]
    \Big\} \\
    &{}+
    \mathcal{O}(\varepsilon t^{3}, \varepsilon^{2})
    \,.
  \end{split}
\end{align}

Notice that when $\vec{\beta}^{(0)}\neq 0$, the components of
$\vec{V}$ at a point $\lambda$ depend on components of derivatives of
$\vec{\alpha}^{(1)}$ and $\vec{\beta}^{(1)}$.  Namely, to this order,
we need all of the values
\begin{align}
  \alpha^{(1)i}\ ,\ 
  \beta^{(1)i}\ ,\ 
  \beta^{(0)k}\alpha^{(1)j}_{,k}\ ,\ 
  \beta^{(0)k}\beta^{(1)j}_{,k} \,, \nonumber\\
  \beta^{(0)i}\beta^{(0)k}_{,i}\alpha^{(1)j}_{,k} \ ,\ 
  \beta^{(0)k}\beta^{(0)i}\alpha^{(1)j}_{,ki}
  \,,
\end{align}
where we have introduced the notation of the parameter ``comma derivative'',
$f_{,i} \equiv \pd_{i} f = \pd f/\pd \lambda^{i}$.  We emphasize here
that all of these are simply constant coefficients in a Taylor
expansion at a background point $\vec{\lambda}_{0}$.

It is convenient to collect all of these yet-to-be-determined constant
coefficients in a vector of vectors, $\psi^{j(\mu)}$, where $\mu$
labels the collection of coefficients to be extracted.  We collect the
remaining dependence on time and the background flow in the vector of
matrices $\mathcal{T}^{i(\mu)}_{j}$,
\begin{equation}
\label{eq:comp_flow}
V^{i}=\varepsilon \mathcal{T}^{i(\mu)}_{j}\psi^{j(\mu)}
\,.
\end{equation}
For this example, the vector $\psi^{j(\mu)}$ contains the flows
$\psi^{(0)}\equiv\vec{\beta}^{(1)}$ and
$\psi^{(1)}\equiv\vec{\alpha}^{(1)}$,
and the others are given by
\begin{align}
\psi^{j(2)}\equiv{}& \beta^{(0)k}\beta^{(1)j}_{,k},\label{eq:aux_1}\\
\psi^{j(3)}\equiv{}& \beta^{(0)k}\alpha^{(1)j}_{,k},\label{eq:aux_2}\\
\psi^{j(4)}\equiv{}& \beta^{(0)i}\beta^{(0)k}_{,i}\alpha^{(1)j}_{,k},\label{eq:aux_3}\\
\psi^{j(5)}\equiv{}& \beta^{(0)k}\beta^{(0)i}\alpha^{(1)j}_{,ki},\label{eq:aux_4}\\
\mathcal{T}^{i(0)}_{j} \equiv{}& (t-t_0)\delta^i_j+\frac{1}{2}(t-t_0)^2\beta^{(0)i}_{,j},\\
\mathcal{T}^{i(1)}_{j} \equiv{}& \delta^i_j+(t-t_0)\beta^{(0)i}_{,j}\\
&{}+\frac{1}{2}(t-t_0)^2\bigg(\beta^{(0)k}_{,j}\beta^{(0)i}_{,k}-\beta^{(0)k}\beta^{(0)i}_{,jk}\bigg),\nonumber\\
\mathcal{T}^{i(2)}_{j} =& -\mathcal{T}^{i(4)}_{j}= -\mathcal{T}^{i(5)}_{j}\equiv -\frac{1}{2}(t-t_0)^2\delta^i_j,\\
\mathcal{T}^{i(3)}_{j} \equiv{}& -(t-t_0)\delta^i_j-(t-t_0)^2\beta^{(0)i}_{,j}
\,.
\end{align}
Notice that in the special case where there is no background flow,
$\vec{\beta}^{(0)}=0$, there is a great simplification: higher order
terms in the Taylor expansion would not be needed.

We insert this into the matching procedure of
Eq.~\eqref{eq:rec_phi1R}, which we repeat here for convenience.
With the infinitesimally shifted flow
$\vec{\lambda}_{\Ren}=\vec{\lambda}+\vec{V}+\mathcal{O}(\varepsilon^2,\varepsilon
t^3)$, the two different ways to write the first order solution
are
\begin{align}
\label{eq:rec_ren_full}
&\varphi^{(0)A}(\vec{\lambda}) + \varepsilon \varphi^{(1)A} =
  \varphi^{(0)A}(\vec{\lambda}_{\Ren}) + \varepsilon \varphi_{\perp}^{(1)A} \,, \\
& \varphi^{(1)A} -\left[\mathcal{T}^{i(\mu)}_j\psi^{j(\mu)}\right] \frac{\delta \varphi^{(0)A}}{\delta \lambda^{i}}
  = \varphi_{\perp}^{(1)A}\label{eq:difference_op}\,.
\end{align}
To determine the coefficients in $\psi$,
we propose minimizing the norm of $\varphi^{(1)A}_{\perp}$ by
defining a ``cost function''
\begin{equation}
\label{eq:cost_function}
I=\left\|\varphi^{(1)A}_{\perp}\right\|^2=\int\left|\varphi^{(1)A}-
\left[\mathcal{T}^{i(\mu)}_j\psi^{j(\mu)}\right]\frac{\delta\varphi^{(0)A}}{\delta\lambda^i}\right|^2 dt
\,,
\end{equation}
using a Euclidean norm for the components labeled by $A$, and
which may also involve a spatial integration in the case of solving PDEs.
Let us define
\begin{equation}
\label{eq:new_bas_func}
e^{(\mu)A}_j\equiv \mathcal{T}^{i(\mu)}_j \frac{\delta\varphi^{(0)A}}{\delta\lambda^i}
\end{equation} 
as a convenient linear combination of the basis functions $\delta\varphi^{(0)}/\delta\lambda$,
and time/background dependence in $\mathcal{T}$. With respect to these
vectors, the cost function becomes a quadratic form,
\begin{align}
I={}&\int\left|\varphi^{(1)A}-\psi^{j(\mu)}e^{(\mu)A}_j\right|^2 dt\nonumber,\\
={}&\mathbb{M}^{(\mu)(\nu)}_{ij}\psi^{i(\mu)}\psi^{j(\nu)}-2\mathbb{V}^{(\nu)}_i\psi^{i(\nu)}+\mathbb{D}
\label{eq:quad_form}\,,
\end{align}
where the coefficients $\mathbb{M}^{(\mu)(\nu)}_{ij}$,
$\mathbb{V}^{(\nu)}_i$ and $\mathbb{D}$ read
\begin{align}
\label{eq:coeffs_quad}
\mathbb{M}^{(\mu)(\nu)}_{ij}&=\int \left(e^{(\mu)A}_ie^{(\nu)A}_j\right) dt\,,\\
\mathbb{V}^{(\nu)}_{i}&=\int \left(e^{(\nu)A}_i\varphi^{(1)A}\right) dt\,,\\
\mathbb{D}&=\int \left(\varphi^{(1)A}\varphi^{(1)A}\right) dt\,,
\end{align}
where summation is implied on repeated $A$ indices.
If $\mathbb{M}^{(\mu)(\nu)}_{ij}$ is an invertible and positive definite matrix,
then the optimization
\begin{equation}
\label{eq:minim_quad}
\frac{\pd I}{\pd \psi^{i(\mu)}}=0\,, 
\end{equation}
minimizes the cost functional for a fixed value of $(\mu)$. Such minimization  
only needs the inversion of $\mathbb{M}^{(\mu)(\nu)}_{ij}$, which yields
\begin{equation}
\label{eq:inv_psi}
\psi^{i(\mu)}=\left(\mathbb{M}^{-1}\right)^{(\mu)(\nu)ij}\mathbb{V}^{(\nu)}_j
\,.
\end{equation}
At every point in the background parameter space, performing this
minimization yields values of $\vec{\alpha}^{(1)}$,
$\vec{\beta}^{(1)}$, and possibly higher derivative corrections from
Eqs.~(\ref{eq:aux_1}--\ref{eq:aux_4}).  If higher derivatives are
extracted, these must be consistent with the $\lambda$-dependence of
$\vec{\alpha}^{(1)}$ and $\vec{\beta}^{(1)}$.  It is essential to
mention that not all of the extracted components of $\psi^{i(\mu)}$ are 
relevant to provide a ``good fit'' of the perturbative solutions. Hence, it is worthwhile to 
assess how each component of the flow affects the quality of the fit. In the hypothetical
case in which the fit of the perturbative solution fails, it is important to revise the 
expansion order kept [e.g.\ in Eq.~\eqref{eq:vec_v}], and potentially
include more terms in the fit.

We can extend the perturbative scheme to consider higher-order corrections in 
$\varepsilon$, recalling that (as every perturbative scheme) it is necessary to solve for 
all the parameters, flows, and derivative corrections at lower
perturbative orders, as they are sources for higher orders.
Even though extending our procedure to higher perturbative
orders is not an objective of this paper, we will try to explain how this procedure might work. 
There are two alternative approaches one could follow.  The first one repeats the method
described above, expanding order-by-order, and extracting $\vec{\alpha}$,
$\vec{\beta}$ (including their corresponding auxiliary higher-derivative corrections) up to the
perturbative order required. The perturbative equations of motion in Eq.~\eqref{eq:pert_eq_mov}
need to be expanded to higher order.  The second option is identical
except, at each order, replacing the flow of $\vec{\lambda}(t)$ with
the renormalized flow $\vec{\lambda}_{\Ren}(t)$ [thus replacing the
background solution in Eq.~\eqref{eq:pert_eq_mov} with the
renormalized solution $\varphi^{(0)}(\vec{\lambda}_{\Ren})$] built
from all lower orders.
In either of these cases, it is always essential to construct the naive perturbative solution
in order to understand how it diverges in powers of $(t-t_0)$. Knowing this ensures that
the renormalized parameter flow contains sufficient terms to reconstruct the solutions
at every perturbative order. 

\figbackfull

To close this section, we summarize the algorithm one follows to build
the renormalized solution at first order in $\varepsilon$:
\begin{enumerate}
\item Compute the differentials $\delta\varphi^{(0)A}/\delta\lambda^i$
  for all parameters $\lambda^{i}$ whose flows you may attempt to
  renormalize.  These differentials may be computed analytically, if
  an analytical solution is available, or numerically.  Notice what
  powers of $(t-t_{0})$ appear in each basis function.
\item Solve the equations of motion in Eq.~\eqref{eq:pert_eq_mov} and
  evaluate the naive perturbative solution, at many background points
  $\vec{\lambda}_{0}$ in the parameter space.  Notice what powers of
  $(t-t_{0})$ appear in the naive perturbative solution
  $\varphi^{(1)A}$.
\item \label{item:candidates}
  Consider a candidate set of parameters to try to fit.  This will
  have to be determined individually for each problem, either by
  understanding the phenomenology of this problem, or by examining the
  features of $\varphi^{(1)A}$ and
  $\delta\varphi^{(0)A}/\delta\lambda^i$, and the different powers of
  $(t-t_{0})$ that appear in each.  This will inform what order needs
  to be kept in expanding Eq.~\eqref{eq:exp-composition} using the BCH
  theorem [an example being Eq.~\eqref{eq:vec_v}].
\item Build the cost function in Eq.~\eqref{eq:cost_function} and
  extract $\psi$ for every simulation (each corresponding to a point
  $\vec{\lambda}_{0}$ in the parameter space).
  Examine $\varphi_{\perp}^{(1)A}$ to assess the quality of
  the fit of the perturbative solution as a
  combination of the basis function and the extracted flows.  If the
  residual $\varphi^{(1)A}_{\perp}$ still exhibits secularly-growing
  features, go back to
  item~\ref{item:candidates} and consider more parameters, or
  expanding $\vec{V}$ to higher order in $(t-t_{0})$.
\item Once the fit has captured all the secularly-growing features, we
  can trust the extracted values of $\vec{\alpha}^{(1)}$ and
  $\vec{\beta}^{(1)}$, which can then be interpolated over the
  $\Lambda$ space.  Solve the flow equations in
  Eq.~\eqref{eq:flow_equations} using
  $\vec{\lambda}_{\Ren}(t_0)=\vec{\lambda}_{0}+\varepsilon\vec{\alpha}^{(1)}(\vec{\lambda}_{0})$
  as initial conditions for the renormalized parameters. The
  renormalized solution is $\varphi^{(0)}(\vec{\lambda}_{\Ren})$,
  where $\vec{\lambda}_{\Ren}$ solves the flow equations.
\end{enumerate}

\section{Perturbing the KdV equation\texorpdfstring{\\*}{}
  and DRG extraction procedure}
\label{sec:KdV}

We now proceed to our main example, which is treating the Korteweg-de\
Vries-Burgers (KdVB) equation fully, using naive perturbation theory
(which suffers from secular divergence), and with the numerical
dynamical renormalization group approach.  The one-dimensional KdVB
equation~\cite{CANOSA1977393} can be written as
\begin{equation}
  \label{eq:kdvb}
  \partial_t\varphi=-6\varphi\partial_x\varphi-\partial^3_x\varphi+\varepsilon\partial^2_x\varphi
  \,.
\end{equation}
Dropping the third derivative term gives Burgers' equation, while
setting $\varepsilon=0$ gives the KdV equation.  The KdV equation
is integrable and admits soliton solutions.  Throughout, we will treat
the $\varepsilon$ term as a deformation of the KdV equation.
This is a dissipation or diffusion term acting on a soliton, as seen in
Fig.~\ref{fig:kdv_sols}.  It does not modify the principal part of the
PDE, and thus does not affect the well-posedness of the problem.

\fignaivepert

From now on, we use $\varphi_{\mathrm{full}}$ when
referring to the full solution of the KdVB equation in Eq.~\eqref{eq:kdvb}. 
Using the naive perturbation ansatz, $\varphi =
\varphi^{(0)} + \varepsilon \varphi^{(1)}$, we expand the solution 
up to the first order in $\varepsilon$. The background equation of
motion is the well-known KdV equation,
\begin{equation}
\label{eq:kdv_back}
  \partial_t\varphi^{(0)}=-6\varphi^{(0)}\partial_x\varphi^{(0)}-\partial^3_x\varphi^{(0)}
  \,.
\end{equation} 
We are interested in background solutions which are a single soliton,
of the form
\begin{align}
  \varphi^{(0)} &= \frac{v}{2}\sech^2\left[\frac{\sqrt{v}}{2}(x-x_{0}-vt)\right]
  \,,\\
  \label{eq:kdv_soliton}
  \varphi^{(0)} &= \frac{v}{2}\sech^2\left[\frac{\sqrt{v}}{2}(x-x_{\mathrm{c}}(t))\right]
  \,.
\end{align}
This solution is parameterized by the two-dimensional parameter space
of $\vec{\lambda}=(x_{0}; v)$, where $x_{0}$ is the \emph{initial}
peak position, or $\vec{\lambda}=(x_{\text{c}}; v)$,
where the \emph{instantaneous} peak position $x_{\mathrm{c}}$ is
given by
\begin{align}
\label{eq:peak_pos}
  x_{\mathrm{c}}(t) = x_0+\int_{t_0}^t v(t')\, dt'
  \,.
\end{align}
Using $x_{0}$ or $x_{\text{c}}$ as a coordinate choice in parameter space
affects whether the background beta function vanishes or not.
In the $x_{0}$ coordinate, $\vec{\beta}^{(0)}=0$.
However, the time derivative of Eq.~\eqref{eq:peak_pos} shows that
\begin{align}
\label{eq:back_flow_kdv}
\frac{dx_{\text{c}}}{dt} = v,\quad \vec{\beta}^{(0)}=v\frac{\pd}{\pd x_{\text{c}}}\,,
\end{align}
i.e., the parameter $v$ determines the zeroth-order beta function for the 
peak position $x_{\text{c}}$.
This is analogous to how $\Omega_0$ generates the flow
of $\phi_{\circ}(t)$ in Sec.~\ref{sec:analytical-example}.
Throughout we will use $x_{\text{c}}$, since at first order $\varepsilon$, we
will anyway develop a non-zero beta function.

Since the KdVB equation is translation invariant, the dynamics can not
depend on $x_{0}$, except for a trivial translation.  Aside from the
initial position, the one-soliton solution of Eq.~\eqref{eq:kdv_back}
is determined by the velocity $v$, which simultaneously controls the
amplitude ($v/2$), the width (proportional to $v^{-1/2}$), and the
motion of the peak position (found by the integral in
Eq.~\eqref{eq:peak_pos}).

We obtained numerical solutions for Eq.~\eqref{eq:kdvb} and
the perturbation equation [Eq.~\eqref{eq:kdv_pert}, below]
using a pseudospectral method for space and the method of lines for
time integration.  We provide full details of the numerical method in
Appendix~\ref{app:numerical}, the space and time scales
involved, and the sources of error in the extraction of the beta
functions (discussed in Sec.~\ref{sec:results}).

In Fig.~\ref{fig:kdv_sols}, we plot the solutions for
Eqs.~\eqref{eq:kdvb} and \eqref{eq:kdv_back}, using a KdV soliton as an
initial condition released at $x_0=0.0$ and $v=2.0$ in both of
them. In the KdVB equation, the perturbative damping coefficient is
$\varepsilon=0.1$. A key observation of the full solution is
that, in principle, it is reasonable to build an approximate solution
by modifying all the shape parameters of $\varphi^{(0)}$ -- i.e., the
amplitude, the width, the position, and velocity of the solitonic
peaks -- by time-dependent
functions. This intuition provides us with a motivation to build such
an approximate solution, which we will call $\varphi_{\mathrm{ren}}$
from now on, where the ``bare'' shape parameters are promoted to 
become functions of time. The main idea is that the initial conditions 
and the flow in time of the promoted parameters can be found by the 
renormalization procedure shown in Sec.~\ref{sec:RG_PT}. 
Later in Sec.~\ref{sec:results}, we will also show that it is
consistent to build $\varphi_{\mathrm{ren}}$ by renormalizing the bare
$(x_{c};v)$ in the analytic KdV soliton of Eq.~\eqref{eq:kdv_soliton},
rather than promoting the amplitude and width to independent
parameters.

It is reasonable to expect that the renormalized solution does not contain 
all the information of the KdVB solution, such as the small step growing 
horizontally behind the decaying peak. These deviations, shown in a rectangle 
in the right panel of Fig.~\ref{fig:kdv_sols}, become smaller as the damping 
parameter $\varepsilon$ reduces. We will show below that such deviations 
can be tabulated by computing the residual $\varphi^{(1)}_{\perp}$ as defined 
in Eq.~\eqref{eq:difference_op}. We defer to future work the problem
of refining the renormalized solution with these small deviations.

We now proceed to (naive) first-order perturbation theory, where the
equation of motion reads
\begin{equation}
\label{eq:kdv_pert}
  \partial_t\varphi^{(1)} = \mathrm{KdV}^{(1)}\left[\varphi^{(1)}\right] + P
  \,.
\end{equation}    
Here the linear operator $\mathrm{KdV}^{(1)}$ acting on $\varphi^{(1)}$, and the source term $P$,
are background-dependent, given by
\begin{align}
  \mathrm{KdV}^{(1)}\left[\varphi^{(1)}\right]&
  \equiv
  \left[
    -6\varphi^{(0)}\partial_x - 6\left(\partial_x\varphi^{(0)}\right)
    -\partial^3_x
  \right]\varphi^{(1)}
  \,,\nonumber\\
  P &\equiv \partial^2_x\varphi^{(0)}
  \,.
\end{align}
It is important to keep in mind the explicit space/time dependence of
$\varphi^{(0)}$ when solving this PDE for $\varphi^{(1)}$.
In Fig.~\ref{fig:kdv_pert}, we show the solution $\varphi^{(1)}$ of Eq.~\eqref{eq:kdv_pert}, and
the reconstruction $\varphi^{(0)}+\varepsilon \varphi^{(1)}$ in naive perturbation
theory, using $\varepsilon=0.01$. In this case, the perturbative
solution has initial conditions $\varphi^{(1)}(t=0, x)=0$, and the background
solution is taken to be a KdV soliton with $v=2.0$ started at
$x_0=300.0$ at time $t=0$.

At early times, linear perturbation theory captures the effects of the
damping term in slowing down the soliton.  However, the perturbative
solution eventually grows to an amplitude of $1/\varepsilon$,
signaling the breakdown of naive perturbation theory.  Similarly, the
peak position differs by $\mathcal{O}(1)$ (in units of the soliton
width) on a secular timescale $T_{\text{sec}}\sim \varepsilon^{-1/2}$.
The first step to build an improved solution
is to find which parameters need to be renormalized, by studying how
the naive solution grows in time.
We can compare the diverging features of $\varphi^{(1)}$ and the differentials
$\delta \varphi^{(0)}/\delta \lambda^{i}$ to determine the vectors
$\vec{\alpha}^{(1)}$ and $\vec{\beta}^{(1)}$ in order to renormalize
the solution.

Space and time-translational invariance, as is the case of all
symmetries, play an important role in determining the structure of
the beta functions and, consequently, the renormalized parameters'
dependence. The solitonic solutions of Eqs.~\eqref{eq:kdvb}
and \eqref{eq:kdv_back} are translational invariant since none of
the terms contained in the equations of motion have an explicit spatial
dependence. We expect, therefore, that the renormalized
parameters do not depend on the peak position $x_{\text{c}}$.
We also make the choice that the background kinematic relationship
between $v$ and $x_{\text{c}}$ [Eq.~\eqref{eq:back_flow_kdv}] continues to
hold at higher orders in perturbation theory.
Therefore we assume that the alpha and beta vectors take the form
\begin{align}
\label{eq:alpha_beta_1D}
\vec{\alpha}^{(1)}=\alpha^v(v)\frac{\pd}{\pd v},\quad \vec{\beta}^{(1)}=\beta^v(v)\frac{\pd}{\pd v}
\,,
\end{align}
i.e.\ that we only renormalize the velocity.  In principle, we can also add
the shift $\alpha^{x_{\text{c}}}(v)\pd_{x_{\text{c}}}$, but this does
not change our results drastically.

As in Sec.~\ref{sec:RG_PT}, we construct the first-order solution in
two ways: using naive perturbation theory, and with renormalized
(flowing) parameters,
\begin{equation}
\label{eq:kdv_match}
\varphi=\varphi^{(0)}+\varepsilon\varphi^{(1)}\quad\text{and}\quad \varphi=\varphi^{(0)}(\vec{\lambda}_{\Ren}(t))+\varepsilon\varphi^{(1)}_{\perp}
\,.
\end{equation}
Here $\varphi^{(0)}$ is the one-soliton KdV solution in
Eq.~\eqref{eq:kdv_soliton}, $\varphi^{(1)}$ is the perturbative
solution of Eq.~\eqref{eq:kdv_pert}, and $\varphi^{(1)}_{\perp}$ is the residual
to be minimized.  For short times, the renormalized parameters
$\vec{\lambda}_{\Ren}(t)\equiv\left(x^{\Ren}_{\text{c}}(t);v_{\Ren}(t)\right)$
can be computed by using the BCH formula as in Eq.~\eqref{eq:vec_v},
\begin{align}
\vec{\lambda}_{\Ren} &= \vec{\lambda}+\vec{V}\,,\nonumber\\
\vec{V}&=\varepsilon\left(\alpha^v\Delta t+\frac{\beta^v}{2}\Delta t^2\,;~\alpha^v+\beta^v\Delta t\right)
\,,
\label{eq:kdv_single_param}
\end{align}
where we made use of the background flow of Eq.~\eqref{eq:back_flow_kdv}.
We find that the renormalized peak position $x_{\text{c}}(t)$ (derived from the composite flow in
Eq.~\eqref{eq:exp-composition}) is consistent with our physical intuition of a point particle
moving with constant acceleration.
Here $\varepsilon\alpha^{v}$ gives an initial velocity shift, and 
$\varepsilon \beta^{v}$ gives an acceleration (the background moves at
constant velocity).  The form of the background
flow and the dependence in Eqs.~\eqref{eq:back_flow_kdv} and \eqref{eq:alpha_beta_1D}
have canceled all of the derivative corrections in
Eqs.~(\ref{eq:aux_1}--\ref{eq:aux_4}).

In analogy to the procedure in Eq.~\eqref{eq:difference_op}, we match
the two expressions in Eq.~\eqref{eq:kdv_match} at first order in
$\varepsilon$,
\begin{align}
  \label{eq:kdv_part}
  \varphi^{(1)}_{\perp}={}&\varphi^{(1)}-d\varphi^{(0)}(\vec{V})
  \,,
\end{align}
where we must use \eqref{eq:kdv_single_param}, and the differential map,
\begin{align}
d\varphi^{(0)}(\vec{V})\equiv{}&\frac{\delta\varphi^{(0)}}{\delta x_{\mathrm{c}}}
\left(\alpha^v\Delta t+\frac{\beta^v}{2}\Delta t^2\right)\nonumber\\
&+\frac{\delta\varphi^{(0)}}{\delta v}\left(\alpha^v+\beta^v\Delta t\right)\label{eq:kdv_fit}\,.
\end{align}
The components of the differential map appearing are
\begin{align}
\frac{\delta \varphi^{(0)}}{\delta v}&=\frac{1}{2}\sech^2\xi\left[1+\xi\tanh\xi\right]\,,\label{eq:proj_coeff_v}\\
\frac{\delta \varphi^{(0)}}{\delta x_{\mathrm{c}}}&=\frac{v^{3/2}}{2}\sech^2\xi\tanh\xi \,,\label{eq:proj_coeff_xc}
\end{align}
where $\xi \equiv \sqrt{v}(x-x_{\mathrm{c}})/2$ and
$x_{\mathrm{c}}=x_0+v\Delta t$.
Now we have all the elements necessary
extract $\alpha^{v}$ and $\beta^{v}$, for any value of $v$, by optimizing
the cost functional in Eq.~\eqref{eq:quad_form}. In general, by
dimensional analysis, the coefficients multiplying
$\vec{\alpha}$ will always have a smaller power of $\Delta t$ than the
corresponding coefficients multiplying $\vec{\beta}$. This means that at
longer integration times, the optimization routine is more sensitive
to $\vec{\beta}$ than it is to $\vec{\alpha}$.

\section{Results}\label{sec:results}

This section presents the results of the numerical DRG extraction procedure described 
in Sec.~\ref{sec:KdV} for the KdVB problem. We use the extracted values of $\alpha^v$ 
and $\beta^v$ to build the renormalized solution $\varphi_{\mathrm{ren}}$. Our 
approach is not restricted to the original 2D parameterization of the one-soliton 
KdV solution, as written in Eqs.~\eqref{eq:kdv_soliton}.
We later consider the case of having additional independent shape parameters, such 
as the amplitude and the width of the soliton peak. We will also test if the renormalized 
solution is a good approximation by comparing it with the single-peaked solution of 
the KdVB equation. 

\figdeltaphirecon

\subsection{Original KdV parameterization}\label{subsec:single}

\tabAlphaBetaExtracted

In the setup described in Sec.~\ref{sec:KdV}, which considers 
a 2D parameter space $\vec{\lambda}\equiv(x_{\mathrm{c}},v)$, 
we made the ansatz to fit only the two components $\vec{\psi}=(\beta^v,\alpha^v)$.  
Consequently, the cost function $I$ to be minimized reduces to a 
2-dimensional quadratic form
\begin{align}
\label{eq:quad_alpha_beta}
I ={}&
\left[{\begin{array}{cc} \beta^v & \alpha^v \end{array}}\right]
\left[\begin{array}{cc} \mathbb{M}^{(0)(0)} & \mathbb{M}^{(0)(1)}\\ \mathbb{M}^{(1)(0)} & 
\mathbb{M}^{(1)(1)}\end{array}\right]
\left[\begin{array}{c} \beta^v \\ \alpha^v \end{array}\right]\nonumber\\
&{}-2 \left[{\begin{array}{cc} \beta^v & \alpha^v \end{array}}\right]
\left[\begin{array}{c} \mathbb{V}^{(0)} \\ \mathbb{V}^{(1)} \end{array}\right]+\mathbb{D}
\,.
\end{align}
Here
$\mathbb{D}$ does not participate in the optimization procedure. We
compute the matrix and vector coefficients of the vector of
differentials $\vec{e}\equiv (e^{\beta}; e^{\alpha})$, as detailed in
Eq.~\eqref{eq:new_bas_func},
\begin{align}
\label{eq:basis_single}
\vec{e}=\bigg(\frac{\delta\varphi^{(0)}}{\delta x_{\mathrm{c}}}\frac{\Delta t^2}{2}+\frac{\delta\varphi^{(0)}}{\delta v}\Delta t;
~\frac{\delta\varphi^{(0)}}{\delta x_{\mathrm{c}}}\Delta t+\frac{\delta\varphi^{(0)}}{\delta v}\bigg)
\,,
\end{align}
yielding the coefficients in Eq.~\eqref{eq:quad_alpha_beta},
\begin{align}
\mathbb{M}^{(\mu)(\nu)}&=\int_L dx\int_{t=t_0}^{t=T_{\max}}dt\:e^{(\mu)}e^{(\nu)}\,,\label{eq:single_mat}\\
\mathbb{V}^{(\mu)}&=\int_L dx\int_{t=t_0}^{t=T_{\max}}dt\:\varphi^{(1)}e^{(\mu)}\,,\label{eq:single_vec}
\end{align}
where $L$ is the length of the simulation box, and $T_{\max}$ is the 
total evolution time of the perturbative solution. We find the vector of 
optimum values containing $\beta^v$ and $\alpha^v$ by performing 
the same matrix inversion introduced in Eq.~\eqref{eq:inv_psi}; since 
the matrix is just $2\times 2$, this is
\begin{align}
  \label{eq:minimum_alpha_beta_v}
  \begin{bmatrix}
    \beta^{v} \\
    \alpha^{v}
  \end{bmatrix} =
  \frac{1}{\det \mathbb{M}}
  \begin{bmatrix}
    \mathbb{M}^{(1)(1)} & -\mathbb{M}^{(0)(1)} \\
    -\mathbb{M}^{(1)(0)} & \mathbb{M}^{(0)(0)}
  \end{bmatrix}
  \mathbb{V}
  \,.
\end{align}
Once we have the extracted values of $\alpha^v$ and $\beta^v$, it is crucial to test the 
quality of the linear decomposition of $\varphi^{(1)}$ in terms of the basis functions in 
Eqs.~\eqref{eq:proj_coeff_v} and \eqref{eq:proj_coeff_xc}. From the definition of 
$\varphi^{(1)}_{\perp}$ in Eq.~\eqref{eq:kdv_part}, we compare $\varphi^{(1)}$ and 
$d\varphi^{(0)}(\vec{V})$ in Fig.~\ref{fig:fitting_pert}, showing a good fit of the 
perturbative solution as a linear decomposition in basis functions 
for $v=2.0$.
The quality of the fit is due to both the correct choice of basis
functions and the correct values of $\alpha^v$ and $\beta^v$. 
The quality of the fit also shows if we have considered an appropriate time-dependence 
of the infinitesimal shift $\vec{V}$. We define the relative difference
\figlnpart
\begin{equation}
\Delta\varphi^{(1)}_{\mathrm{rel}}(t, x)\equiv \frac{\varphi^{(1)}_{\perp}(t,x)}{\max_{x'}
\varphi^{(1)}(t,x')},
\label{eq:delta_pert_rel}
\end{equation}
to corroborate the goodness of the fit even at late times. In Fig.~\ref{fig:fitting_error_pert},
we observe that the relative difference is never greater than $10^{-5}$ for $T_{\max}=50$.
Interestingly, the residual $\varphi^{(1)}_{\perp}$ (plotted in white) has the
same shape as the ``diluting tails'' shown in the right panel of Fig.~\ref{fig:kdv_sols}, 
suggesting that it is possible to also recover the ``instantaneous''
perturbative features (those not captured by renormalization)
of the full solution with a refinement of this method.

\figextinvTmax
\figalphabetav

The upper limit in the integration time ($T_{\max}$
in Eqs.~\eqref{eq:single_mat} and \eqref{eq:single_vec}) plays a significant
role in evaluating the stability of the extracted alpha and beta
functions. If $T_{\max}$ is too short, the specific choice of initial
conditions becomes a dominant feature of the solution. Therefore, 
it is prudent to evaluate the perturbative solution $\varphi^{(1)}$ for a sufficiently 
long time. A minimal consistency condition for the evolution of the
system is that, knowing the that the width of the soliton is roughly given by
$v^{-1/2}$, $T_{\max}\gg v^{-3/2}$ ensures that the perturbative solution has displaced a 
distance much larger than a single soliton width.
We perform the integration for a variety of values of $T_{\max}$, and
then use (quadratic) Richardson
extrapolation (RE)~\cite{10.2307/90994, Press:1992zz}
in powers of $T_{\max}^{-1}$ to find the generators $\beta^v$ and
$\alpha^v$ in the limit $T_{\max}\to \infty$, and estimate their corresponding 
errors, for different values of $v$.
In Fig.~\ref{fig:2d_alpha_beta}, we show the way the RE works, finding
the values of $\alpha^v$ and $\beta^v$ (in colored squares) reported in 
Table~\ref{tab:beta_v}.  From this figure, we notice how the
extracted values of $\alpha^v$ and $\beta^v$ smoothly converge to the
extrapolated values as $T_{\max}\to\infty$. Moreover, it is
clear that the extracted values of $\beta^v$ and $\alpha^v$ (empty
circles) do not show large variations as $T_{\max}$ grows. We observe
that all of the values of $\beta^v$ are negative, in agreement with the
notion of a decelerating peak, as shown by the non-linear solution
plotted the in the right panel of Fig.~\ref{fig:kdv_sols}.
We estimate the errors arising from RE (denoted $\sigma\text{(RE)}$)
from the difference between the extrapolated values and the extracted
values from the largest $t=T_{\max}$ of each simulation.

Table~\ref{tab:beta_v} shows the values of $\alpha^v$ and $\beta^v$
extrapolated from each of the simulations, and their corresponding
errors.  The main sources of error are (a)~the numerical calculation
of the perturbative solution $\varphi^{(1)}$, and (b)~the fact that
simulations are evaluated at finite (but large) values of $T_{\max}$.
Even when these sources of error can be combined, we chose to
treat them independently. The numerical convergence error is further 
discussed in Appendix~\ref{app:numerical}.

For a general application of the numerical DRG, one would have to rely
on interpolation of $\vec{\alpha}^{(1)}, \vec{\beta}^{(1)}$ over the
$\vec{\lambda}$ parameter space, in order to numerically solve the
$\beta$ function equations.  For our particular problem of the KdVB
equation, we can make an argument that $\alpha^{v}(v)$ and
$\beta^{v}(v)$ will be pure power laws in $v$ (this analytical
argument only came after our numerical explorations).  The background
($\varepsilon=0$) KdV equation has a scaling symmetry, such that if
$\varphi(t,x)$ is a solution, then so is
$\gamma^{2} \varphi(\gamma^{3} t, \gamma x)$.  This corresponds to a
simultaneous change of length and time units, under which velocity
should change to be $\gamma^{-2}v$; and dimensional analysis shows
that $\varepsilon$ should change to be $\gamma^{-1} \varepsilon$.  We
expect the renormalized solution will inherit the background's
symmetry.  We allow some undetermined transformations
$\alpha^{v} \to \gamma^{c} \alpha^{v}$ and
$\beta^{v} \to \gamma^{d} \beta^{v}$. Applying the scaling
transformation to the infinitesimal flow, we have
\begin{align}
  v &\to v + \varepsilon(\alpha^{v} + \beta^{v} \Delta t) \,, \\
  \gamma^{-2} v &\to \gamma^{-2}v + \gamma^{-1}\varepsilon( \gamma^{c} \alpha^{v}
  + \gamma^{d}\beta^{v} \gamma^{3}\Delta t) \,.
\end{align}
We find the powers $c$ and $d$ in order to make this homogeneous
in $\gamma$, namely, $c=-1$ and $d=-4$.  This is satisfied with
$\alpha^{v} \propto v^{1/2}$ and $\beta^{v} \propto v^{2}$.
As we were completing this manuscript, we learned of
Ref.~\cite{PhysRevE.66.046625}, whose results also imply a power-law
for $\beta^{v}$.  Therefore,
instead of using interpolating functions, we fit power laws
for $\alpha^{v}$ and $\beta^{v}$.

Given the extrapolated 
values in Table~\ref{tab:beta_v},
we use the ansatz
\begin{equation}
  \ln |\beta^v| = m \ln v + b
  \,,
\label{eq:fit_beta_v}
\end{equation}
and similarly for $\alpha^{v}$.  The best fit power law is plotted
in the upper panels of Fig.~\ref{fig:beta_v}.
The fractional error bars are plotted in the bottom panels, which are
too small to see without magnification in the top panels; we omit
error bars in later plots.
The quality of the fit needs to be as good as possible, since errors
in the fit will incur secular errors in the renormalized solution.
Later, in Sec.~\ref{subsec:num_kdvb}, we will compare the renormalized
solution against the full solution of the KdVB equation.

We used the standard non-linear fit routine \texttt{curve\_fit} in 
\texttt{scipy}~\cite{2020SciPy-NMeth}, with weights coming from the
estimated RE errors (the convergence errors are much smaller, as seen
in the lower panels of Fig.~\ref{fig:beta_v}; see
Appendix~\ref{app:numerical} for full details on convergence testing).
The \texttt{curve\_fit} routine returns the optimal fits, in
Table~\ref{tab:alpha_beta_fit}, and covariance matrix estimates on
the two parameters $(m,b)$ for each of the two fits,
\begin{align}
\label{eq:num_cov_beta}
\Sigma^2_{\beta^v}&= \begin{bmatrix}
  8.7\times10^{-9} & -5.5\times 10^{-9} \\
  -5.5\times 10^{-9} & 3.6\times 10^{-9}
\end{bmatrix},\\
\label{eq:num_cov_alpha}
\Sigma^2_{\alpha^v}&= \begin{bmatrix}
  4.8\times10^{-6} & 5.1\times 10^{-6} \\
  5.1\times 10^{-6} & 1.3\times 10^{-5}
\end{bmatrix}\,.
\end{align}
These fits agree (very well for $\beta^{v}$, less so for $\alpha^{v}$)
with the scaling argument for the power laws $\beta^{v}\propto v^{2}$
and $\alpha^{v}\propto v^{1/2}$. The quality of both fits improves
(more substantially for $\alpha^v$) if we omit the point with $v=0.0625$.

\tabAlphaBetaFit

With these $\alpha$ and $\beta$ functions in hand, we can now
construct the renormalized solution $\varphi_{\text{ren}} =
\varphi^{(0)}(\vec{\lambda}_{\Ren})$ as a semi-analytic expression of
the form
\begin{equation}
\varphi^{(0)}(\vec{\lambda}_{\Ren})= 
\frac{v_{\Ren}(t)}{2}\sech^2
\left[\frac{\sqrt{v_{\Ren}(t)}}{2}(x-x^{\Ren}_{\mathrm{c}}(t))\right]
\,.
\label{eq:ren}
\end{equation}
This approximate solution is simply constructed by replacing
$v\rightarrow v_{\Ren}(t),~x_{c}(t)\to x_{c}^{\Ren}(t)$ in the background
solution of Eq.~\eqref{eq:kdv_soliton}, where these components of
$\vec{\lambda}_{\Ren}$ satisfy the flow equations
\begin{align}
  \label{eq:flow_eq_v}
  \frac{dv_{\Ren}}{dt} &=0 + \varepsilon\beta^v(v_{\Ren})=-\varepsilon \, e^{b} \, v_{\Ren}^m
  \,, \\
  \label{eq:new_peak}
  \frac{dx_{c}^{\Ren}}{dt} &= v_{\Ren} + \varepsilon \beta^{x_{c}} = v_{\Ren}
  \,,
\end{align}
subject to the initial condition $v_{\Ren}(t_0)=v_{0}+\varepsilon\alpha^v(v_{0})$
reparameterized by $\alpha^v$, whereas $x_{c}^{\Ren}(t_{0})$ is not
shifted, according to the argument above Eq.~\eqref{eq:alpha_beta_1D}.
The values in Table~\ref{tab:alpha_beta_fit} suggest that the
analytical $\beta$ function is
\begin{align}
\beta^v = -\frac{4}{15}v^2\,.
\label{eq:analytic_beta_v}
\end{align}
As we were completing this manuscript,
we learned of an analytical calculation in~\cite{PhysRevE.66.046625}
which implies this same $\beta$ function.  If one takes a time
derivative of their Eq.~(53), and performs some algebra, one can
recover our Eq.~\eqref{eq:analytic_beta_v}.

\figphirenres

In Fig.~\ref{fig:phi_ren_vs_full}, we show the 
renormalized solution for $\varepsilon=0.1$, and compare it with the one-soliton solution of the KdVB equation. 
From the left panel of this figure, it is clear that the amplitude of the renormalized 
solution does not increase as the naive perturbative solution plotted in the right 
panel of Fig.~\ref{fig:kdv_pert}. Moreover, this evolving solution is not substantially different 
from the KdVB solution depicted in the right panel of Fig.~\ref{fig:kdv_sols}, 
except for the absence of the small step dubbed as ``diluting tails'' in the evolution 
of $\varphi_{\text{full}}$. In the right panel, we depict the difference
between the renormalized expression and the full solution by introducing a
fractional difference variable, $\Delta\varphi_{\mathrm{rel}}$, defined as
\begin{align}
\Delta\varphi_{\mathrm{rel}}(t,x)\equiv \frac{\varphi_{\mathrm{full}}(t,x)-
\varphi_{\mathrm{ren}}(t,x)}{\max_{x'} \varphi_{\mathrm{full}}(t,x')},
\label{eq:delta_rel}
\end{align}
where the expression in the denominator corresponds to the decreasing
amplitude of the peak at each instant of time.  The main differences
between the full and the renormalized solutions are the presence of
diluting tails in the solution (a horizontally-growing ``bump'' to the
left of the two peaks, which was also
visible in the black rectangle of the right panel of
Fig.~\ref{fig:kdv_sols}), and the secular position error between the
peaks.  It is interesting to note that the magnitude of the ``diluting
tails'' coincides with the residual $\varphi^{(1)}_{\perp}$ multiplied
by $\varepsilon$, as plotted in Fig.~\ref{fig:fitting_error_pert}.

The initial shift $v\rightarrow v+\varepsilon\alpha^v$ changes the initial soliton's 
amplitude, velocity, and width by a small amount compared to the original shape
parameters.  At time $t=0$, this small change only amounts to
around 3\% of the amplitude.  However, if we had omitted the
$\alpha^v$ reparameterization, the maximum fractional error
$\Delta\varphi_{\mathrm{rel}}$ secularly grows, increasing by $10\%$
by $t=50$, due to starting with the wrong initial velocity.

The main question about the renormalized solution is whether it
captures secular effects of the true solution, which has several
secular timescales.  For quantities whose background flow vanishes,
the estimate $T_{\text{sec}}\sim(\varepsilon\beta^{(1)})^{-1}$
of Sec.~\ref{sec:general-formalism} is still valid.  These quantities
include the velocity, and derived features such as the width and
amplitude.  Investigating our solution, we find that the velocity
error is bounded, even on much longer times, $t \gg
(\varepsilon\beta^{(1)})^{-1}$.  The amplitude and width follow the
same behavior.  These differences will be further detailed in
Sec.~\ref{subsec:num_kdvb}.

However, the peak position is
sensitive to an even shorter timescale, due to the ``deceleration'' of
the peak, and this leads to the dominant error.  The background flow
Eq.~\eqref{eq:back_flow_kdv} generates the acceleration-like term
$\varepsilon\Delta t^2$ in the infinitesimal generator $\vec{V}$ in
Eq.~\eqref{eq:kdv_single_param}.  This leads to the shorter secular
time scale
\begin{align}
  T_{\text{sec}} \sim \frac{1}{\sqrt{\varepsilon\beta^v}}
  \,.
\label{eq:new_time}
\end{align}
Notice that the difference between the renormalized peak position and
the full solution is only half the peak's width at $t=50$.  This time
$t=50$ is vastly longer than the secular time
$(\varepsilon\beta^v)^{-1/2} \approx 1.9$ for $v=2$ and
$\varepsilon=0.1$.
Naive perturbation theory had already failed by this time
$T_{\text{sec}}$.
Thus, the renormalized solution presented in
Eq.~\eqref{eq:ren} represents $\varphi_{\text{full}}$ far better than
the naive perturbative expression
$\varphi^{(0)}+\varepsilon\varphi^{(1)}$ in Eq.~\eqref{eq:sol_pert_m}.

In what remains of this section, we will introduce the amplitude,
width, and peak position as additional independent parameters of the
system.
We evaluate their corresponding alpha and beta functions using (a)~our
minimization scheme, and (b)~by direct evaluation of the KdVB solution
(in Sec.~\ref{subsec:num_kdvb}).
We will verify that the renormalized solution can be
written using only two flowing parameters, $(x_{c}(t), v)\rightarrow
(x^{\Ren}_{c}(t), v_{\Ren}(t))$, similar to the
background KdV soliton.  This also allows us to perform a more
detailed comparison of the features between the renormalized and full solutions.
The reader can safely skip these subsections to learn
about different applications of this technique in Sec.~\ref{sec:FW}.

\subsection{Alternative parameterizations:\texorpdfstring{\\*}{}
  the multiparameter case}
\label{subsec:multi_KdV}

One potential unknown in the numerical DRG procedure is whether the
parameterization for the renormalized solution is sufficiently
general.  It can happen that the background problem has one
dimensionality, but upon being perturbed, the dimensionality
increases~\cite{Ei:1999pk}.
In this subsection, we check if this happens in our KdV example by
proposing a higher-dimensional parameterization for the background KdV
soliton.  This allows us to confirm that our previous two-dimensional
parameterization was actually sufficient, by testing the consistency
between different parameters' flows.
We parametrize the zeroth-order solution by labelling the shape
parameters as
\begin{equation}
  \varphi^{(0)}=\mathcal{A}\sech^2\left[\mathcal{M}\left(x-x_{\mathrm{c}}\right)\right]
  \,,
\label{eq:new_param}
\end{equation}
where we can identify the amplitude $\mathcal{A}=v/2$ and inverse
width $\mathcal{M}=\sqrt{v}/2$, in terms of the original parameters of
the KdV solution.  If these background relationships are maintained
upon renormalization, then we would have the relationships
\begin{align}
  \alpha^{\mathcal{A}} &=\frac{\alpha^v}{2}
  \,, &
  \alpha^{\mathcal{M}} &=\frac{\alpha^v}{4\sqrt{v}}
  \,,\label{eq:alpha_a_m_v}\\
  \beta^{\mathcal{A}} &=\frac{\beta^v}{2}
  \,,&
  \beta^{\mathcal{M}} &=\frac{\beta^v}{4\sqrt{v}}
  \,.\label{eq:check_1}
\end{align}
If these relationships are maintained, then the flows are tangent to a
2-dimensional solution manifold described by $(x_{\mathrm{c}};v)$, within
the ambient 4-dimensional space $\vec{\lambda}=(\mathcal{A};
\mathcal{M}; x_{c}; v)$.
This is similar to the example in Sec.~\ref{sec:analytical-example},
where we saw that the $R_{\Ren}$ and $\Omega_{\Ren}$ components of the
$\beta$ function in Eq.~\eqref{eq:beta-G-R} are not independent --
they are related by preserving the form of Kepler's law.

To check the dimensionality, we calculate the
first-order beta functions for $\mathcal{A}$ and $\mathcal{M}$ following
the same renormalization-based scheme suggested in Sec.~\ref{sec:RG_PT},
as well as in previous instances of the current subsection. To do so, first we 
pose the $\vec{\alpha}, \vec{\beta}$ ansatz
\begin{align}
\label{eq:flow_multi}
&\vec{\beta}^{(0)}=v\frac{\pd}{\pd x_{\text{c}}}\,,\\
&\vec{\alpha}^{(1)}=\alpha^{\mathcal{A}}(v)\frac{\pd}{\pd\mathcal{A}}+
\alpha^{\mathcal{M}}(v)\frac{\pd}{\pd\mathcal{M}}+\alpha^v(v)\frac{\pd}{\pd v}\,,\\
&\vec{\beta}^{(1)}=\beta^{\mathcal{A}}(v)\frac{\pd}{\pd\mathcal{A}}+
\beta^{\mathcal{M}}(v)\frac{\pd}{\pd\mathcal{M}}+\beta^v(v)\frac{\pd}{\pd v}\,,
\end{align}
where the addition of a shift in the initial peak position $\alpha^{x_{\text{c}}}(v)
\partial_{x_{\text{c}}}$ does not alter our results significantly. 
This gives the flow equations
\begin{align}
  \frac{d\mathcal{A}_{\Ren}}{dt}&=0+\varepsilon\beta^{\mathcal{A}}(v_{\Ren})
  \,,&
  \frac{d\mathcal{M}_{\Ren}}{dt}&=0+\varepsilon\beta^{\mathcal{M}}(v_{\Ren})\,,\\
  \frac{dv_{\Ren}}{dt}&=0+\varepsilon\beta^{v}(v_{\Ren})
  \,,&
  \frac{dx^{\Ren}_{\text{c}}}{dt}&=v_{\Ren}+\varepsilon\beta^{x_{\text{c}}}=v_{\Ren}\,,
\end{align}
that is, $(\mathcal{A}; \mathcal{M}; v)$ have vanishing background
flows, and the flow of $x_{c}$ maintains its kinematic meaning.

\figalphasbetasTs

Now we compute the first-order deformation flow $\vec{V}$ using the
BCH theorem in Eq.~\eqref{eq:BCH3}, which in the
$(\mathcal{A}; \mathcal{M}; x_{\text{c}}; v)$ coordinates is given by
\begin{align}
\label{eq:reparam_multi}
\vec{\lambda}_{\Ren}={}&\vec{\lambda}+\vec{V}\,,\nonumber\\
\vec{V}={}&\varepsilon\big(\alpha^{\mathcal{A}}+\beta^{\mathcal{A}}\Delta t;~\alpha^{\mathcal{M}}+\beta^{\mathcal{M}}\Delta t;\nonumber\\
&\quad\alpha^v\Delta t+\frac{\beta^v}{2}\Delta t^2;~\alpha^v+\beta^v\Delta t\big)\,.
\end{align}
Next we compute the differentials by taking partial derivatives of Eq.~\eqref{eq:new_param}, 
which are given by
\begin{align}
&\frac{\delta\varphi^{(0)}}{\delta\mathcal{A}}=\sech^2\xi,\label{eq:conn_coeff_A}\\
&\frac{\delta\varphi^{(0)}}{\delta\mathcal{M}}=-4\mathcal{M}\xi\sech^2\xi\tanh\xi,\label{eq:conn_coeff_W_inv}\\
&\frac{\delta \varphi^{(0)}}{\delta x_{\mathrm{c}}}=4\mathcal{M}^3\sech^2\xi\tanh\xi,\label{eq:conn_coeff_xc_inv}\\
&\frac{\delta\varphi^{(0)}}{\delta v}=0
\,.\label{eq:conn_coeff_v_inv}
\end{align}
Here $\xi\equiv\sqrt{v}(x-x_{\mathrm{c}})/2$, and
$x_{\mathrm{c}}=x_0+v\Delta t$ follows from the background flow definitions in 
Eqs.~\eqref{eq:back_flow_kdv} and \eqref{eq:flow_multi}. It is worth mentioning
that the new parameterization splits the original dependence in $v$, seen in 
Eq.~\eqref{eq:kdv_soliton}, between $\mathcal{A}$ and $\mathcal{M}$.
The only $v$ dependence is in the implicit background flow of the peak position
$x_{c}$ -- there is no explicit $v$ dependence.  This causes the
differential $\delta \varphi^{(0)}/\delta v$ to vanish in Eq.~\eqref{eq:conn_coeff_v_inv}. 

To extract $\vec{\alpha}^{(1)}$ and $\vec{\beta}^{(1)}$, we
reuse the previous numerically-computed naive first-order solutions of
Eq.~\eqref{eq:kdv_pert}, with the same velocities as before.
We calculate $\varphi^{(1)}_{\perp}$ as the residual after fitting the
perturbative solution $\varphi^{(1)}$ as a linear combination of basis
functions, in the same way detailed in Sec.~\ref{sec:RG_PT}, and
build a new cost function $I$ for this case,
\begin{align}
I={}&\int_L dx\int_{t_0}^{T_{\max}}dt\:\bigg\{\varphi^{(1)}-\frac{\delta\varphi^{(0)}}{\delta x_{\mathrm{c}}}
\left(\alpha^v\Delta t+\frac{\beta^v}{2}\Delta t^2\right)\nonumber\\
&-\left[\frac{\delta\varphi^{(0)}}{\delta\lambda^j}\left(\alpha^{j}+
\beta^{j}\Delta t\right)\right]\bigg\}^2\,.\label{eq:multi_cost}
\end{align}
Here $j$ sums over the parameters in the subspace $\vec{\lambda}=(\mathcal{A};
\mathcal{M}; v)$.  As before we minimize $I$, looking for the critical
point $\delta I/\delta \vec{\psi}=0$, giving the six-dimensional
vector $\vec{\psi}=(\vec{\alpha}^{(1)}, \vec{\beta}^{(1)})$.

Examining the extracted values $\vec{\alpha}^{(1)}$ and
$\vec{\beta}^{(1)}$, as functions of $T_{\max}$,
is crucial to verify if the time dependence and perturbative order proposed in 
Eq.~\eqref{eq:reparam_multi} is sufficient to capture the parameter flows.
In Fig.~\ref{fig:all_alpha_beta}, we show that all the beta functions smoothly
converge to fixed values when $T_{\max}$ is sufficiently long, which provides
clear evidence of finding the correct time dependence in $\vec{V}$.
Interestingly, there are no significant differences between (a)~the values of
$\alpha^v$ and $\beta^v$ extracted from the minimization procedure in the 2D 
parameter case (in Sec.~\ref{subsec:single}), and (b)~the ($\alpha^v,\beta^v$) values
extracted using the 4D parameterization in this section. The dashed
lines represent values of $T_{\max}$ longer than the simulation
time, where the alpha and beta functions have not been extracted.

In all of the cases we plotted, the beta functions have converged to a
stable value within the simulation time. Meanwhile, even though
$\alpha^v$ is stable with $T_{\max}$, the same cannot be said about
all of the alpha functions: $\alpha^{\mathcal{A}}$ and
$\alpha^{\mathcal{M}}$ have not converged to stable values by a time
$T_{\max} \approx 10^3$.
This may be due to the minimization of $I$ being more sensitive to
$\beta^{i}$ than to $\alpha^{i}$, since, by dimensional analysis,
there is one more factor of $\Delta t$ in front of $\beta^{i}$.
It is possible that this parameterization is insufficient, or that
going to higher order in $\varepsilon$ or
$\Delta t$ would improve the convergence of these $\alpha$'s.

\figtesttang
\figsolparamevol

Since $\alpha^{\mathcal{A}}$ and $\alpha^{\mathcal{M}}$ have not
converged, we can not check the consistency conditions in
Eq.~\eqref{eq:alpha_a_m_v}.  But, we can check the $\beta$ function
tangency conditions in Eq.~\eqref{eq:check_1}.
In the upper panel
of Fig.~\ref{fig:a_w_tang}, we show $|\beta^v|$ and
what should be two equivalent expressions, if tangency is satisfied:
$2|\beta^{\mathcal{A}}|$, and
$4\sqrt{v}|\beta^{\mathcal{M}}|$. In the lower panel, we plot the fractional
errors $|(\beta^v/2-\beta^{\mathcal{A}})/\beta^{\mathcal{A}}|$ and
$|(v^{-1/2}\beta^v/4-\beta^{\mathcal{M}})/\beta^{\mathcal{M}}|$, finding
that the deviations from a tangent flow are very small.
The errors in the tangency conditions
for $v<0.5$ can be reduced by increasing the
resolution (though this is computationally expensive, since we must
increase $T_{\max}$ as $v$ becomes smaller).

The conclusion seems to be that the $\beta$ functions for the
\emph{flow} are consistent with being tangent to the two-dimensional
submanifold.  Meanwhile, the $\alpha$ functions setting the initial
conditions can not be tested for consistency, since only $\alpha^{v}$
has converged.  This type of test would be prudent when applying the
numerical DRG, unless one knows \emph{a~priori} the functional form of
the renormalized solutions.

\subsection{Comparing DRG
  \texorpdfstring{$\alpha$}{alpha} and
  \texorpdfstring{$\beta$}{beta} functions\texorpdfstring{\\*}{}
against the full KdVB solution}\label{subsec:num_kdvb}
In this subsection, we extract the amplitude, width, position, and
velocity of the peak from the single-peaked solution of the full KdVB
equation of Eq.~\eqref{eq:kdvb}, as well as their evolution in time. The
study of the full solution enables us to explore the
$\varepsilon$-dependence of DRG, and the accuracy of the
$\alpha$ and $\beta$ functions extracted using the procedure described in Section
\ref{sec:RG_PT}. To do so, we first need to determine the peak
position, amplitude, and width at each time of both
$\varphi_{\text{ren}}$ and $\varphi_{\mathrm{full}}$. For the
renormalized solution, we numerically integrate the
flow Eqs.~\eqref{eq:flow_eq_v} and \eqref{eq:new_peak} using our
numerical fits.  This
immediately gives $v_{\Ren}(t)$ and $x^{\Ren}_{c}(t)$.
We get the renormalized amplitude $\mathcal{A}_{\Ren} = v_{\Ren}/2$
from the background relationship.  For the width, we use the full
width at half max (FWHM): the difference in $x$ values where the value
of $\varphi(x)$ is half of its peak value.  From the form of the
soliton solution, this is given by
\begin{align}
  \label{eq:w_ren}
  \mathcal{W}_{\Ren}=\frac{4\cosh^{-1}\sqrt{2}}{\sqrt{v_{\Ren}(t)}}\approx\frac{3.525}{\sqrt{v_{\Ren}(t)}}
  \,.
\end{align}
We caution that although the symbol $\mathcal{M}$ used throughout
Sec.~\ref{subsec:multi_KdV} has units of inverse width, it is not
exactly the reciprocal of the FWHM $\mathcal{W}$ that we use in this
section -- they differ by a multiplicative constant.

To find the same parameters from the full solution, we use Fourier
interpolation~\cite{Boyd1989ChebyshevAF} to evaluate
$\varphi_{\text{full}}$ at points not tabulated in the collocation
grid. We use Newton's method to root-solve for the peak location
$x^{\text{full}}_{c}(t)$, determined by
\begin{align}
  \label{eq:xc_det}
  \frac{\pd\varphi_{\text{full}}}{\pd x}\bigg{|}_{x=x^{\text{full}}_{\mathrm{c}}}=0
  \,.
\end{align}
We can calculate the instantaneous velocity $v^{\text{full}}$
of the peak for $\varphi_{\mathrm{full}}$ by calculating the numerical
time derivative of the peak position. In our implementation, we used a
fourth-order accurate finite difference
(we only evaluate at interior times so that we only need to implement
the central finite difference).
Once we find the peak position, we obtain the amplitude of the
peak at each time,
\begin{align}
\mathcal{A}_{\mathrm{full}}(t) = \varphi_{\text{full}}(t, x^{\text{full}}_{c}(t))
\,,
\end{align}
again using spectral interpolation.

Calculating the FWHM
of the peak $\mathcal{W}_{\text{full}}$ from $\varphi_{\mathrm{full}}$ requires finding the
set of two points $x_{1/2(>)}$ and $x_{1/2(<)}$, to the right and
left of the peak, satisfying
\begin{align}
\varphi_{\mathrm{full}}(t,x_{1/2(\lessgtr)})=\frac{\mathcal{A}_{\mathrm{full}}(t)}{2}
\end{align}
at each instant of time.  We again use Fourier interpolation and
Newton root-polishing.  Then the FWHM at a given instant of time is
$\mathcal{W}_{\mathrm{full}}=x_{1/2(>)}-x_{1/2(<)}$.  Notice that the
peak is asymmetric due to the presence of the ``diluting tails'' in
the full solution.

Once we have calculated the values of the shape parameters from both
the full and the renormalized solutions, we compare our results by
defining the difference
\begin{align}
  \label{eq:W_error}
  \Delta \mathcal{W} \equiv \mathcal{W}_{\mathrm{full}}-\mathcal{W}_{\Ren}
  \,,
\end{align}
and similarly for the amplitude $\mathcal{A}$, the peak velocity $v$,
and the peak position $x_{c}$.  In Fig.~\ref{fig:error_param_unif}, we
show the evolution of the errors of all of these quantities, at
four
different values of the perturbation parameter $\varepsilon$, for
simulations with $v=2$.  We see that $\Delta v$, $\Delta \mathcal{A}$,
and the relative error $\Delta \mathcal{W}/\mathcal{W}_{\text{full}}$
are bounded in time. We scale these three quantities by $\varepsilon^{-1}$,
showing that each is proportional to $\varepsilon$.  Similarly we used
$\varepsilon t$ as the time axis for these three panels, showing that
our solutions are valid at secularly-large times, $t \gg \varepsilon^{-1} \gg
\varepsilon^{-1/2}$. For the values of $\varepsilon$ reported in the figure,
we observe that the difference $\Delta \mathcal{W}$ is never
larger than 4.5\% of the FWHM $\mathcal{W}_{\text{full}}$.

Meanwhile, the position error in units of width 
$\Delta x_{c}/\mathcal{W}_{\text{full}}$ is proportional to
$\varepsilon$ but growing linearly in time, due to error in the
initial velocity (we discuss this more below).  Still, the position
error is at most half a width by time $t=50$, as seen in
Fig.~\ref{fig:error_xcs}.

From the extracted position $x^{\text{full}}_{c}$, we can also try to
directly reconstruct $\beta^{v}$.  Still using the kinematic intuition
of the flow of $x_{c}$, we use another finite difference to compute
\begin{align}
\label{eq:beta_v_cent}
  \frac{1}{\varepsilon}\frac{d^2 x^{\text{full}}_{\mathrm{c}}}{dt^2}
  =\frac{1}{\varepsilon}\frac{dv^{\text{full}}}{dt}
  =\beta^v_{\text{full}}
  \,.
\end{align}
In Fig.~\ref{fig:beta_const}, we compare the reconstructed beta
functions using the acceleration of the peak position for different
values of $\varepsilon$, with the numerical DRG beta function plotted
in the left panel of Fig.~\ref{fig:beta_v}.
We solved the full KdVB equation for each $\varepsilon$ to a maximum
time of $T_{\max}=50$.
Therefore, the length of the curves
increases as the damping $\varepsilon$ grows: the larger the value of
$\varepsilon$, the wider the range of velocities explored before the
fixed $T_{\max}$.
The curves with different values of $\varepsilon$ converge towards the
numerical DRG curve as $\varepsilon\to 0$, confirming the validity of
our procedure.  Figure~\ref{fig:beta_const} uses the kinematical
velocity $v$ as the horizontal axis, rather than the $v$ coordinate,
which is related by the $\alpha^{v}$ diffeomorphism.  From this
dataset alone we can not perform this reparameterization; however, if
we use $\alpha^{v}_{\text{full}}$ found below, and plot $v-\varepsilon
\alpha^{v}_{\text{full}}(v)$ on the horizontal instead of $v$, then
all of the curves coincide.

\figbetaconv

\figbetaspread

We can similarly numerically extract $\beta^{v}$ from time derivatives
of $\mathcal{W}_{\text{full}}$ and $\mathcal{A}_{\text{full}}$.
From Eq.~\eqref{eq:w_ren}, we should have
\begin{align}
\label{eq:beta_v_w}
\beta_{\text{full}}^v \approx -\frac{\varepsilon^{-1}}{2 \cosh^{-1}\sqrt{2}} v_{\text{full}}^{3/2} \frac{d\mathcal{W}_{\mathrm{full}}}{dt}
\,.
\end{align}
Here the approximate equality is due to the peaks of the full solution
not being symmetric.
Similarly, from the original parameterization of the amplitude, we
should have
\begin{align}
\label{eq:beta_v_a}
\beta_{\text{full}}^v=\frac{2}{\varepsilon} \frac{d\mathcal{A}_{\mathrm{full}}}{dt},
\end{align}
which is also seen in Eq.~\eqref{eq:check_1}.

In Fig.~\ref{fig:beta_spread},
we plot the spread of the curves representing all of the ``equivalent'' forms of $\beta^v$
[from Eq.~\eqref{eq:beta_v_cent}, \eqref{eq:beta_v_w},
and Eq.~\eqref{eq:beta_v_a}]. For each choice of $\varepsilon$,
we shaded the areas containing all of the curves with a different color. 
Notice that the curves tend to spread more as $\varepsilon$ becomes 
larger, and therefore the colored regions grow in the same manner. 
As before, the case $\varepsilon=0.01$ limits the range
of velocities, and determines the size of the horizontal axis in which all the regions
can be compared.
The DRG-extracted value of $\beta^v$ (yellow curve in Fig.~\ref{fig:beta_const})
overlaps with the region shaded in black in
Fig.~\ref{fig:beta_spread} after reparameterizing the horizontal axis
with $\alpha^{v}_{\text{full}}$.  Therefore this result is consistent
with the system being described by the two-dimensional flow, as seen
in Sec.~\ref{subsec:multi_KdV}.

\figalphadiv

We can also determine $\alpha_{\text{full}}^v$ from
$\varphi_{\text{full}}$.  To do so, we parameterize each time snapshot
of the full solution of Eq.~\eqref{eq:kdvb} as
\begin{equation}
\varphi_{\mathrm{full}}=\frac{v^{\mathrm{fit}}}{2}\sech^{2}\left[\frac{\sqrt{v^{\mathrm{fit}}}}{2}(x-x^{\mathrm{fit}}_{\mathrm{c}})\right]
\,,
\end{equation}
and fit $v^{\mathrm{fit}}$ and $x^{\mathrm{fit}}_{\mathrm{c}}$
using \texttt{curve\_fit}.  We can reconstruct the reparameterization
$\alpha_{\text{full}}^{v}$ by computing the difference
\begin{align}
\label{eq:alpha_num_extract}
|\alpha_{\mathrm{full}}^v|=\frac{1}{\varepsilon}\bigg|v^{\mathrm{fit}}-
\frac{dx^{\mathrm{fit}}_{\mathrm{c}}}{dt}\bigg|
\,.
\end{align}
That is, the reparameterization captures the difference between the
kinematical velocity $d x^{\text{fit}}/dt$ versus the shape parameter
named $v^{\text{fit}}$.
In Fig.~\ref{fig:alpha_err}, we plot this quantity versus
the DRG-extracted value of $\alpha^{v}$ (in the right panel of Fig.~\ref{fig:beta_v}).
Our results show that there is no convergence towards the renormalized
alpha function as $\varepsilon\rightarrow 0$, as in the case for the beta function 
$\beta^v$. The slope seems to be correct, but the value of the alpha function
extracted using our renormalization procedure is smaller than
$|\alpha^v_{\mathrm{full}}|$ by a factor of 2. If we use $\alpha^v_{\text{full}}$ 
as a shift in the initial velocity to reconstruct the renormalized solution, we 
notice a slight reduction in the relative error, compared to what is reported 
in the right panel of Fig.~\ref{fig:phi_ren_vs_full}. We do not currently 
understand the origin of this difference. We leave the resolution of this 
mismatch for a future project.

\section{Potential applications}
\label{sec:FW}

As already surveyed in several textbooks~\cite{goldstein,
  jose_saletan_1998, bender1999advanced, kevorkian1981perturbation} and
articles~\cite{Chen:1995ena, Kunihiro:1995zt, Ei:1999pk}, there are a
wide range of physical problems where secular effects need to be
captured properly.  The DRG unifies several approaches to secular
perturbation theory and can thus be applied to any such secular
problem.  We expect our addition of a numerical formulation of DRG
will further extend its applicability to include problems which can only 
be solved numerically. One potential application is to compute
the beta functions for long-lived cosmological solutions, such as
oscillons~\cite{Amin:2011hj, Lozanov:2019ylm, Amin:2018xfe,
Olle:2020qqy, Zhang:2020bec, Cyncynates:2021rtf}.  It might be
possible to produce oscillons
(quasibreathers) from continuous deformations of the sine-Gordon
breather~\cite{Ablowitz:1973fn}. Concretely, our method can be applied
to find an estimated lifetime of such oscillons. 
  
Our original motivation to implement the numerical DRG arose from a
certain problem in gravitational physics.
We are interested in the gravitational waves emitted by black hole
binary systems.  As we have already seen in
Sec.~\ref{sec:analytical-example}, the post-Newtonian regime (where
$1/c$ is a perturbation parameter) can be treated using the DRG
analytically. As an update of the results in~\cite{Galley:2016zee}, the work 
of Yang and Leibovich uses the DRG to include spin-orbit effects in 
the inspiral~\cite{Yang:2019oqm}.  
The extreme mass-ratio inspiral (EMRI)
problem~\cite{Poisson:2011nh}, which is treated perturbatively in
powers of the small mass ratio, should also be amenable to the
DRG.  There are several secular timescales in the EMRI problem, all of
which need to be controlled (see e.g.~\cite{Miller:2020bft}). 

The specific problem of interest is in how the inspiral and resulting
gravitational waves are modified by the presence of corrections to
Einstein's theory of general relativity~\cite{Berti:2015itd}.  For
most beyond-GR theories, the status of the initial value problem is
open, though it is expected that most of these theories lack a good
initial value formulation~\cite{Lehner:2014asa}.  Instead, the only
sound way to treat such a theory is as a perturbation around GR, where
a parameter $\varepsilon$ controls the strength of the deformation
away from GR; this is the viewpoint of effective field theory
(however, for a different nonperturbative proposal,
see~\cite{Cayuso:2017iqc, Allwright:2018rut, Cayuso:2020lca}).
This is also in line with observations, which to date are consistent
with the predictions of general
relativity~\cite{LIGOScientific:2019fpa}.

Indeed, treating beyond-GR theories as perturbations to GR has been
successful for finding stationary
solutions~\cite{Yunes:2011we,Yagi:2013mbt}, where there are no secular
effects; and even for addressing the post-Newtonian regime of the
binary inspiral problem, which does suffer from secular effects
(Refs.~\cite{Yagi:2011xp,Yagi:2013mbt} treated these secular effects
with traditional secular perturbation theory, rather than the DRG).

The challenge now is to handle the late inspiral and merger phase of a
binary black hole system in a beyond-GR theory such as dynamical
Chern-Simons gravity (dCS)~\cite{Alexander:2009tp}.  The merger phase
can only be treated with full numerical relativity, not by any
analytical means.  The beyond-GR perturbation is similarly treated
numerically, expanded about the nonlinear GR
solution~\cite{Okounkova:2017yby, Okounkova:2018pql,
  Okounkova:2019zjf, Okounkova:2020rqw}.
This perturbative solution however suffers from secular growth, as
predicted in~\cite{Okounkova:2017yby} and confirmed
in~\cite{Okounkova:2018pql, Okounkova:2019zjf, Okounkova:2020rqw}.
Similarly, secular growth appears when applying naive perturbation
theory to Einstein-dilaton-Gauss-Bonnet gravity~\cite{Okounkova:2020rqw}.

The origin of this secular growth is easy to understand.  The
background (GR) solution inspirals at a particular rate; the
correction to GR includes a change in the energy radiated, thus
changing the rate of inspiral.  This also has a simple analogy with
the KdV equation, which motivated this study.  Both the KdV problem
and the binary black hole inspirals in GR have nonlinear background 
solutions, and these background solutions both have non-vanishing flows
$\vec{\beta}^{(0)}$.  In both cases, the perturbation removes energy
from the system, causing the true speed (of the soliton or inspiral)
to deviate from the background speed.

It is this secular growth that we seek to control.  As a reminder, the
initial-value problem for most beyond-GR theories can only be
formulated in naive perturbation theory.  Below we sketch this naive
perturbation theory approach, which breaks down, and then how the
numerical DRG will be used to renormalize the secularly-growing
solutions.  We do not claim that this is the only or the best approach
to this problem.  Indeed if there was another viable approach
available (e.g.\ that proposed in~\cite{Cayuso:2017iqc,
  Allwright:2018rut, Cayuso:2020lca}), it would be prudent to compare
the independent methods to assess their merits.  However, no other
general-purpose approach is available which has been shown to simulate
arbitrary beyond-GR theories.

The equations of motion of such beyond-GR theories can be cast as a
deformation of Einstein's field equations,
\begin{align}
  \label{eq:GR-deformed}
  G_{ab} + \varepsilon C_{ab} = 8\pi T_{ab}
  \,,
\end{align}
where $C_{ab}$ is generating the correction to GR, and is controlled
by the parameter $\varepsilon$.  The metric is expanded as an ordinary
perturbation series,
\begin{align}
  g_{ab} &= g^{(0)}_{ab} + \varepsilon g^{(1)}_{ab} + \mathcal{O}(\varepsilon^{2})
  \,,\label{eq:g_exp}
\end{align}
and similarly for any other degrees of freedom.  The background
solution $g^{(0)}_{ab}$ satisfies the nonlinear Einstein field
equations (and already contains gravitational waves).  The correction
due to beyond-GR effects, $g^{(1)}_{ab}$, satisfies the linearization
of Eq.~\eqref{eq:GR-deformed} and can be integrated alongside
$g^{(0)}_{ab}$, as was first demonstrated
in~\cite{Okounkova:2018pql}.  It is this $g^{(1)}_{ab}$ which suffers
from secular growth, as seen in~\cite{Okounkova:2019zjf}.

Let us review the ingredients needed to implement the numerical DRG in
this case.  The first necessary condition is that the problem can be
described by a finite-dimensional attractor manifold.  This may not be
clear since GR is a field theory and thus has an infinite number of
degrees of freedom.  But, as long as the initial data is close to a
binary of black holes, any small gravitational fluctuations will
radiate away rapidly, leaving a system with a finite-dimensional
solution manifold, parameterized by the two black holes' masses,
spins, and separation (here we ignore eccentricity).  In this
finite-dimensional parameter space $\Lambda$, so-called surrogate
models~\cite{Blackman:2017pcm, Varma:2018mmi} have been highly
successful in giving a faithful numerical model for the asymptotic
waveform at infinity, which we will denote as simply
\begin{align}
  \label{eq:h-surr}
  h^{\text{Surr}}[\vec{\lambda}]
  \,,
\end{align}
where $\vec{\lambda}\in\Lambda$ are the system parameters.  This
quantity may come from a spline interpolant (or other reduced order
model), and therefore we also have access to the differentials
\begin{align}
  \label{eq:2}
  \frac{\delta h^{\text{Surr}}}{\delta \vec{\lambda}}
  \,,
\end{align}
which are then also spline interpolants.  The parameters $\vec{\lambda}$ 
already experience a background flow, $d\vec{\lambda}/dt = \vec{\beta}^{(0)}(\vec{\lambda})$, 
since the binary inspirals, and the spins (and orbit) precess.  Using this background flow,
we can build the infinitesimal flow $\vec{V}$ using the BCH theorem,
and thus have a model for $h^{(1)}_{\parallel}$, the secularly-growing
part of $g^{(1)}_{ab}$, in terms of the first-order $\vec{\alpha}^{(1)}$
and $\vec{\beta}^{(1)}$.  Finally, since we have access to the
numerical first-order solution $g^{(1)}_{ab}$, we fit the model
$h^{(1)}_{\parallel}$, getting numerical $\vec{\alpha}^{(1)}$ and
$\vec{\beta}^{(1)}$ as fit parameters, for some individual beyond-GR
simulation.  After fitting, we also have the residuals $h^{(1)}_{\perp}$.

If we repeat the fit for many beyond-GR simulations, we can then
interpolate $\vec{\alpha}^{(1)}$ and $\vec{\beta}^{(1)}$.  Finally, we can
solve the deformed flow equations
\begin{align}
  \frac{d\vec{\lambda}_{\Ren}}{dt} = \vec{\beta}^{(0)} + \varepsilon \vec{\beta}^{(1)}
  \,,
\end{align}
to find $\vec{\lambda}_{\Ren}(t)$.  This renormalized flow captures
the different rate of inspiral due to the beyond-GR effects.
Finally, we can evaluate the renormalized asymptotic waveforms,
\begin{align}
  h_{\Ren} = h^{\text{Surr}}[ \vec{\lambda}_{\Ren}(t) ]
  \,,
\end{align}
which do not suffer any secular effects.  Although this captures most
of the beyond-GR effects, there are still $\mathcal{O}(\varepsilon)$
``instantaneous'' corrections in $h^{(1)}_{\perp}$, which should also be
incorporated.

\section{Discussion}
\label{sec:conclusion}

In this paper, we proposed a systematic numerical method to applying
the dynamical renormalization group to finite- or infinite-dimensional
dynamical systems, even in situations when analytical perturbation
theory is not possible.
To make this possible, we formulated the DRG in the language of
differential geometry, in Sec.~\ref{sec:general-formalism}.
From the geometric point of view, naive perturbation theory finds
tangent vectors in solution space, which are then integrated together
to find the whole DRG flow.
This geometric formulation is general enough that the DRG can be
applied to systems that already have a background flow in parameter
space, so that the DRG may be iterated to higher order.

As a proof of concept, in Sections~\ref{sec:KdV} and
\ref{sec:results}, we applied this method to the Korteweg-de\ Vries
equation, deforming it to the Korteweg-de\ Vries-Burgers equation.  We
used naive perturbation theory, numerically, and as expected, found
secularly-growing solutions.  We fit these solutions using appropriate
basis functions, which are computed from derivatives of the background
solution with respect to parameters,
$\delta\varphi^{(0)}/\delta \vec{\lambda}$, along with knowledge of
the background flow $\vec{\beta}^{(0)}$.  By minimizing an appropriate
cost functional, we extract values of the generators
$\vec{\alpha}^{(1)}$ and $\vec{\beta}^{(1)}$, for each numerical
simulation.  Just finding these values already gives deep information
about the structure of parameter space.  Now one can numerically solve
the deformed parameter flow
$d\vec{\lambda}_{\Ren}/dt = \vec{\beta}^{(0)} + \varepsilon
\vec{\beta}^{(1)}$, for example by interpolating through parameter
space.  Finally we find the renormalized solution
$\varphi^{(0)}[\vec{\lambda}_{\Ren}(t)]$.

This example highlights a number of key features of the numerical DRG
approach.  Most importantly, we have controlled the secular divergence
on the shortest timescale in the problem,
$T_{\text{sec}} \sim \varepsilon^{-1/2}$.  The numerical beta function
we extracted is highly suggestive of a power law $\propto v^{2}$, in
agreement with an analytical calculation suggested
by~\cite{PhysRevE.66.046625} (a numerical fit to the power law index
differs only in the fourth decimal place).  Despite the excellent
agreement in the beta function, the reparameterization generator
$\alpha^{v}$ seems to disagree with an independent check we performed
in Sec.~\ref{subsec:num_kdvb}.  Nonetheless, even using this wrong
value of $\alpha$ gives results that are better than using no $\alpha$
reparameterization at all, and the solution is still valid on
secularly long times.  Finally, in Sec.~\ref{subsec:multi_KdV} we
demonstrated how to test if the perturbation has increased the
dimensionality of the parameter space, by considering a more general
parameterization to the one-soliton KdV solution.  For the KdV
problem, we found that solutions lie in a submanifold with the
original dimensionality, so dimensionality was not increased.

There are a plethora of potential applications for the numerical DRG.
We discussed some possibilities in Sec.~\ref{sec:FW}, including
finding oscillon lifetimes, secular divergences in extreme mass-ratio
inspirals, and gravitational waves from beyond-GR theories.  There are
also still unanswered questions raised by the KdV example of this
work.  For example, we do not yet understand the apparent factor of
two discrepancy in the $\alpha$ function, found in
Sec.~\ref{subsec:num_kdvb}.  Our example demonstrated control of the
secular effects, which are most important for the breakdown of
perturbation theory, but there are also ``instantaneous'' perturbative
effects that we did not include.  We noticed in
Sec.~\ref{subsec:single} that both the residual
$\varphi^{(1)}_{\perp}$ and the true solution $\varphi_{\text{full}}$
contain ``diluting tails.''  However in this work we restricted
attention to the renormalization procedure and the solution
$\varphi_{\text{ren}} = \varphi^{(0)}(\vec{\lambda}_{\Ren}(t))$, so we
made no effort to capture the $\mathcal{O}(\varepsilon)$ instantaneous
effects.  The formalism to include information from
$\varphi^{(1)}_{\perp}$ still needs to be developed.

\acknowledgments
The authors would like to thank Jonathan Braden, Cliff Burgess, Andrei Frolov,
Chad Galley, Nigel Goldenfeld, Igor Herbut, Luis Lehner,
Jordan Moxon, Maria Okounkova, Ira Rothstein, Sashwat Tanay, Alex
Zucca, and two anonymous referees,
for many fruitful conversations and their valuable
feedback in earlier versions of this paper.
Some computations were
performed on the Sequoia cluster at the Mississippi Center for
Supercomputing Research (MCSR) at the University of Mississippi.
The work of JG was partially supported by the Natural 
Sciences and Engineering Research Council of 
Canada (NSERC), funding reference \#CITA 490888-16, \#RGPIN-2019-07306.
The work of LCS was partially supported by Award No.\ 80NSSC19M0053
to the MS NASA EPSCoR RID Program, and by NSF CAREER Award PHY--2047382.

\appendix

\section{Numerical setup and errors in the alpha and beta functions}\label{app:numerical}
In this Appendix, we describe the numerical setup to evolve the
first-order perturbations of the KdV equation in Eq.~\eqref{eq:kdv_pert}
and the full KdVB equation in Eq.~\eqref{eq:kdvb}. We followed a
standard algorithm described by Boyd~\cite{Boyd1989ChebyshevAF}
to solve the KdVB equation, and depicted it in Fig.~\ref{fig:scheme}.
This is pseudospectral in space, using the Fourier basis, and the
method of lines for time evolution.
The essence of this algorithm is to compute the time derivative by
finding real-space operations (like products) in the collocation basis, but
computing any derivatives in the spectral domain, and afterwards
transform back to the collocation domain for time evolution. We use
\texttt{fftw3}~\cite{Frigo:2005zln} in our implementation to perform
fast Fourier transforms (FFTs) and their inverses. As every
spectral approach, working in the Fourier domain has several benefits:
\begin{itemize} 
\item It simplifies the representation of spatial derivatives as
  multiplying by powers of the wavevector $k$.
\item It does not require any explicit preparation of boundary conditions since these are periodic by definition.
\item The output data allows spectrally accurate operations, such as
  spatial differentiation, integration along the x-axis, and
  interpolation.
\end{itemize}
We exploited all of these advantages during the post-processing phase
of our simulation results.
It is also possible to build a code with perfectly
matched layers, in addition to periodic boundary conditions. 
This procedure prevents the re-entry of fast/high frequency modes in
the simulation box after transforming oscillatory modes into 
decaying modes by analytic continuation~\cite{Frolov:2017asg}.
We will use such an implementation in a future project. 

\figevolscheme
\figlownoise

We use the method of lines for the collocation data to evolve in the
time domain.
Our time integration routine is an explicit eighth-order accurate
Gauss-Legendre integrator~\cite{10.2307/2003405}, which is A-stable
and symplectic for Hamiltonian problems.
In order to understand the time scales
involved, we write the space and time Fourier transform of the linear operator
$\mathrm{KdV}^{(1)}$, with ``frozen coefficients,''
to derive the high-frequency dispersion relation from the homogeneous
part of Eq.~\eqref{eq:kdv_pert},
\begin{equation}
\omega\varphi^{(1)}_k= \left[6k\varphi^{(0)} + 6i\partial_x\varphi^{(0)}-k^3\right]\varphi^{(1)}_k,
\label{eq:disp_rel}
\end{equation}
where the background and its derivatives are smooth and bounded
functions.
For $v < 10$, the dominant contribution to the dispersion relation
comes from the third spatial derivative (the contribution of the two
other terms is comparable to $k^3$ only when $v\geq 10$).
Therefore we find the time step for evolution is
limited by the Courant-Friedrichs-Lewy condition, in this case,
\begin{align}
\Delta t_{\mathrm{CFL}}\approx\frac{1}{k_{\max}^3}.
\label{eq:time_step}
\end{align}

As there are real and imaginary parts of the
frequency, we can observe the presence of attenuated
oscillatory modes propagating to the left with a phase velocity
proportional to $k^2$. It is important to notice their presence since
it is possible for these modes to travel and propagate through the
periodic domain and deform the solitonic peak and the perturbative
solution. If it is not controlled, the propagation of these
oscillatory modes introduces oscillations in all the evolution plots
of the shape parameters shown in Figs.~\ref{fig:error_param_unif}
and \ref{fig:beta_const}. In order to avoid or minimize those effects,
we use a large simulation domain with length $L=2560$ to allow the
attenuation of oscillatory modes as these propagate.

From the solitonic initial condition in
Eq.~\eqref{eq:kdv_soliton} we notice that as the parameter $v$ grows,
the peak becomes more acute and hence the solution has more power in
higher frequencies. This not only results in smaller time steps
for resolving the system correctly, but the solution becomes prone to
develop high-frequency instabilities. Thus, the selection of the range
of parameter values for $v$ demands us to proceed with caution.
The largest $v$ probed in our study is $v=2$.
In the left panel of Fig.~\ref{fig:spec_conv}, we plot the power
spectrum of the final snapshot at $t=50$ of the full KdVB solution.
Here we used the KdV soliton in Eq.~\eqref{eq:kdv_soliton} with $v=2$
as the initial condition.  In the right panel, we plot the power
spectrum at $T_{\max} = 300$ for the perturbative solution, where the
background is a KdV soliton with $v=2$, and the initial conditions for
the perturbation vanish at time $t=0$.
The floor at high-frequencies corresponds to
round-off error at the level of machine precision, showing that our
results are free of high-frequency instabilities. Still, the
convergence error grows with $v$, as can be seen in
Table~\ref{tab:beta_v}.

\tabResolutions

Depending on the resolution, we used either $N=2^{12}$,
$N=2^{13}$, or $N=2^{14}$ collocation nodes on an equally spaced
Fourier grid.  We calculated the perturbative solution
$\varphi^{(1)}$ at four different resolutions specified in
Table~\ref{tab:resolutions} with the purpose of finding the
convergence errors for all the extracted alpha and beta
values.  In the range of $v<0.5$, where the solution peaks are
wider and have a slower propagation, it is convenient to shift the 
resolutions to also consider $2^{12}$ collocation points with a time-step 
$\Delta t=0.01$.  As can seen in the table,
in the range of $v<0.5$, such a new configuration becomes the ultra low 
resolution, the ``u-low'' case for $v\geq 0.5$ is now the ``low'' resolution 
and each of the remaining resolutions for $v\geq 0.5$ are promoted to 
be the next highest resolution for $v<0.5$. 

\figconvalphabetav

Extracting the values of the beta functions 
requires spatial integration of the coefficients in
Eqs.~\eqref{eq:single_mat} and \eqref{eq:single_vec} 
along the full simulation domain, which can be
computed spectrally accurately by using the Fourier transform,
\begin{align}
  \label{eq:spect_x}
  \int_{-L/2}^{L/2}f(x)dx=L\, \tilde{f}(k=0)
  \,,
\end{align}
where $\tilde{f}(k)$ is the Fourier transform of the integrand.
For the time integrals we used the standard Simpson integration rule,
which is second-order accurate.

All the derivatives of the background solution 
(i.e., $\delta \varphi^{(0)}/\delta \lambda^i$) reported in this paper 
are computed from analytic expressions, and thus do not introduce 
errors in the extraction. From the evaluation of the expressions in 
Eq.~\eqref{eq:minimum_alpha_beta_v} at different resolutions, it is 
possible to calculate the alpha and beta functions using Eq.~\eqref{eq:inv_psi} 
at each resolution for every tabulated value of the varying parameter. 
The convergence errors reported in Table~\ref{tab:beta_v} (dubbed as $\sigma_
{\alpha^v,\beta^v}~(\text{conv.})$) were computed as follows.  First,
we compute $\alpha$ and $\beta$ at all the resolutions in Table~\ref{tab:resolutions}. 
Secondly, we calculate the differences of the values of $\alpha^v$ and $\beta^v$ extracted 
at the highest resolution with the corresponding values at the other
three lower resolutions.  These differences are plotted in
Fig.~\ref{fig:conv_errors}.  Notice that the errors
decrease as the resolution increases, forming a clear convergence
pattern for all values of $v$.
To be extremely conservative, we used the difference hi-low, plotted in red in
Fig.~\ref{fig:conv_errors}, as the convergence errors.
The vast majority of the values shown in Fig.~\ref{fig:conv_errors}
are significantly smaller than the RE error reported in Table~\ref{tab:beta_v}. 
These values were plotted as a complement to the results visible in the lower
panels of Fig.~\ref{fig:beta_v}, showing in detail the convergence errors
$\sigma_{\alpha^v}~\text{(conv.)}$ and $\sigma_{\beta^v}~\text{(conv.)}$ 
evaluated at different resolutions.  In the future, we may instead
perform independent Richardson extrapolation at each resolution; and
then check convergence of the Richardson extrapolants across
resolutions.

\figlnalpha
\figpeakevolren

The values of $\alpha^{v}$, $\beta^{v}$, and their Richardson
extrapolation errors all entered into the power law fits, producing
the optimal values in Table~\ref{tab:alpha_beta_fit} and estimated
``covariance matrices'' in Eqs.~\eqref{eq:num_cov_beta} and 
\eqref{eq:num_cov_alpha}.  These are not statistical
covariances, all being due to systematic errors; nonetheless we can
interpret them as Gaussian distributions in order to show the region
of the $(\ln v, \ln \alpha^{v})$ plane where the true $\alpha^{v}$ may
be found.  This is plotted in Fig.~\ref{fig:beta_props}, using the
package \texttt{fgivenx}~\cite{2018JOSS....3..849H}.  To produce a visible 
output, it was necessary to multiply the covariance matrix by a factor of
$500$ for $\ln\alpha^v$.  The same can be applied to find the possible
region for $\beta^{v}$ in the $(\ln v,\ln |\beta^v|)$ plane, in the
context of a pure power law $\beta$ function.

In the right panel of Fig.~\ref{fig:phi_ren_vs_full}, for
$\varepsilon=0.1$, we observe that the greatest difference between the
renormalized and the full solution comes from a shift in the peak
position $x_{\mathrm{c}}$.  Here we test whether this is due to errors
in our fits for
$\alpha^{v}$ and $\beta^{v}$, or due to truncation errors in the
time integration when solving the numerical DRG equations for
$(v_{\Ren}(t), x^{\Ren}_{c}(t))$.
To assess the importance of truncation error in the DRG time
integration, we performed the integration with $\Delta t = 0.5,~0.01$
and $0.001$, using the same underlying power-law fits for $\alpha^{v}$
and $\beta^{v}$ as reported in Table~\ref{tab:alpha_beta_fit}.  We did
not observe any visible differences between the outcomes for the
different time step choices. This is a clear indication of the
subdominance of the integration error.
To assess if the power law fit errors are under-reported, we changed
the number of points for fitting $\ln|\beta^v|$ as
a linear function of $\ln v$.  The left panel of Fig.~\ref{fig:beta_v}
already provides enough evidence of the linear
relation between these variables.  Therefore, in principle we only
require two points of the sample in Table~\ref{tab:beta_v} to
determine the coefficients $m$ and $b$.  We reconstructed $\beta^v$ as
a function of $v$ choosing two, six, and ten of the points in the table
(the points $v=0.125$ and $v=2$ are considered in all three cases, 
we did not include the extremum $v=0.0625$ in the case with 2 points 
since it has the largest relative error). We then integrated the coupled 
system in Eqs.~\eqref{eq:flow_eq_v} and \eqref{eq:new_peak} for all 
of the beta functions generated by each of these choices, and plot the 
results in Fig.~\ref{fig:error_xcs} (normalized by the width 
$\mathcal{W}_{\text{full}}$, computed as in Sec.~\ref{subsec:num_kdvb}).
The results are not affected by changing the number of points included
in the fit, again suggesting that the fits are not the responsible for
the error $\Delta x_{c}$.
Having ruled out either of these possibilities, we suspect that the
main source of this error comes from the mismatch between $\alpha^{v}$
and $\alpha^{v}_{\text{full}}$ discussed in
Sec.~\ref{subsec:num_kdvb}.

\bibliography{Bibnotes}

\end{document}